# Open-Source Benchtop Magnetophotometer (MAP) for Characterizing the Magnetic Susceptibility of Suspended Nanoparticles

*Alexis Scholtz[1,2], Jack Paulson[2,3], Victoria Nuñez[4], and Andrea M. Armani[1,2,3,*]*

[1]Alfred E. Mann Department of Biomedical Engineering, University of Southern California, Los Angeles, CA 90089, USA

[2]Ellison Medical Institute, Los Angeles, CA 90064, USA

[3]Mork Family Department of Chemical Engineering and Materials Science, University of Southern California, Los Angeles, CA 90089, USA

[4]Thomas Lord Department of Computer Science, University of Southern California, Los Angeles, CA 90089, USA

*aarmani@emila.org

**Abstract:** Magnetic nanoparticles (MNPs) form the foundation of many technologies, frequently serving to purify proteins or cells from a biological sample or to remove environmental contaminants. Their success relies on their magnetic response, which allows them to be easily controlled in a liquid or solution. Therefore, the magnetic susceptibility provides one metric for assessing the suitability of a MNP for a given application. Unfortunately, conventional methods for measuring the magnetic susceptibility relies on instrumentation that characterizes the MNPs as a dry powder. Because MNPs are typically used in suspension, the measured value may be different from their behavior in suspension, thus providing inaccurate readings. Here, we present the design and validation of a magnetophotometer (MAP), an instrument that characterizes the effective magnetic susceptibility of suspended MNPs via differential optical spectroscopy, providing a more relevant measure of MNPs' magnetic properties. As part of this work, we developed a mathematical model to calculate the effective magnetic susceptibility from the MAP data and validated the model using control measurements with iron oxide nanoparticles. Finally, we demonstrate that MAP testing is non-destructive by successfully characterizing bioconjugated particles without damaging the bioactivity of the surface bioconjugation, providing a path for in-line quality control assessment.

## Introduction

Magnetic nanoparticles (MNPs) serve as the foundation of many crucial nanotechnologies in fields such as medicine,[1–4] energy,[5–7] and environmental sciences.[8,9] Typically, MNPs consist of two to three layers, including a core magnetic material, a protective shell, and a surface modification, such as a chemical functional group or biological moiety. The combination of the magnetic core and surface functional group allows for selective and remote manipulation within a larger system. For example, MNPs are used in applications such as cell isolation, targeted drug delivery, and water purification.

When characterizing a batch of MNPs, there are key properties that should be considered, including physical and chemical characteristics. Typically, if batches are synthesized by academic researchers, these are analyzed after synthesis. Otherwise, researchers or industry members rely on the reported values from MNP manufacturers. However, despite the utility of MNPs relying on their magnetic properties, the magnetic characteristics are rarely measured or reported.

One magnetic property relevant to many applications of MNPs is the magnetic susceptibility, which quantifies how much of a material is magnetized by an external magnetic field.[10] Currently, the most common methods of magnetic characterization are using either a vibrating sample magnetometer (VSM)[11] or a superconducting quantum interference device (SQUID).[12] However, these instruments are expensive, require extensive training and expertise, and are difficult for the average researcher to access. They also require milligram-scale quantities of MNPs. In biomedical and therapeutics applications involving surface chemistries, this quantity can represent the entire aliquot volume. More importantly, these instruments analyze dried MNPs, measuring the $\chi_0$ of the bulk material.[10,12,13]

Past work has established that the magnetic susceptibility of a MNP can be influenced by the MNP packing fraction or the proximity of adjacent MNPs. As a result, powdered particles with a high packing fraction have different values than suspended particles in a dilute concentration. To bridge the gap, the effective magnetic susceptibility $\chi_{eff}$ was defined. This quantity scales with packing fraction by volume of magnetic material, $\phi$, such that $\chi_{eff} = \phi\chi_0$.[14,15] However, despite $\chi_{eff}$ providing a more relevant metric of suspended MNP behavior for many applications, there are not currently available instruments that can measure this quantity.

Here, we describe the design, theory, and validation of a magnetophotometer (MAP), a benchtop instrument that measures the $\chi_{eff}$ of suspended MNPs in a non-destructive manner using sub-milligram quantities of MNPs. The MAP quantifies $\chi_{eff}$ based on a modified differential optical spectroscopy method. To confirm the MAP operation, we measure a series of distinctly synthesized iron oxide nanoparticles and compare the values to SQUID control measurements. In addition, a second measurement using bioconjugated nanoparticles are performed and the non-destructive nature of the MAP is confirmed.

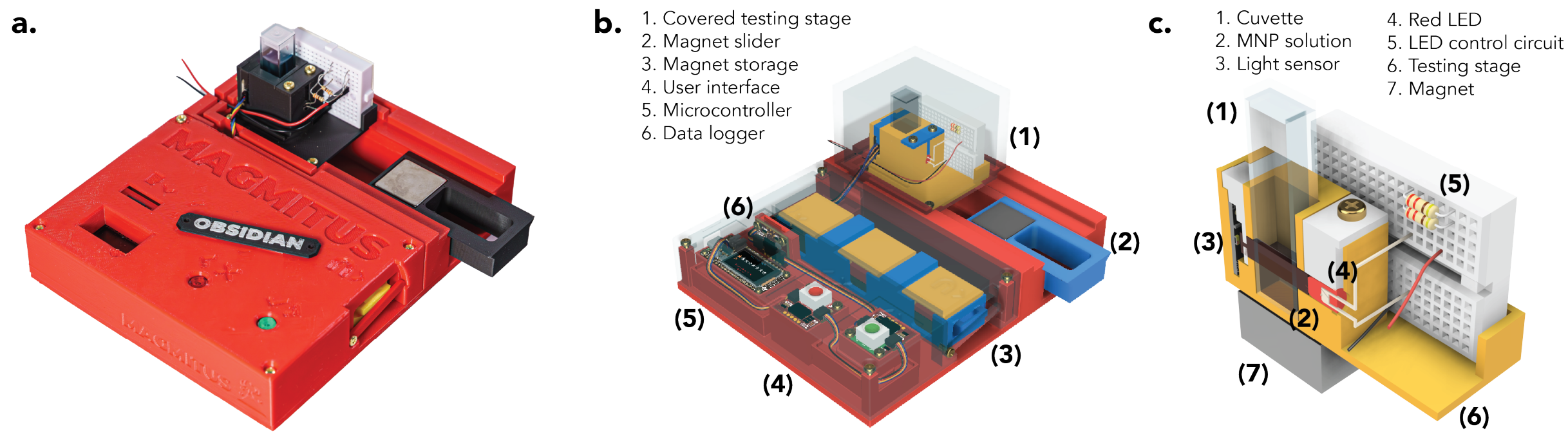

**Figure 1. a.** Picture of the actual MAP with the cover over the testing stage removed. **b.** Labeled rendering of the MAP with transparent covers. **c.** A cross-section rendering of the testing stage with key components labeled.

**Theory**

The detection principle of the MAP is single-wavelength time-dependent differential spectroscopy. Similar to conventional photospectroscopy methods, like UV-Vis spectroscopy, the change in optical intensity transmitted through a sample is correlated to the sample's concentration[16–18]. However, to simplify the footprint of the system, the MAP uses only a single wavelength of light. In addition, to determine the magnetic susceptibility, the system measures irradiance, which is related to optical intensity, continuously over time while a magnetic perturbation is applied. A brief overview of the theoretical principle follows.

When the sample is exposed to a static magnetic field, the concentration of MNPs in the measurement region decreases as the MNPs are pulled towards the magnet. Thus, we can directly correlate the optical signal and the movement of the MNPs. To use the system to analyze $\chi_{eff}$ of the MNPs, we derived a mathematical model that relates the irradiance over time to $\chi_{eff}$.

Beginning with first principles, the net force $F_{net}$ in the downward $z$-direction on a suspended MNP with an applied static magnetic field is a combination of the gravitational force $F_g$, drag force $F_d$, buoyant force $F_b$, and magnetic force $F_m$:

$$F_{net} = F_g + F_d + F_b + F_m. \qquad (1)$$

Expanded, this net force can be expressed as

$$F_{net} = mg - 6\pi\eta r\dot{z} - V\rho_s g + \frac{2a^2\chi_{eff}V}{\mu_0(1+\chi_s)}z + \frac{2ab\chi_{eff}V}{\mu_0(1+\chi_s)} + m_0 a, \qquad (2)$$

where $z$ and $\dot{z}$ are the position and velocity of the MNP in the $z$-direction. The particle's physical properties are represented by $m$, $r$, $V$, $\chi_{eff}$, and $m_0$ which are the mass, radius, volume, effective magnetic susceptibility, and initial magnetic moment of the MNP. The solvent's characteristics are represented by $\eta$, $\rho_s$, and $\chi_s$ which are the viscosity, density, and magnetic susceptibility of the solvent. The magnetic field (B) of the block magnet is approximated by the expression $B = az + b$, where $a$ and $b$ are the geometric and magnetic field parameters.[19–21] Lastly, fundamental constants include gravitational acceleration ($g$), and permeability of free space ($\mu_0$).

This expression can be easily converted into a time-dependent expression for the concentration $c(t)$ of particles in the sampling region, and we can apply three boundary conditions: (1) the MNP concentration in the sampling region will drop to zero over time, (2) the MNP concentration in the sampling region is some initial $c_0$, and (3) the net velocity of the MNPs is initially zero. This yields the fit equation:

$$c(t) = \kappa e^{\delta_1 t} + (c_0 - \kappa)e^{\delta_2 t}, \qquad (3)$$

where $\kappa$, $\delta_1$, and $\delta_2$ encapsulate the physical parameters of the system.

By converting the measured irradiance data to concentration using the Bouguer-Beer-Lambert Law[22] and fitting to eq 3, $\chi_{eff}$ in the SI unit system is obtained. Notably, a simple inspection of eq 2 reveals that there are several experimental variables that can be explored to validate this model. These include the magnetic field properties, the particle physical properties, and the solvent properties. The full derivation of the model, including a discussion of operational bounds[13,23–27] and details of the data processing, are in the SI.

## Experimental Materials and Methods

### Iron Oxide ($Fe_3O_4$) Nanoparticle Synthesis

Three batches of $Fe_3O_4$ MNPs were synthesized using a standard coprecipitation method (Batches 1, 2, and 3). Briefly, 1 g of iron(II) chloride and 0.4 g of iron(III) chloride were added to 20 mL of deionized (DI) water under argon. The solution was heated to 80°C, 5 mL of ammonium hydroxide was added dropwise, and the reaction was left to occur for 1 hour. Once cooled, the MNPs were isolated via magnetic separation and washed with DI water three times. A detailed protocol is in the SI.

### Bioconjugated MNPs (Dynabeads)

Dynabeads (Thermo Fisher Dynabeads M-280 Streptavidin) are MNPs with a streptavidin surface coating, and they are optimized to remove or isolate biotin from the solution. They were diluted to 100 μg/mL in Dulbecco's phosphate-buffered saline (DPBS) at a volume of 1 mL and washed once in DPBS. Otherwise, they were used as received.

### Fluorescein-labeled Biotin

Biotin-4-fluorescein (fl-biotin, Thermo Fisher) was dissolved in dimethylsulfoxide (DMSO) but otherwise used as received.

### $Fe_3O_4$ MNP Sample Preparation and Measurements

A summary of all conditions studied is presented in Table 1. MNP solutions were brought to room temperature to avoid temperature-related effects on $\chi$.[10] $Fe_3O_4$ MNPs were suspended in DI water at the concentration listed or serial dilutions are used with a starting concentration of 100 μg/mL. Each sample was sonicated for 5 minutes and vortexed for 1 minute to produce a uniform dispersion of MNPs[23] before adding 1 mL of sample into a standard 3.5-mL spectroscopy cuvette and inserting it into the MAP.

Irradiance measurements were taken continuously for 3 minutes and then converted to a concentration time series, which was fit as detailed in the SI to determine $\chi_{eff}$. Each sample was tested in triplicate and resuspended via aspiration between trials. Two controls were performed: water with no MNPs and MNPs with no applied magnetic field. Additional signal processing details are included in the SI. All magnets are referred to by their surface field for convenience, and a full list of magnets used in this work is in the SI.

**Dynabeads MAP and Biotin-binding Measurements**

Aliquots of Dynabeads were tested in the MAP following the standard protocols outlined in previous sections. Fl-biotin was added to the aliquots at one of four quantities (0, 1, 10, or 100 pmol) and incubated for 1 hour under gentle rotation. Magnetic separation using the 4933 G magnet was used to isolate the Dynabeads and bound fl-biotin, and 100 µL of the supernatant (10% of the total sample) was removed from the cuvette. The removed supernatant was tested for fluorescence from unbound fl-biotin at 523 nm with excitation at 494 nm using a Molecular Devices SpectraMax M2e. Control measurements included solvent only (a mixture of DPBS and DMSO), fl-biotin solution without Dynabeads, and Dynabeads incubated with fl-biotin without the initial MAP testing step. All measurements were performed in triplicate, and full experimental details are in the SI.

**Table 1.** Summary of experimental variables tested with the MAP.

| **Material** | **Concentration** | **Magnetic Field** |
|---|---|---|
| $Fe_3O_4$ batch 1, 2, and 3 | 0, 100 µg/mL | 0, 4933 G |
| $Fe_3O_4$ batch 1 | 0-500 μg/mL | 0, 4933 G |
| $Fe_3O_4$ batch 1 | 0, 100 μg/mL | 0-4933 G |
| Dynabeads | 100 μg/mL | 4933 G |

**SQUID Sample Preparation and Measurements**

$Fe_3O_4$ MNPs from the same three batches were tested in a SQUID. A few milligrams of dried MNPs were fixed in an eicosane matrix in the sample holder and tested on a Quantum Design MPMS 3. Magnetic moment of the sample was measured during a 5-quadrant applied magnetic field sweep between ± 5000 Oe. The linear fit at low fields (20-250 Oe) normalized to volume was used to calculate $\chi$, reported in this work in the SI unit system. The full data and analysis are available in the SI.

**Results and Discussion**

**System Design**

The MAP was designed as an open-source, benchtop instrument to measure the $\chi_{eff}$ of suspended MNPs. A 620-nm red LED and a light sensor form an optical path through a cuvette, with a pinhole restricting the sampling window to a 1 mm x 1 mm region centered around a height of 9.5 mm from the bottom of the cuvette (Fig. 1). A neodymium block magnet mounted on a linear slide rail applies the external magnetic field directly under the cuvette in a reproducible manner. The MAP has an embedded control and data acquisition system with an integrated user interface, allowing the MAP to run as a standalone instrument, although data analysis must occur on an external computer.

A fully packaged system was created with custom 3D-printed housing, giving the integrated system a footprint of approximately 6” x 7”. The full details of the MAP design can be found in the SI.

**Effective Magnetic Susceptibility Measurements**

To validate the MAP's performance, we first compared the $\chi_{eff}$ values of the suspended sample measured by the MAP to the $\chi$ values of the powdered sample obtained from SQUID measurements on three different synthesized batches of $Fe_3O_4$ MNPs. Representative data from the MAP are shown in Figure 2a, including MNPs with an applied magnetic field and two controls: water and MNPs with no applied field.

The control measurements provide the upper and lower bound values for the experiment. The control solution with no MNPs provides information on the fundamental noise and potential signal drift of the system over the course of the measurement. For this data set, the linear system drift is $-1.15 \times 10^{-3}$ $W/m^2 \cdot s$, and the $3\sigma$ noise is 0.0506 $W/m^2$. These values set the minimum detection limits of the system, and the absolute signal value serves as an upper bound, or largest possible signal change achievable, for the MAP system.

The second control measurement analyzed a solution with MNPs but no magnetic field. This measurement provides information on the noise that nanoparticles could induce in the signal. Additionally, it is expected that the signal would increase over the course of the measurement due to the effect of gravity on the particles. For this data set, the change in signal over time is $1.27 \times 10^{-2}$ $W/m^2 \cdot s$, and the $3\sigma$ noise is 0.243 $W/m^2$. This higher noise relative to the MAP test with just water is expected due to the presence of particles in this sample. Thus, gravitational effects are clearly observed, confirming one component in the model. This absolute value of this signal is the lower bound. Additional instrument-level characterization was performed, and the results are included in the SI[28].

In the solution with the MNPs exposed to the magnetic field, we observe that the measured irradiance spans from the lower to the upper bound (Figure 2a), indicating that the magnetic field removes nearly all the MNPs from the sampling region. This change is also evident upon visual inspection of the cuvette before and after testing (Figure 2a inset). The detection signal exceeds the system noise and drift values obtained by the control measurements by at least one order of magnitude. The raw time-dependent irradiance data from the MAP was transformed to time-dependent concentration and then fit to eq 3 via processes detailed in the SI (Figure 2b). This fit yielded a prediction of $\chi_{eff} = 5.15$ for this batch of MNPs with an $r^2 = 0.987$, indicating a strong agreement with the theoretical model.

The $\chi_{eff}$ for all MNP batches were determined using this approach, and the results are shown in Figure 2c, with the exact values given in Table S1.1 in the SI. The SQUID results of $\chi$ for each batch are also presented in Figure 2c for direct comparison. Crucially, the $\chi$ measured by the SQUID and $\chi_{eff}$ reported by the MAP both fall within the expected range for bulk magnetite ($Fe_3O_4$).[29] Notably, as expected, there were variations in $\chi$ and $\chi_{eff}$ between the different synthetic batches that were detected by both the SQUID and the MAP.[30,31] For both instruments, the measured $\chi$ or $\chi_{eff}$ trended in the same direction between batches, with batch 1 displaying the strongest magnetic response and batch 3 the lowest. This demonstrates one application of the MAP for use in quality control of different synthetic batches of the same type of MNP.

As previously discussed, we do not expect the SQUID and MAP measurements to match. Instead, the SQUID measurements of $\chi$ should be viewed as an upper bound on the MAP measurements of $\chi_{eff}$ because the $\chi$ measured by the SQUID represents the situation in which the packing fraction of the magnetic material is approximately 1.[14] In contrast, the MAP measures the $\chi_{eff}$ of suspended MNPs, which has a significantly lower packing ratio and can also be subject to the known demagnetization effects of MNP suspensions.[14] Thus, because of the differing states of the tested particles, we expect that MAP will measure a lower value, as seen in Figure 2c. However, for many applications, the particles are used in suspension, and thus, the MAP provides a more accurate characterization of the MNP response.

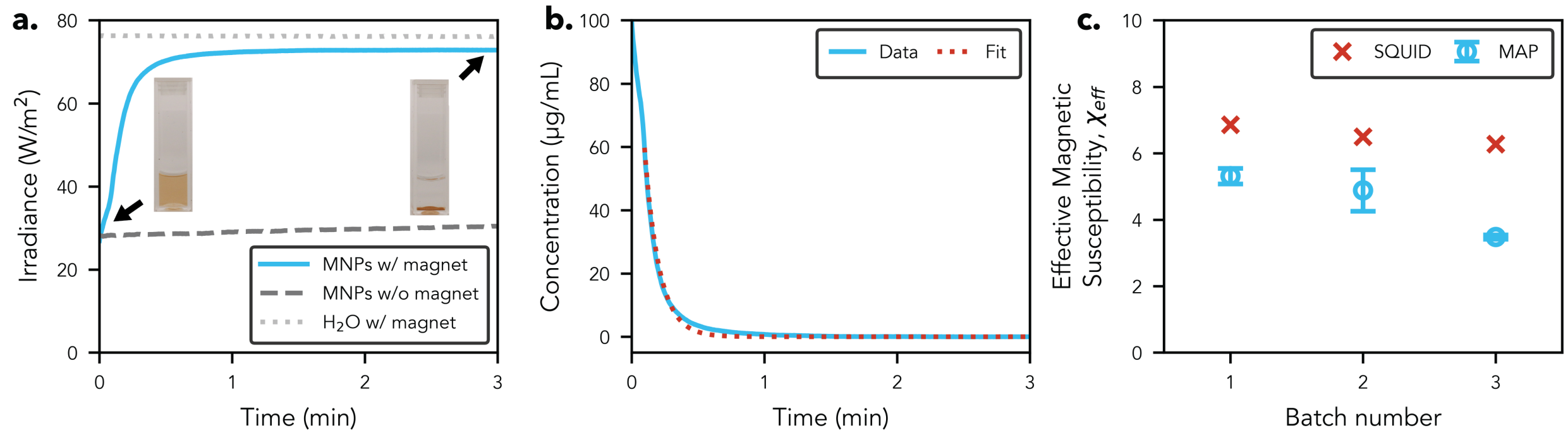

**Figure 2. a.** MAP irradiance data from $Fe_3O_4$ MNPs from batch 1 in water tested both with (blue) and without (red dashed) an applied magnetic field. The results from a control of water tested with an applied magnetic field (dotted gray) is also shown. **(inset)** Pictures of the 3.5 mL cuvettes at the beginning and end of MAP testing, corresponding to initial and final sensed concentrations of MNPs. **b.** An example of calibrated and transformed concentration data from part a (blue line) and its fit to Eq. (6) (red dotted line) after pre-processing, as described in the SI. **c.** The $\chi_{eff}$ measured by the MAP and the $\chi$ measured by the SQUID for each particle batch, all in the SI system.

**Measurements of Varied Concentrations**

To investigate the effects of concentration on $\chi_{eff}$, we tested a range of concentrations from 1 to 500 μg/mL of batch 1 $Fe_3O_4$ MNPs. A subset of the data is shown in Figure 3a, with the full dataset including all concentrations and replicates included in the SI. Both the initial irradiance as well as the overall change in irradiance during a MAP test was proportional to the MNP concentration, as expected from the Bouguer-Beer-Lambert law. Furthermore, the irradiance stabilized to approximately the same irradiance measured from just water with no MNPs for all trials except one (31 μg/mL). This difference could be due to scattering due to cuvette placement or a fluctuation in LED power. However, should this become a significant problem, simple modifications to the instrument could be made to correct these changes, as described in the SI. This outlier emphasizes the importance of analyzing the data in a differential manner and indicates that our calibration and fitting processes are successful because the $\chi_{eff}$ calculated from this trial was still comparable to the other measurements (Figure 3b-c).

This concentration sweep facilitated further sensor characterization of the MAP. Figure 3b shows the overall change in irradiance measured during a MAP test relative to the initial MNP concentration. Looking first at the limits of the system, the average change in irradiance for 1 µg/mL suspensions was 1.15 W/m$^2$, three-fold greater than the 3$\sigma$ noise, indicating that the limit of detection of the MAP is below 1 µg/mL. Saturation appeared to occur between 125 and 250 µg/mL, with aggregation of the MNPs occurring at concentrations greater than 250 µg/mL, causing the dip in the overall change in irradiance above this concentration. By fitting this data to a sigmoid, we established the linear working range of the MAP to be approximately one order of magnitude from about 16 µg/mL to 125 µg/mL. Although these metrics will vary depending on the individual MAP and MNPs, this analysis provides a guideline for optimizing other MAP systems and types of MNPs.

The data from the working range of the MAP was analyzed to determine $\chi_{eff}$ for each concentration (Figure 3c). The SQUID $\chi$ measurement for this batch of MNPs is shown as a dotted line, because it is independent of concentration. There was an increasing trend in $\chi_{eff}$ as the concentration increases, approaching the $\chi$ measured by the SQUID, consistent with the scaling of $\chi_{eff}$ with packing fraction.[14] This insight into $\chi_{eff}$ could facilitate optimization of MNP performance in suspensions based on concentration, information that existing magnetometers cannot provide. These results further reinforce the idea that measuring the magnetic properties of the MNPs in their final application environment is essential.

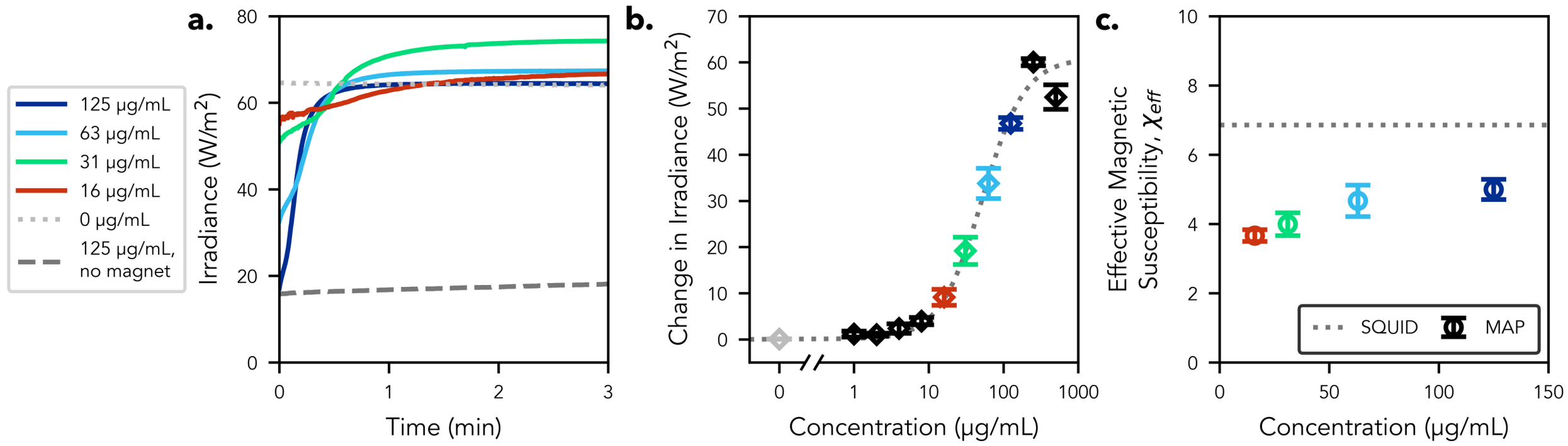

**Figure 3.** A sweep of multiple concentrations from 1-500 μg/mL of batch 1 $Fe_3O_4$ MNPs, tested with the 4933 G surface field magnet, each experiment was run in triplicate. **a.** The MAP irradiance data from one trial of selected concentrations from the linear range for this particle batch and of both control measurements: water and MNPs with no applied magnet. **b.** The sensor response curve showing the measured change in irradiance relative to the initial MNP concentration, fit with a sigmoid function (dashed gray). **c.** The $\chi_{eff}$ measured by the MAP compared to the $\chi$ measured by a SQUID, all calculated using the SI unit system, for several concentrations in the linear working range.

### Measurements with Varied Applied Magnetic Fields

In addition to varying the concentration, we also varied the applied magnetic field to validate our theoretical model. Solutions of 100 μg/mL of $Fe_3O_4$ MNP batch 1 were tested by applying 5 different magnets with surface fields ranging from 2108 G to 4933 G as well as no magnet (0 G) as a control. Representative data is shown in Figure 4a, with the full dataset available in the SI. For all non-zero applied fields, the irradiance signal stabilized to approximately the same value, which was expected since the same concentration is used in each trial.

To quantify the effect of the magnetic field on the response, we determined the response time for each field strength or the time for the signal to reach 90% of the overall change. The response time directly increased with magnetic field, consistent with the model (Figure 4b). As the magnetic field strength increases, a stronger magnetic force is experienced by the MNPs, corresponding to a more rapid increase in signal. Notably, the 90% response times were all less than 1 minute, and the full signal change was captured in the 3-minute test duration used in this work. A 3-minute test represents a time savings of 99.8% relative to the typical 24-hour protocol used for a SQUID or VSM, reducing the time to measurement result by orders of magnitude.

Figure 4c shows the results of the MAP analysis. The SQUID $\chi$ is shown with a dotted line since the SQUID measures $\chi$ by sweeping over applied magnetic fields, and $\chi$ is independent of the applied magnetic field for the low field strengths used in this work. The $\chi_{eff}$ measured by the MAP did not vary in a statistically significant manner with the applied magnetic field (1-way ANOVA with p = 0.619), indicating that our theoretical model correctly accounts for this variable.

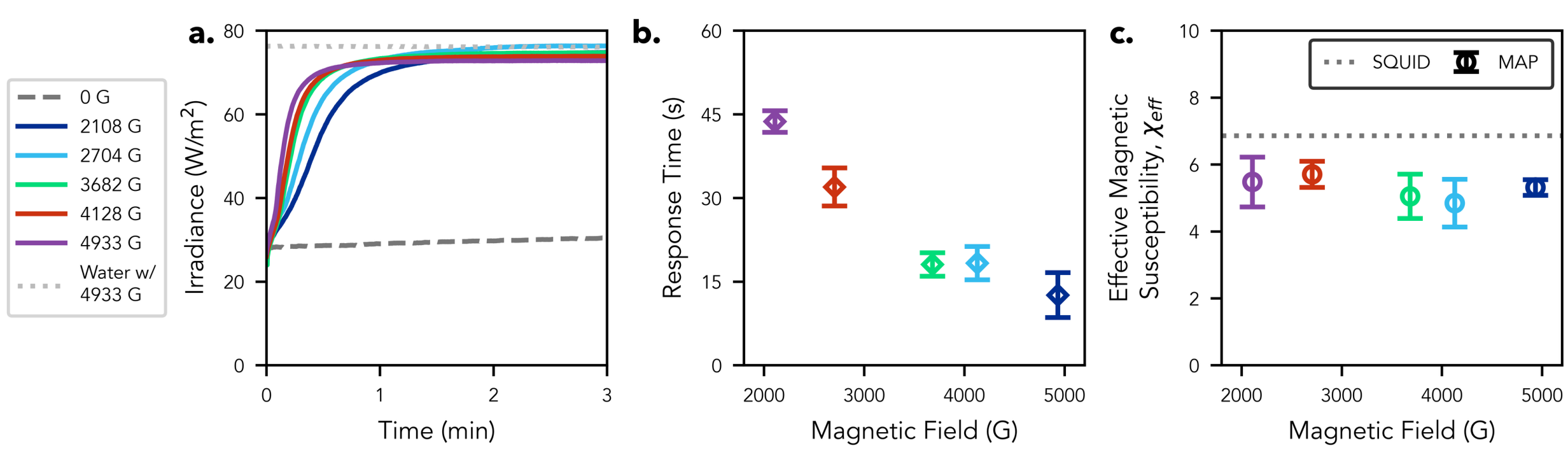

**Figure 4.** A sweep of multiple applied magnetic fields ranging from 0 to 4933 G on batch 1 $Fe_3O_4$ MNPs prepared at a concentration of 100 µg/mL in water; each measurement was run in triplicate. **a.** The MAP irradiance data from one trial of all magnetic field strengths, including the control with no magnetic field and with just water. **b.** The response time (the time for signal to reach 90% of its overall change in value) relative to magnetic field strength, showing a decreasing response time with increasing field strength. **c.** The $\chi_{eff}$ determined by the MAP compared to the $\chi$ measured by a SQUID, all calculated using the SI unit system, for all magnetic fields strengths.

**Non-Destructive Testing of Surface-Coated MNPs**

To confirm that MAP testing does not damage biological surface coatings during a measurement, we analyzed the $\chi_{eff}$ of streptavidin-coated Dynabeads and then assessed their ability to bind and remove fluorescent biotin (fl-biotin) after testing. An overview of the approach is illustrated in Figure 5a.

Aliquots of Dynabeads were either initially tested in the MAP using the standard protocol (MAP-tested) or remained untested and then incubated with fl-biotin. Then, the Dynabeads along with any bound fl-biotin were removed from the solution via magnetic separation with the 4933 G magnet and the fluorescence of the supernatant was measured. As long as the MAP does not change the bioactivity of the bioconjugation, both sets of Dynabeads will bind the fl-biotin with the same efficiency, and the concentration of fl-biotin in the solution will decrease.

The fluorescence of the supernatent should be lower than that of a solution containing only fl-biotin since the Dynabeads are designed to isolate and remove fl-biotin. This behavior was demonstrated at each quantity of added fl-biotin, indicating that both MAP-tested and untested Dynabeads successfully bound the fl-biotin. At the two higher quantities of fl-biotin (10 and 100 pmol), the fluorescence of the Dynabead samples decreased by an average of 57% and 78% for MAP-tested Dynabeads and 57% and 79% for untested Dynabeads when compared to the control fl-biotin solution. These measured fluorescent values were significantly greater than the noise floor of 2 a.u., which was established based on the fluorescence of the PBS/DMSO solvent mixture (the control), indicating there was still residual unbound fl-biotin. Thus, the streptavidin binding sites on the Dynabeads were saturated at the higher quantities of fl-biotin. In contrast, for the lowest quantity of fl-biotin, the fluorescence in the Dynabeads samples were much closer to the fluorescence level of the noise floor, indicating that the Dynabeads successfully bound all the fl-biotin added to the sample.

Importantly, the performance of the Dynabeads was not affected by initial MAP testing. As can be seen in Figure 5b, the fluorescence of samples with untested Dynabeads and with MAP-tested Dynabeads were not statistically significant at any of the three quantities of added fl-biotin, indicating that MAP testing does not affect the ability of Dynabeads to bind proteins and remove them from solution.

Lastly, using the MAP data taken during this measurement series (Figure 5a and Figure S1.10), the average effective magnetic susceptibility of this batch of Dynabeads was calculated to be 2.63 ± 0.09. This value is comparable to our previously measured values for iron oxide which is the reported magnetic core material in Dynabeads. Importantly, the manufacturer does not report a value, and other methods of characterizing particles, such as VSM and SQUID magnetometers, require drying the particles and adding an immobilizing matrix to the MNPs during sample preparation. This sample preparation process both destroys the surface coatings of the MNPs and prevents further use of the MNPs. For the many commercial MNPs sold with pre-functionalized surfaces for ease of use, the MAP provides a method to perform initial and nondestructive quality control measurements prior to using the particles.

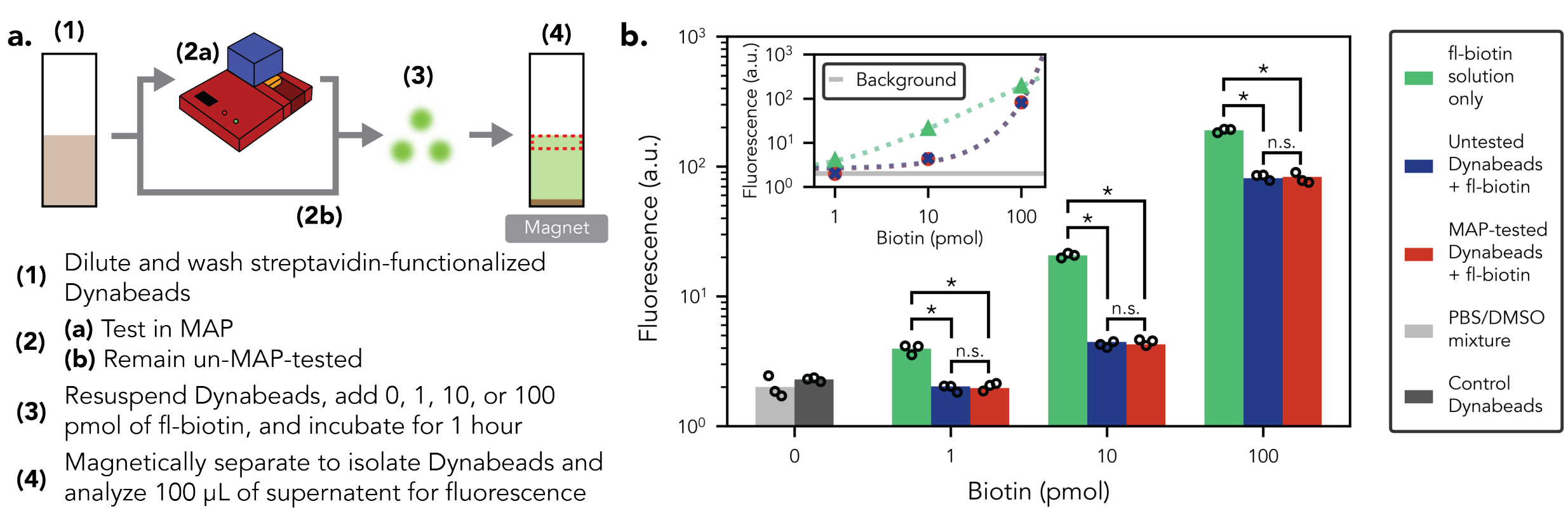

**Figure 5. a.** To confirm that MAP testing does not damage the surface coatings of Dynabeads, (1) 1 mL aliquots of 100 μg/mL Dynabead solutions were either (2a) MAP-tested in triplicate according the protocol used in previous measurements or (2b) not tested (untested). (3) Both sets of particles were then incubated with 1, 10, or 100 picomoles of fl-biotin. (4) Magnetic separation isolated the Dynabeads

bound to fl-biotin from the free fl-biotin; 100 µL of supernatant was removed and its fluorescence measured in arbitrary units (a.u.). **b.** Average fluorescence of free fl-biotin solutions (teal), the supernatant of MAP-tested Dynabeads (red), and the supernatant of untested Dynabeads (blue). Controls of solvent (light gray) and Dynabeads alone (dark gray) were also tested. Student's t-test results: *statistically significant difference between group averages with $p < 0.01$; n.s. no significant difference found. **(inset)** The average fluorescence at each biotin quantity plotted with a trendline. The noise floor as established from the pair of control measurements is also plotted as a gray line.

## Conclusions

Here, we have described the design, development, and validation of an open-source, benchtop MAP capable of measuring the effective magnetic susceptibility of suspended MNPs. We derived a mathematical model to extract $\chi_{eff}$ from continuous irradiance measurements while perturbing a suspension of MNPs with an external magnetic field. Using synthesized $Fe_3O_4$ MNPs, we demonstrated that the MAP provides relevant measurements of $\chi_{eff}$ for many applications of MNPs used in suspensions. Finally, we confirmed that MAP testing does not damage surface coatings of MNPs by measuring the protein-binding performance of bioconjugated Dynabeads, a standard commercial MNP.

Furthermore, the design of the MAP gives it inherent advantages over existing magnetometers: the MAP is portable, does not require external cooling, requires orders of magnitude less sample, has orders of magnitude shorter testing times, and is significantly easier to use, requiring only minimal knowledge of electromagnetics. The MAP was also designed following open-source principles and the fully published design facilitates easy modification or adaption of the MAP to meet emerging and evolving needs. The increased accessibility and utility of the instrument, especially in the significant reduction of cost relative to existing magnetometers and its ability to characterize MNPs in scientifically-relevant environments, will enable researchers to perform valuable quality control measurements prior to using them in downstream applications, improving the rigor of scientific research in a variety of fields.[32]

**Supporting Information**

- Full theoretical model derivation, experimental details, full datasets for all validation data, and full sensor characterization (PDF)
- MAP design documentation, including a bill of materials, description of all components, user interface description, assembly information, code installation instructions, and troubleshooting guide (PDF)

**Notes**

All design materials and documentation required to assemble a MAP system are available at: https://github.com/armanilab/MAP.

The authors declare the following competing financial interest(s): A.M.A. serves as the Senior Director of Engineering and Physical Sciences for the Ellison Medical Institute (paid position).

**Acknowledgments**

The authors acknowledge the Office of Naval Research (N00014-24-1-2296), the National Science Foundation (DBI-2414158), and the Ellison Medical Institute. The authors would like to thank Armando Urbina and Ruojiao Sun for synthesizing and characterizing the iron oxide nanoparticles and Dr. Kylie Trettner for advice on statistical methods.

# Supplementary Information 1: Theory, Validation, and Characterization

# Open-Source Benchtop Magnetophotometer (MAP) for Characterizing the Magnetic Susceptibility of Nanoparticles

*Alexis Scholtz[1,2], Jack Paulson[2,3], Victoria Nuñez[4], and Andrea M. Armani[1,2,3,*]*

[1]Alfred E. Mann Department of Biomedical Engineering, University of Southern California, Los Angeles, CA 90089, USA

[2]Ellison Medical Institute, Los Angeles, CA 90064, USA

[3]Mork Family Department of Chemical Engineering and Materials Science, University of Southern California, Los Angeles, CA 90089, USA

[4]Thomas Lord Department of Computer Science, University of Southern California, Los Angeles, CA 90089, USA

*aarmani@emila.org

## Table of Contents

## 1.1. First Principles Model Derivation

The primary goal of the theoretical model is to develop a mathematical framework to understand how changes in measured light over time caused by magnetic nanoparticle (MNP) motion in a fluid due to a static magnetic field can be used to determine the magnetic properties of the MNP. Specifically, we want to calculate the effective magnetic susceptibility of our particles, $\chi_{eff}$. Thus, we derived a first principles model, beginning with the forces on a single MNP and then scaling to the multi-particle system present in the MAP (Figure S1.1).

### 1.1.1 Force on a Single Particle

We began with a simple free-body diagram of a single particle (Figure S1.1b). The free-body diagram includes all forces present in the system: buoyancy $\vec{\boldsymbol{F}}_{\boldsymbol{b}}$, drag $\vec{\boldsymbol{F}}_{\boldsymbol{d}}$, gravity $\vec{\boldsymbol{F}}_{\boldsymbol{g}}$, and applied magnetic force $\vec{\boldsymbol{F}}_{\boldsymbol{m}}$. The total force that a single particle experiences can thus be written generally as:

$$\vec{\boldsymbol{F}}_{\boldsymbol{total}} = \vec{\boldsymbol{F}}_{\boldsymbol{T}} = \vec{\boldsymbol{F}}_{\boldsymbol{b}} + \vec{\boldsymbol{F}}_{\boldsymbol{d}} + \vec{\boldsymbol{F}}_{\boldsymbol{g}} + \vec{\boldsymbol{F}}_{\boldsymbol{m}}. \tag{S1.1}$$

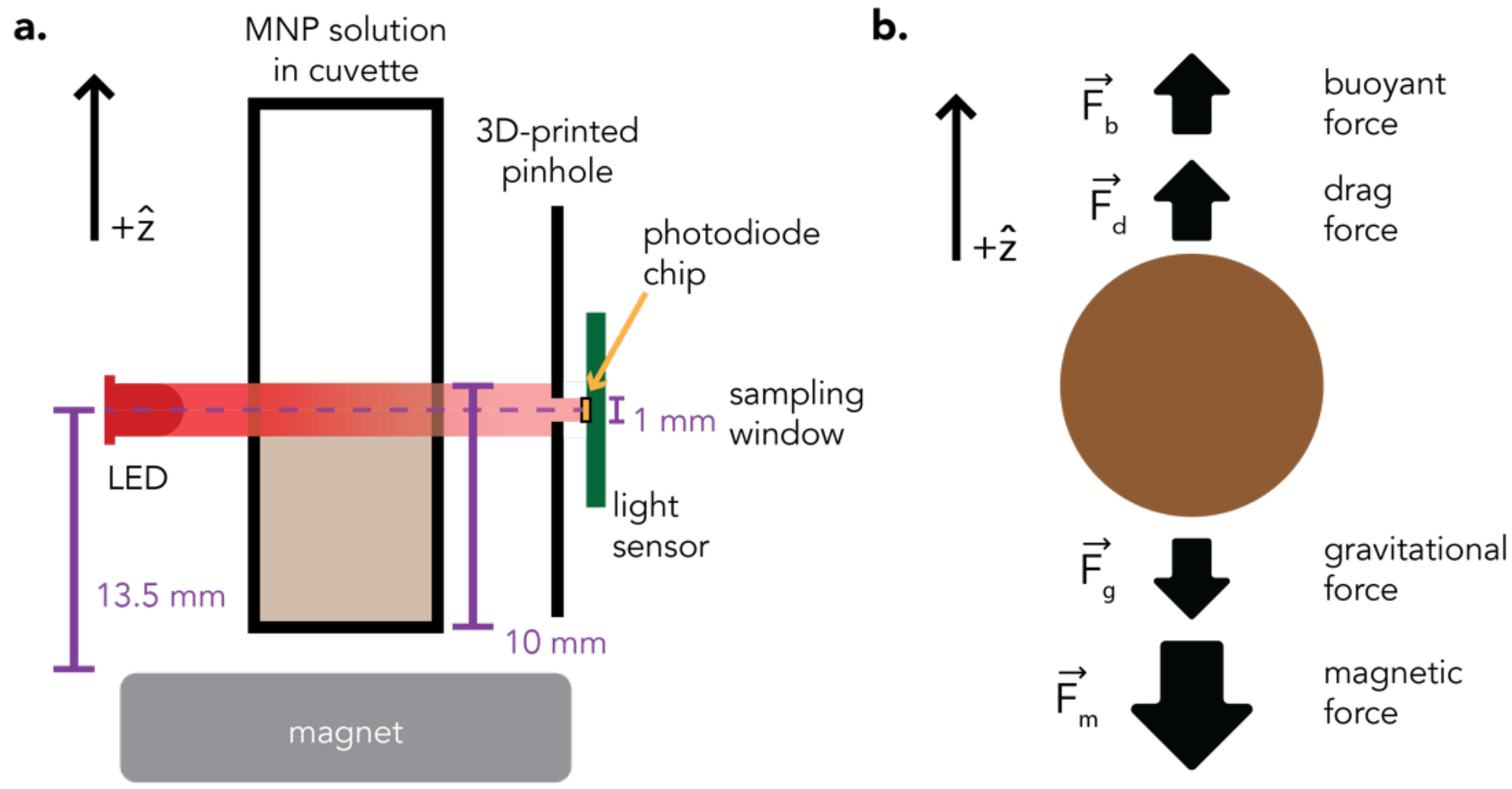

**Figure S1.1. a.** A simplified schematic showing the testing setup with the optical components and the magnet. Key dimensions are labeled in purple. **b.** Free-body diagram of all the forces acting on a single nanoparticle.

In the present system, the magnetic force is only applied along the $z$-axis (Figure 1, main text). By selecting magnets with a surface (2.54 cm x 2.54 cm) significantly larger than the cuvette (1 cm x 1 cm), any stray magnetic forces due to non-uniformities in the magnet would be uniform in the $x$- and $y$-directions. However, the dominant gradient force is in the $z$-direction. Additionally, given the sensor configuration, the system can only detect changes in $z$-axis position. Therefore, this expression can be simplified to a single axis (the $z$-axis). To improve readability, we will drop the vector notation.

The buoyant, drag, and gravitational forces are well defined. The force due to gravity, acting in the negative $z$-direction, is given by:

$$F_g = -m_p g = -\rho_p V g, \tag{S1.2}$$

where $m_p$ is the mass of the particle, which is equivalent to the density of the particle $\rho_p$ multiplied by the volume of the particle $V$, and $g$ is the gravitational acceleration. Similarly, the buoyant force is given by:

$$F_b = \rho_s V g, \tag{S1.3}$$

and will act in the positive z-direction on the particle, where $\rho_s$ is the density of the solvent.

The MNPs are assumed to be spherical with diameters with sub-micron diameters. Given the velocities involved, their motion falls in the low Reynolds number regime, and the drag force can be assumed to be Stokes' drag:

$$F_d = 6\pi\eta r\dot{z}, \tag{S1.4}$$

where $\eta$ is the dynamic viscosity of the solvent, $r$ is the effective radius of the particle, and $\dot{z}$ is the velocity of the particle moving in the $z$-direction.[1] Note even if the particles form chains or cylindrical clumps instead of remaining as spherical particles, Stokes drag will be assumed to apply because of the extremely low Reynolds number. Drag will always oppose overall motion of the particle. In our system, gravitational force will exceed buoyant force because the particles will be much denser than the solvent and the neodymium magnet will apply a strong downwards magnetic force. As a result, there is a strong net force in the negative $z$-direction. Thus, we can expect the drag force to be in the positive $z$-direction.

Therefore, we can write the net force as:

$$F_{net} = \rho_s V g + 6\pi\eta r\dot{z} - \rho_p V g - F_m. \tag{S1.5}$$

The magnetic force $F_m$ will act in the downwards $z$-direction towards the magnet. The derivation of the magnetic force is nontrivial and will be discussed in the next section in detail.

### 1.1.2. Magnetic Force

To understand the force acting on a nanoparticle by an external magnetic field, we must first understand how the force relates to the magnetic field. Beginning from first principles, the magnetic force $\vec{\boldsymbol{F}}_{\boldsymbol{m}}$ by an applied magnetic field $\vec{\boldsymbol{B}}$ on an MNP may be defined as:

$$\vec{\boldsymbol{F}}_{\boldsymbol{m}} = \vec{\nabla}(\vec{\boldsymbol{m}} \cdot \vec{\boldsymbol{B}}), \tag{S1.6}$$

where $\vec{\boldsymbol{m}}$ is the magnetic dipole moment of the particle.

Therefore, we need to write an expression for the magnetic moment $\vec{\boldsymbol{m}}$ of the MNP. We also know that the magnetization $\vec{\boldsymbol{M}}$ of the MNP can be defined in two ways:

$$\vec{\boldsymbol{M}} = \frac{\vec{\boldsymbol{m}}}{V} = \chi_{eff}\vec{\boldsymbol{H}} + \vec{\boldsymbol{M}}_{\boldsymbol{0}}, \tag{S1.7}$$

where, again, $V$ is the volume, $\chi_{eff}$ is the effective magnetic susceptibility, and $\vec{\boldsymbol{M}}_{\boldsymbol{0}}$ is the initial magnetization of the MNP. $\vec{\boldsymbol{H}}$ can be thought of as the auxiliary magnetic field. Note the magnetic field inside the magnet will be neglected because it is irrelevant to this application. Rearranging this equation gives us an expression for $\vec{\boldsymbol{m}}$:

$$\vec{\boldsymbol{m}} = \chi_{eff}V\vec{\boldsymbol{H}} + V\vec{\boldsymbol{M}}_{\boldsymbol{0}}. \tag{S1.8}$$

By defining the initial magnetic moment $\vec{\boldsymbol{m}}_{\boldsymbol{0}}$ as $\vec{\boldsymbol{m}}_{\boldsymbol{0}} = V\vec{\boldsymbol{M}}_{\boldsymbol{0}}$, we can simplify this equation to:

$$\vec{\boldsymbol{m}} = \chi_{eff}V\vec{\boldsymbol{H}} + \vec{\boldsymbol{m}}_{\boldsymbol{0}}. \tag{S1.9}$$

Note that we do not know $\vec{\boldsymbol{m}}_{\boldsymbol{0}}$. For example, if the particles are superparamagnetic, we can assume that they have no initial magnetization (i.e. $\vec{\boldsymbol{M}}_{\boldsymbol{0}} = 0$) and thus no initial magnetic moment (i.e. $\vec{\boldsymbol{m}}_{\boldsymbol{0}} = 0$), but to keep this derivation general, we will not make this assumption.

The auxiliary magnetic field $\vec{H}$ must propagate through the solvent to reach the MNPs. To be thorough, we must consider how any magnetization of the solvent may affect the effective magnetic field experienced by the MNP. Assuming that our solvent is a linear media, we can write:

$$\vec{\boldsymbol{B}} = \mu_s\vec{\boldsymbol{H}}, \tag{S1.10}$$

where $\mu_s$ is the permeability of the solvent. This permeability can be defined in terms of the permeability of free space $\mu_0 = 4\pi \times 10^{-7}$ as $\mu_s = \mu_0(1 + \chi_s)$ where $\chi_s$ is the magnetic susceptibility of the solvent. Thus, we can write:

$$\vec{\boldsymbol{B}} = \mu_0(1 + \chi_s)\vec{\boldsymbol{H}}. \tag{S1.11}$$

Rearranging eq S1.11 to write an expression for $\vec{\boldsymbol{H}}$ yields:

$$\vec{\boldsymbol{H}} = \frac{\vec{\boldsymbol{B}}}{\mu_0(1+\chi_s)}. \tag{S1.12}$$

Substituting eq S1.12 into the equation for the magnetic moment of the particle eq S1.9, we get:

$$\vec{\boldsymbol{m}} = \chi_{eff} V \frac{\vec{\boldsymbol{B}}}{\mu_0(1+\chi_s)} + \vec{\boldsymbol{m}}_{\boldsymbol{0}}. \tag{S1.13}$$

Returning to our initial definition of magnetic force in eq S1.6, we can now expand $\vec{\boldsymbol{m}}$:

$$\vec{\boldsymbol{F}}_{\boldsymbol{m}} = \vec{\boldsymbol{\nabla}}\left[\left(\frac{\chi_{eff} V}{\mu_0(1+\chi_s)}\vec{\boldsymbol{B}} + \vec{\boldsymbol{m}}_{\boldsymbol{0}}\right) \cdot \vec{\boldsymbol{B}}\right] \tag{S1.14}$$

and distribute the dot product:

$$\vec{\boldsymbol{F}}_{\boldsymbol{m}} = \vec{\boldsymbol{\nabla}}\left[\frac{\chi_{eff} V}{\mu_0(1+\chi_s)}\vec{\boldsymbol{B}} \cdot \vec{\boldsymbol{B}} + \vec{\boldsymbol{m}}_{\boldsymbol{0}} \cdot \vec{\boldsymbol{B}}\right]. \tag{S1.15}$$

Because the magnetic force only applies in the $z$-direction, we may simplify and drop vector notation:

$$F_m = \frac{\partial}{\partial z}\left[\frac{\chi_{eff} V}{\mu_0(1+\chi_s)} B^2 + m_0 B\right]. \tag{S1.16}$$

Note that eq S1.16 matches the form derived in past work on modeling the magnetic force in paramagnetic MNPs done by Gijs[2] and superparamagnetic MNPs done by Shevkoplyas et al.[3]

Now we need to define our applied magnetic field $B$. The magnets used in the MAP design are block magnets sourced from K&J Magnets, Inc. For this block geometry, the magnetic field in the $z$-direction at a distance $z$ away from the surface of the magnet is given by:

$$B = \frac{B_r}{\pi}\left[\arctan\left(\frac{LW}{2z\sqrt{4z^2+L^2+W^2}}\right) - \arctan\left(\frac{LW}{2(z+T)\sqrt{4(z+T)^2+L^2+W^2}}\right)\right]. \tag{S1.17}$$

$B_r$ is the residual flux density, a physical parameter dependent on the grade of neodymium and should be provided by the magnet manufacturer. For two common grades, N42 and N52 neodymium magnets, $B_r$ is 1.32 T and 1.48 T respectively. $L$, $W$, and $T$ are the length, width, and thickness, respectively, of the block magnet. In the system, the magnets are oriented such that the thickness is in the $z$-direction (i.e. the $L$-$W$ plane is perpendicular to the $z$-axis).

Equation S1.17 is illustrated in Figure S1.2a. Note that the horizontal axis is the distance $z$ from the surface of the magnet. Furthermore, the light sensor used to collect the light response behavior only integrates over a small sampling window, as indicated in Figure S1.1a. In this region, the magnetic field is approximately linear (Figure S1.2b).

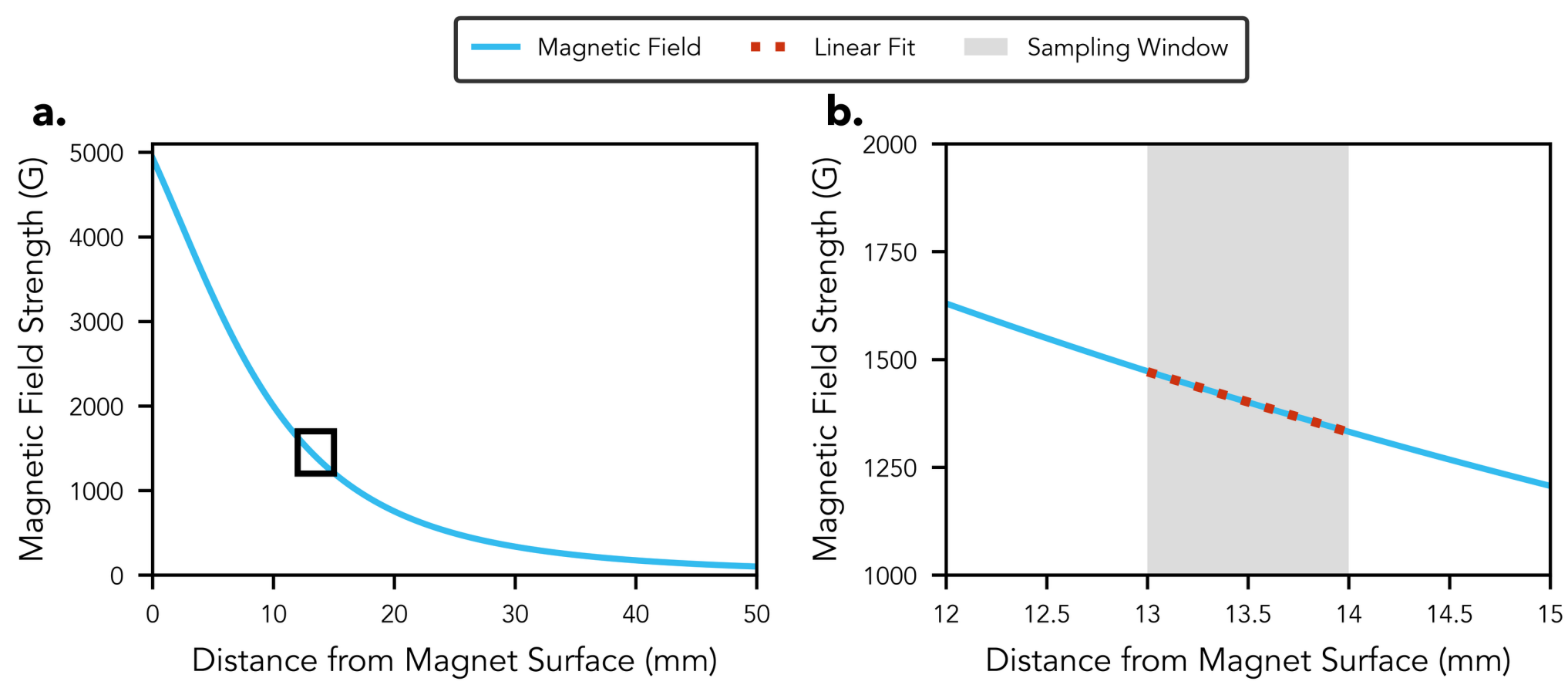

**Figure S1.2. a.** Magnetic field strength as a function of the distance from the magnet surface. **b.** The area highlighted in the black box in part a, demonstrating the linear fit of the magnetic field (red dashed line) within the relevant sampling window (gray box).

Thus, the magnetic field in the sampling region can be approximated by a linear equation in the form:

$$B = az + b, \tag{S1.18}$$

where $a$ and $b$ are the fit parameters that encapsulate the geometric and field parameters of the magnet. In our fit, $z$ is restricted to $z_{bottom} \leq z \leq z_{top}$, where $z_{top}$ and $z_{bottom}$ are the distances to the top and bottom of our sampling region from the center of the $L$-$W$ plane of the magnet. Although a quadratic or cubic fit is slightly better at modeling the sampling region of the magnetic field than the linear fit, this would significantly complicate further analysis. We aim to obtain an analytical solution to a second order, nonhomogeneous differential equation, which will be later used to fit experimentally collected sets of data. Therefore, moving forward, we represent the magnetic field with a linear approximation (Figure S1.2b), which proves to be sufficient for our application.

With eq S1.18 defining the applied magnetic field, we can rewrite the magnetic force expression. From eq S1.16, we get:

$$F_m = \frac{\partial}{\partial x}\left[\frac{\chi_{eff}V}{\mu_0(1+\chi_s)}(az+b)^2 + m_0(az+b)\right]. \tag{S1.19}$$

Taking the partial derivative and distributing the terms gives:

$$F_m = \frac{2a^2\chi_{eff}V}{\mu_0(1+\chi_s)}z + \frac{2ab\chi_{eff}V}{\mu_0(1+\chi_s)} + m_0 a. \tag{S1.20}$$

Therefore, eq S1.20 gives the force an MNP experiences in the sampling window due to a linearly approximated applied magnetic field.

### 1.1.3. Equation of Motion

With the derivation of the magnetic force, we can now write out the full expression for the net force on the MNP in the sampling window:

$$F_{net} = \rho_s V g + 6\pi\eta r\dot{z} - \rho_p V g - \frac{2a^2\chi_{eff}V}{\mu_0(1+\chi_s)}z - \frac{2ab\chi_{eff}V}{\mu_0(1+\chi_s)} - m_0 a. \tag{S1.21}$$

Using Newton's second law, $F = -m_p\ddot{z}$, with the net force $F$ acting in the negative direction. By substituting $m_p = \rho_p V$, we can arrive at our final net force equation:

$$-\rho_p V\ddot{z} = \rho_s V g + 6\pi\eta r\dot{z} - \rho_p V g - \frac{2a^2\chi_{eff}V}{\mu_0(1+\chi_s)}z - \frac{2ab\chi_{eff}V}{\mu_0(1+\chi_s)} - m_0 a. \tag{S1.22}$$

Rearranging eq S1.22 to isolate $\ddot{z}(t)$, we can write it in the standard form of a second order, nonhomogeneous differential equation:

$$\ddot{z} = -\frac{9\eta}{2r^2\rho_p}\dot{z} + \frac{2a^2\chi_{eff}}{\rho_p\mu_0(1+\chi_s)}z + \frac{2ab\chi_{eff}}{\rho_p\mu_0(1+\chi_s)} + \frac{3m_0a}{4\pi r^3\rho_p} + \left(1 - \frac{\rho_s}{\rho_p}\right)g. \quad \text{(S1.23)}$$

By defining some additional variables, we can simplify the representation of this equation into:

$$\ddot{z} = \alpha\dot{z} + \beta z + \gamma, \quad \text{(S1.24)}$$

where we define the following:

$$\alpha = -\frac{9\eta}{2r^2\rho_p}, \quad \text{(S1.25)}$$

$$\beta = \frac{2a^2\chi_{eff}}{\rho_p\mu_0(1+\chi_s)}, \quad \text{(S1.26)}$$

and

$$\gamma = \frac{2ab\chi_{eff}}{\rho_p\mu_0(1+\chi_s)} + \frac{3m_0a}{4\pi r^3\rho_p} + \left(1 - \frac{\rho_s}{\rho_p}\right)g. \quad \text{(S1.27)}$$

The $\alpha$ term can be thought of as the drag effects per unit mass, $\beta$ as magnetic field effects per unit mass, and $\gamma$ as nonhomogeneous field effects.

The second order, nonhomogeneous form in eq S1.24 is a commonly solved differential equation and can be found in many math and physics textbooks. The solution yields the equation of motion for a single particle:

$$z(t) = k_1e^{\delta_1 t} + k_2e^{\delta_2 t} + \frac{\gamma}{\beta}, \quad \text{(S1.28)}$$

where we define rate-limiting constants $\delta_1$ and $\delta_2$ which encapsulate the drag and magnetic properties of the particle, as:

$$\delta_1 = \frac{-\alpha+\sqrt{\alpha^2-4\beta}}{2} \quad \text{(S1.29)}$$

and

$$\delta_2 = \frac{-\alpha-\sqrt{\alpha^2-4\beta}}{2}. \quad \text{(S1.30)}$$

### 1.1.4. Relating Equation of Motion to Concentration

If we consider the sampling region to be a single-compartment model, we can write an expression for the number of particles $n(t)$, and ultimately the concentration of particles $c(t)$, within the sampling region at a given point in time. For a single compartment model, we can relate the change of number of MNPs in our sampling region to the velocity of the MNPs by:

$$\dot{n}(t) = D\dot{z}(t), \tag{S1.31}$$

where $D$ is a constant representing the linear density of MNPs leaving the compartment.

Taking the derivative of eq S1.28 and substituting it into eq S1.31 yields

$$\dot{n}(t) = D\left(k_1\delta_1 e^{\delta_1 t} + k_2\delta_2 e^{\delta_2 t}\right). \tag{S1.32}$$

Integrating eq S1.32 with respect to time to find an expression for $n(t)$ gives us:

$$n(t) = D\left(k_1 e^{\delta_1 t} + k_2 e^{\delta_2 t}\right) + k_3, \tag{S1.33}$$

where $k_3$ is an unknown constant of integration.

To relate the total number of particles in our sampling region to a concentration (mass per volume), we can divide $n(t)$ by the volume of our sampling window $V_w$ and multiply by the mass of a single particle $m_p$:

$$c(t) = \frac{m_p}{V_w} n(t). \tag{S1.34}$$

Expanding $n(t)$ using eq S1.33 yields:

$$c(t) = \frac{m_p}{V_w} D\left(k_1 e^{\delta_1 t} + k_2 e^{\delta_2 t}\right) + \frac{m_p}{V_w} k_3. \tag{S1.35}$$

We can simplify our expression for $c(t)$ by defining some new constants such that $c(t)$ becomes:

$$c(t) = \kappa_1 e^{\delta_1 t} + \kappa_2 e^{\delta_2 t} + \kappa_3, \tag{S1.36}$$

where

$$\kappa_1 = \frac{m_p}{V_w} D k_1, \tag{S1.37}$$

$$\kappa_2 = \frac{m_p}{V_w} D k_2, \tag{S1.38}$$

and

$$\kappa_3 = \frac{m_p}{V_w} k_3. \tag{S1.39}$$

Our unknown quantities can be reduced by applying boundary conditions. First, we will apply the boundary condition describing the final state of the system: by the end of the test, there should be no particles in the sample window because they will all be pulled towards the magnet, so the concentration should be zero (i.e. $\lim_{t\to\infty} c(t) = 0$). Because we have the physical constraints that $\delta_1 < 0$ and $\delta_2 < 0$, the exponential terms of eq S1.36 should go to 0, and we find that

$$\kappa_3 = 0. \tag{S1.40}$$

This leaves us with the expression:

$$c(t) = \kappa_1 e^{\delta_1 t} + \kappa_2 e^{\delta_2 t}. \tag{S1.41}$$

We next apply the boundary condition describing the initial state of the system, when the concentration is a known initial concentration $c_0$ (i.e. at $t = 0$, $c(t) = c_0$). This yields the following relationship between $\kappa_1$ and $\kappa_2$:

$$\kappa_2 = c_0 - \kappa_1. \tag{S1.42}$$

Finally, we can apply a condition related to the change in concentration over time: at the initial time point, because particles are initially assumed to be at rest, there should be no change in concentration (i.e. at $t = 0$, $\dot{c}(t) = 0$). This yields the final relationship needed to rewrite our system without any integration constants. We can define $\kappa_1$ as:

$$\kappa_1 = c_0 \frac{\delta_2}{\delta_2 - \delta_1}. \tag{S1.43}$$

Using eqs S1.42 and S1.43 and redefining $\kappa = \kappa_1$, we can rewrite eq S1.41 to yield our final expression for the concentration of particles in our sampling window:

$$c(t) = \kappa e^{\delta_1 t} + (c_0 - \kappa) e^{\delta_2 t}, \tag{S1.44}$$

where

$$\kappa = \frac{\delta_2}{\delta_2 - \delta_1} c_0. \tag{S1.45}$$

Based on the physical interpretation of this model in our system, we expect $\delta_1$ and $\delta_2$ to be negative, since the MNPs are attracted to the magnet and thus leaving the sample window. Although $\delta_1$ and $\delta_2$ both encompass many variables, most of the variables are physical constants or parameters that are known. There are just two unknown quantities in this equation: the effective radius of the particle $r$ and the effective magnetic susceptibility $\chi_{eff}$.

### 1.1.5. Limitations to the Model

There are several important assumptions made in this model that place operational bounds on the MAP usage. These are discussed in more detail below.

**Optical Linear Working Range**

There must be a balance between the optical properties of the particles and the optical system. These variables must be balanced so that the MAP is operating in the linear working range of the optical sensing system and so the concentration is proportional to the measured transmitted light, as stated by Beer's Law. Outside of this range, this relationship is not valid, and the data cannot be properly transformed to a concentration time series. This range will be dependent on both particle-based and optical-based parameters, including the particle absorptive coefficient, a material property specific to the wavelength of light used in the MAP (620-625 nm in this work), the concentration of particles, the absorptive properties of the solvent, the incident light intensity (based on the optical source used), and the sensitivity of the sensor to this wavelength of light (based on the light sensor). An example of the characterization process to ensure this assumption is true can be found in SI Section 1.1.4: Sensor Characterization.

**Thermal Effects**

The model assumes that the light source does not directly or indirectly heat the nanoparticles or their surrounding fluid. Heating of the solvent would potentially cause a refractive index change that could affect the reliability of the optical measurements. Heating of the nanoparticles could change the magnetic susceptibility of the material or even the oxidative state of metal oxide particles, such as iron oxide nanoparticles. Thus, the light source must be a sufficiently low power such that heating is avoided. As will be shown in SI Section 1.1.4: Sensor Characterization, we observe no heating even over prolonged time periods.

**DC Magnetic Susceptibility**

The model assumes that a non-time-varying magnetic field (DC field) is applied, and as such, it reports the effective DC magnetic susceptibility of the nanoparticles. Because a DC field is used, the MAP cannot measure the AC magnetic susceptibility of the nanoparticles.[4–7]

**Uniform Distribution of Particles**

The particles must be a uniformly distributed suspension at the beginning of a MAP test. This is required both for proper calibration of the MAP and also to properly define the boundary condition for initial concentration used in eqs S1.42 and S1.43. This also avoids known artifacts related to measuring magnetic susceptibility in liquid suspensions.[8,9]

**Suspension of Particles**

The particles must be truly suspended in solution and not form a ferrofluid in the solvent. The particles must be able to drop out of solution for the MAP to properly characterize their effective magnetic susceptibility.

## 1.2. Model Implementation

### 1.2.1. Calibration to Concentration

When implementing this model to fit the recorded optical signal, the data must be calibrated to transform the data from the measured light intensity to concentration of MNPs in the sampling region. To make this conversion, we take advantage of the temporal nature of the data, which requires that the optical signal plateaued by the conclusion of the data collection. This steady state signal is the equivalent of a blank in UV-Vis spectroscopy. It represents all background signal from the solvent, the cuvette, and any non-magnetic solutes that may be present in the solution since the plateaued signal indicates all magnetic materials have been pulled out of sampling window.

First, the light intensity $I(t)$ is converted to transmission $T(t)$ by using the average of the final 100 data points (approximately 10 seconds) in the steady state signal as the initial light transmission $I_0$ (the background signal):

$$T(t) = \frac{I(t)}{I_0}. \tag{S1.46}$$

This transmission can then be converted to optical attenuation $A(t)$ via:

$$A(t) = -\log[T(t)] = -\log\left[\frac{I(t)}{I_0}\right]. \tag{S1.47}$$

Based on the Bouguer-Beer-Lambert law (an expanded version of Beer's law to account for both absorption and scattering), attenuation is defined as:

$$A(t) = (\epsilon l + k_{scatter})c(t) + k_{noise}, \tag{S1.48}$$

where $\epsilon$ is the molar absorptivity, $l$ is the optical path length, $k_{scatter}$ is a scattering coefficient, $c(t)$ is the time-dependent concentration, and $k_{noise}$ is attenuation related to background species and noise.[10] To find this linear relationship, we again use the temporal nature of the collected MAP data. The initial point $A_0$ and the final point of the data $A_f$ correspond to the known initial concentration $c_0$ of MNPs in solution and a zero concentration MNP solution, respectively.

The line fit between these points describes the linear relationship between attenuation and concentration and can be used to transform the time-dependent attenuation $A(t)$ to time-dependent concentration $c(t)$ via:

$$c(t) = \frac{c_0}{A_0 - A_f} \times A(t) + \frac{c_0 A_f}{A_f - A_0}. \tag{S1.49}$$

This concentration time series $c(t)$ found with the transformed data is then fit to eq S1.44 using nonlinear least squares regression to derive the magnetic susceptibility from the transformed MAP data using the two unknown variables $r$ and $\chi_{eff}$ as fit parameters.

### 1.2.2. Data Pre-Processing

After calibrating the data to a concentration time series, a two-phase regime is commonly observed. The first phase is dominated by random MNP motion introduced into the system via sonication, vortexing, aspiration, or some combination thereof as the MNPs are resuspended. The second phase is dominated by motion induced by forces defined in the mathematical model of the system (i.e. unidirectional motion).

Phase one processes introduce a time delay, which is not fit by the model. This phase change can be determined via an inflection point highlighted in the derivative of the signal. We will discard the initial data as this does not represent the particles' pure response to the magnetic field; the random motion, coupled with the particles' inertia, will delay the uniform response to the forces that are included in the model.

Instead, we will truncate the data at the inflection point and use that as the "initial" timepoint for the start of our model. Fitting our model to the second phase of the response results in more accurate fit implementation. This process is illustrated in Figure S1.3.

The discussion justifying this change may be found in the later section discussing the raw data of the multi-magnet measurements.

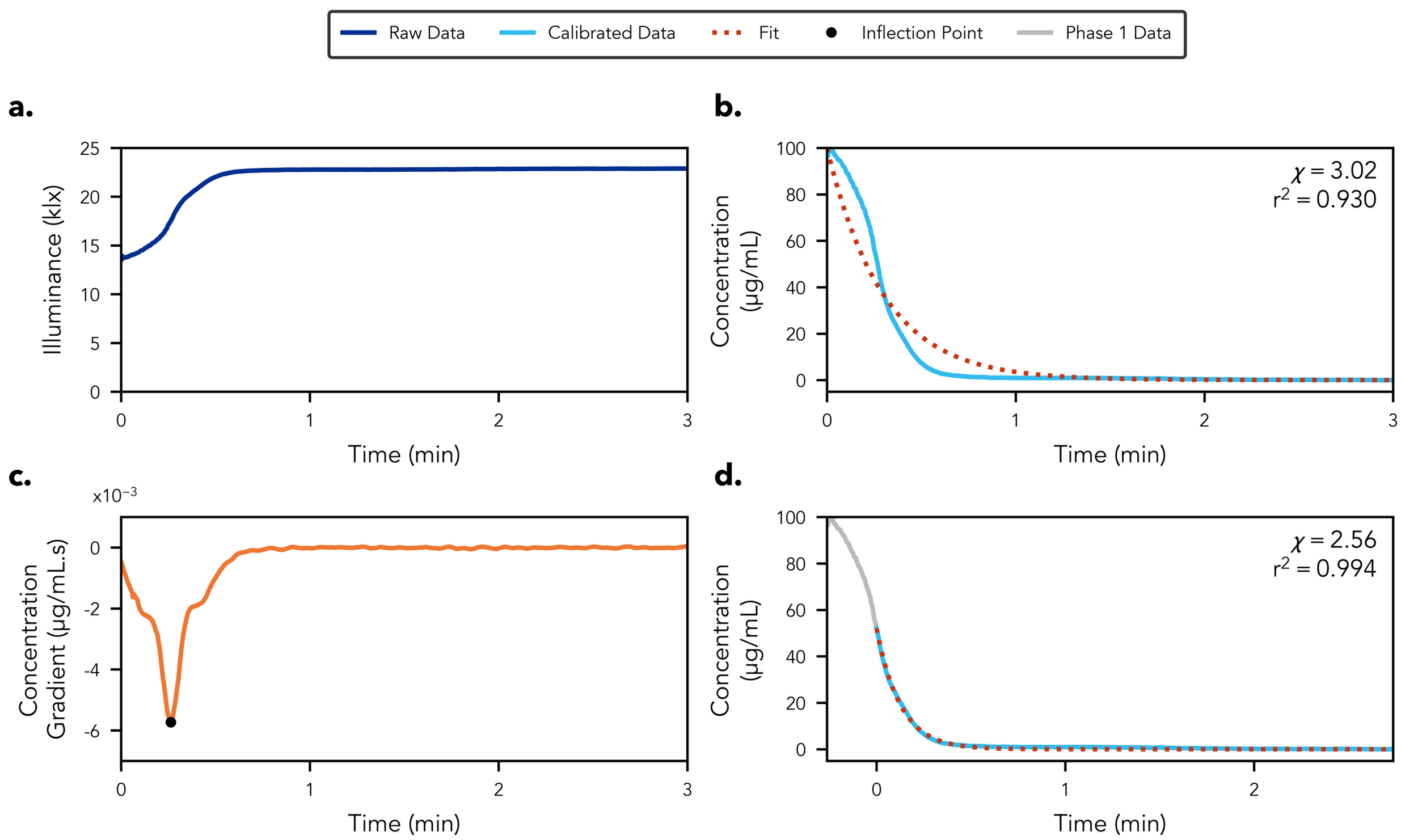

**Figure S1.3.** The pre-processing steps illustrated for the first trial of the Dynabeads MAP tests (section 3.3 of main text). **a.** The raw MAP data, still in photometric units (teal). **b.** The calibrated MAP data (blue) and the fit (red dotted), clearly showing a two-phase response, with the magnetic susceptibility $\chi_{eff}$ and $r^2$ from the fit. **c.** The time derivative (orange) of the smoothed concentration data (the blue line in part b), with the inflection point demarcated with a black dot. **d.** The calibrated data broken into two phases: (1) phase 1 (gray) which is ignored, and (2) phase 2 (blue) which was fit (red dotted) with our theoretical model. The new magnetic susceptibility $\chi_{eff}$ and $r^2$ value clearly indicating a better fit are shown.

## 1.3. Validation Data

### 1.3.1. Iron Oxide Nanoparticle Synthesis Protocol

Iron oxide ($Fe_3O_4$) nanoparticles were synthesized using a coprecipitation protocol under argon. First, purge the Schlenck line with argon; it is important there is no oxygen in the system to avoid oxidizing the iron chloride precursors. Add 20 mL of deionized (DI) water to the reaction flask, seal with a rubber septum, and turn on the condenser. Using a needle through the septa, purge the flask with argon. In a glove box under argon, measure out 1 g of iron(II) chloride and 0.4 g of iron(III) chloride. Remove the argon and exhaust needles and attach argon to the top of the condenser. Quickly remove the rubber septa from the reaction flask and add the solid iron chlorides before replacing the septa.

Heat the iron chloride precursors and water solution to 80°C on a hot plate. Add 5 mL of ammonia hydroxide dropwise while increasing the speed of the stir bar to prevent the reaction from occurring on the stir bar. Ice should be added to the water tank of the condenser and the argon line removed and replaced with foil over the condenser. Let the reaction occur for 1 hour.

After 1 hour, remove foil from the condenser and open the rubber septa, leaving the flask open to cool. The thermometer and condenser can be removed when cool, and a strong magnet should be used to remove the stir bar from the reaction flask. Then, use a magnet to collect the particles at the bottom of the reaction flask, and when the water is clear, remove the supernatant, using the magnet to retain the particles. Repeat 3 times to wash the particles, adding DI water to resuspend the MNPs each time. Then, add a small volume of DI water to the particles to resuspend and use a Pasteur pipette to remove the particle solution for storage. The particles may also be dried and stored as a powder.

### 1.3.2. SQUID and VSM Magnetometry

To confirm the validity of our theoretical model, we compared the analyzed data from the MAP to the magnetic susceptibility value obtained from two commercial magnetometers using MNPs from the same synthesis batch. The same samples were tested on both a vibrating sample magnetometer (VSM) and a superconducting quantum interference device (SQUID) magnetometer. Both the sample preparation, testing protocol, and analysis steps are the same for both the VSM and SQUID magnetometers; they vary only in their mechanism to sense the magnetic moment of the sample. The SQUID used was a Quantum Design MPMS 3 at the University of California Santa Barbara Materials Research Laboratory TEMPO Facility and measurements were performed by their staff scientists. The VSM used was a Quantum Design Physical Property Measurement System (PPMS) DynaCool located at the University of Southern California, and measurements were performed by the authors.

**Sample Preparation**

To prepare the samples for both magnetometers, the iron oxide MNP solution was fully dried until it was a powder. A plastic sample holder was filled with several mg of powder, just below the halfway point of the holder. The exact mass of iron oxide in the samples prepped from each of our three batches is given in Table S1.1. Eicosane (99%, Alfa Aesar) was heated beyond its melting point, until liquid. The remainder of the holder was filled with melted eicosane, the holder was capped, and any residual eicosane cleaned off the outer surfaces of the holder. The eicosane was allowed to cool and set to form a matrix around the iron oxide powder, securing it in place. This prevented the particles from moving within the sample chamber, which would introduce artifacts into the measurements.

**Measurement Protocol**

The sample was loaded into the magnetometer. The magnetometer carried out a 5-quadrant sweep, applying magnetic fields from 0 Oe, +5000 Oe, -5000 Oe, and finally back to +5000 Oe at a set temperature of 293 K to acquire a magnetic hysteresis curve. A measurement point was taken at a resolution of approximately every 8 Oe change in applied magnetic field. The same protocol was run on both the SQUID (Figure S1.4) and VSM (Figure S1.5). It should be noted that batch 3 could not be run on the VSM due to an inability to access the VSM when the lab owning the instrument relocated to a different institution. Each sample was only run once due to the high cost of a single measurement session.

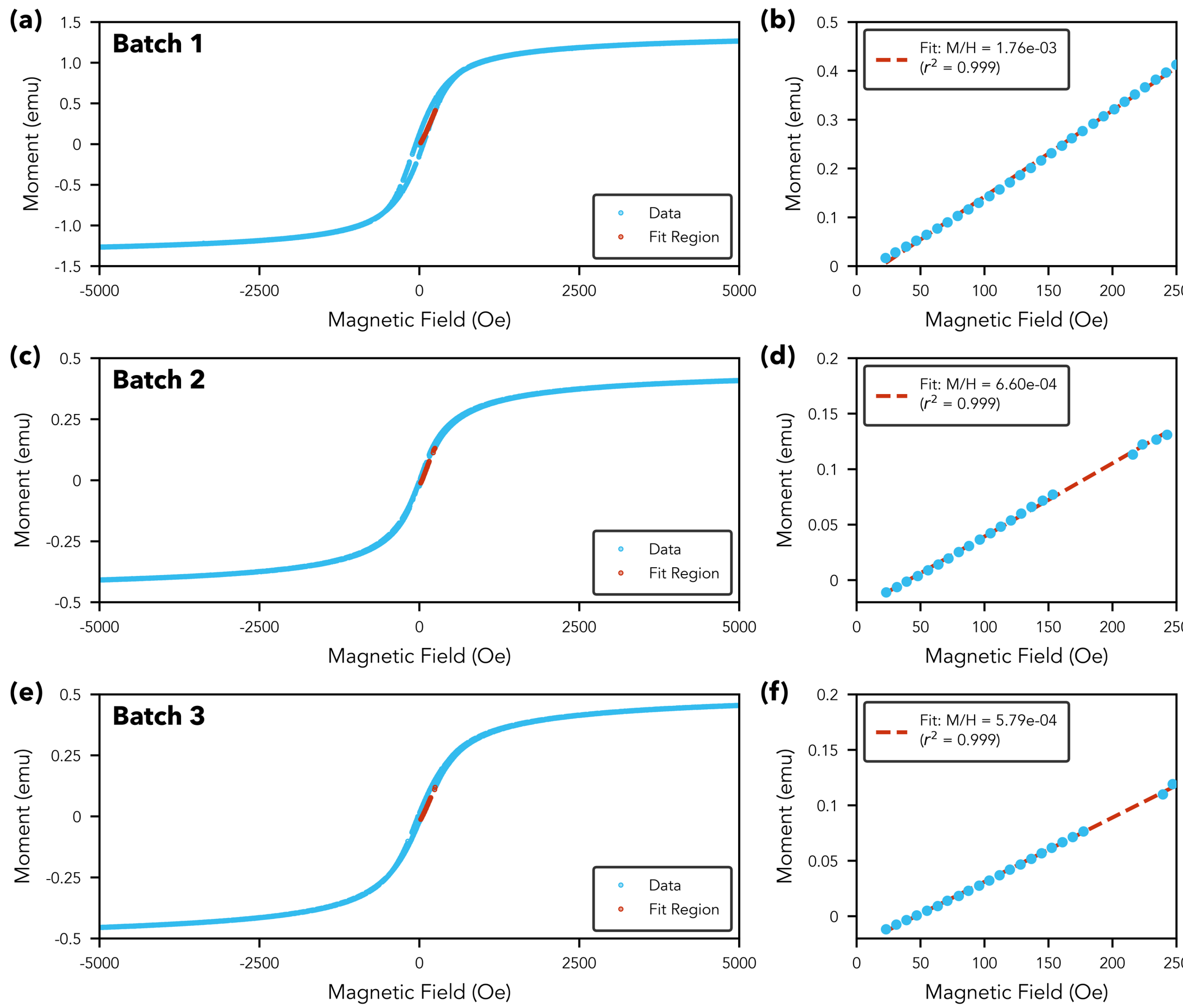

**Figure S1.4. (left)** SQUID data showing the 5-quadrant sweep over a range of applied magnetic field from 0 Oe, to 5000 Oe, to -5000 Oe, and back to 5000 Oe. Data was taken on a Quantum Design MPMS 3. All MNPs showed the expected hysteresis behavior. **(right)** To get the magnetic susceptibility, a line is fit to the initial linear portion when the magnetic field is first applied (shown in red). The slope of that linear fit is the ratio of the magnetization $M$ to the applied magnetic field $H$. **a.** Batch 1 hysteresis curve and **b.** linear fit. **c.** Batch 2 hysteresis curve and **d.** linear fit. **e.** Batch 3 hysteresis curve and **f.** linear fit.

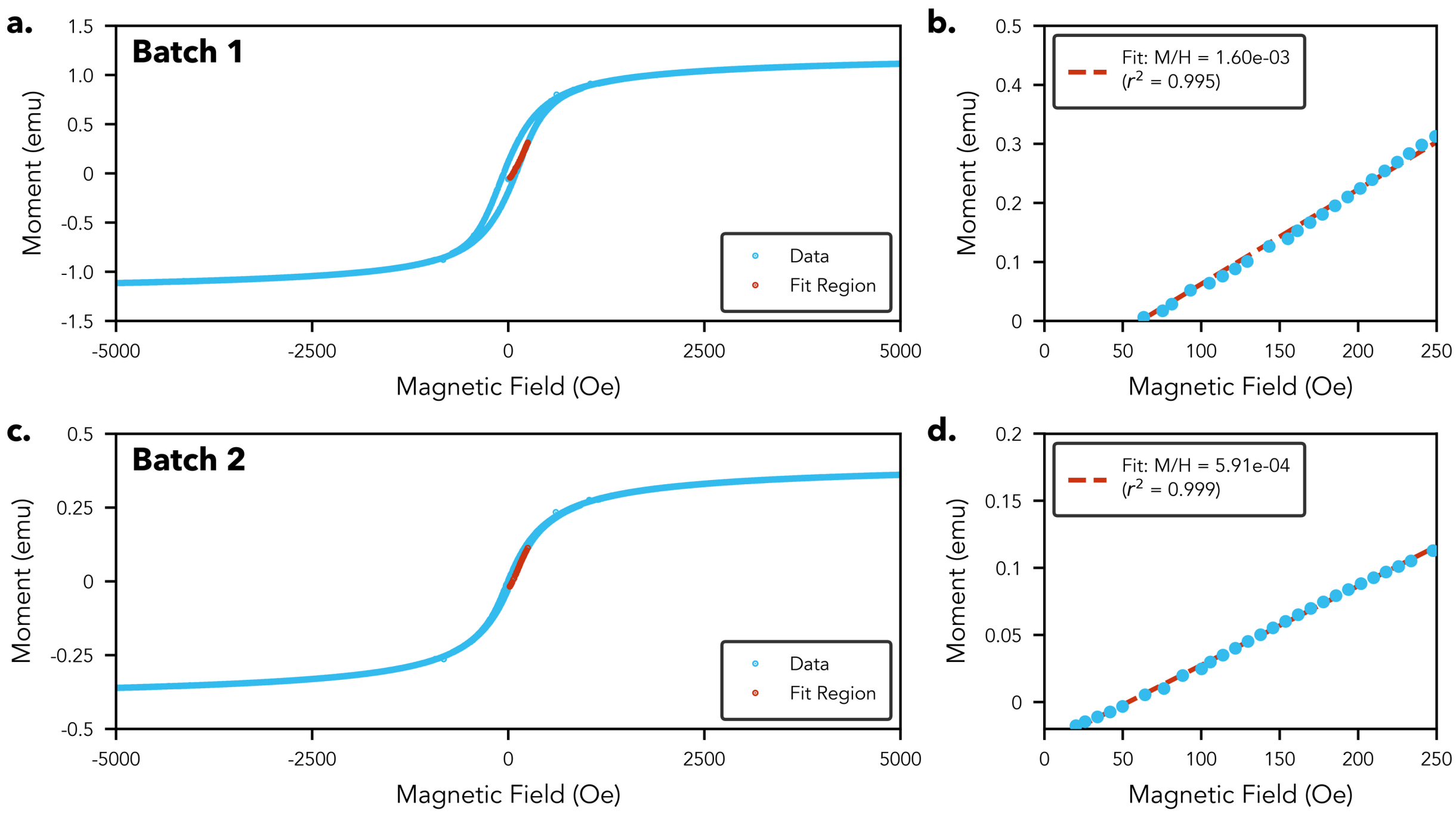

**Figure S1.5. (left)** VSM data showing the 5-quadrant sweep over a range of applied magnetic field from 0 Oe, to 5000 Oe, to -5000 Oe, and back to 5000 Oe. Data was taken on a Quantum Design Dynacool Physical Property Measurement System (PPMS). All MNPs showed the expected hysteresis behavior. **(right)** To get the magnetic susceptibility, a line is fit to the initial linear portion when the magnetic field is first applied (shown in red). The slope of that linear fit is the ratio of the magnetization $M$ to the applied magnetic field $H$. **a.** Batch 1 hysteresis curve and **b.** linear fit. **c.** Batch 2 hysteresis curve and **d.** linear fit.

**Data Analysis**

To find the magnetic susceptibility, the linear portion of the curve was fit to a line using standard linear regression techniques. The slope of that linear fit gives the ratio of the magnetization $M$ relative to the applied magnetic field $H$. For consistency, the linear region was selected to be the portion between 25 to 250 Oe for each sample, and the results had high goodness of fits with each linear model (Figure S1.4b, d, and f, and Figure S1.5b and d).

Magnetic susceptibility is typically defined as the magnetization $M$ per unit volume $V$ divided by the ratio of the applied magnetic field $H$. We converted our measured sample mass to appropriate units of volume using the density of iron oxide. The magnetic susceptibility was calculated in both CGS units, consistent with the units reported by the magnetometers, and SI units, consistent with the units used in our theoretical model derivation. This conversion between the two unit systems is simply:

$$\chi_{SI} = 4\pi\chi_{CGS}. \tag{S1.50}$$

The magnetic susceptibility for each sample on each magnetometer is given in Table S1.1. These fall into the range expected from iron oxide samples.[11] We see good agreement between the VSM and SQUID values; however, it should be noted that the magnetic susceptibilities measured using the VSM and the SQUID vary within the same order of magnitude (9-11%) despite testing the exact same sample.

**Table S1.1.** SQUID and VSM Magnetometry on Three Batches of Iron Oxide Nanoparticles.

| **Batch** | **Sample Mass (mg)** | $\chi_{SQUID,CGS}$ | $\chi_{SQUID,SI}$ | $\chi_{VSM,CGS}$ | $\chi_{VSM,SI}$ | $\chi_{SQUID,SI} - \chi_{VSM,SI}$ |
|---|---|---|---|---|---|---|
| 1 | 16.7 | 0.546 | 6.86 | 0.495 | 6.22 | 0.64 |
| 2 | 6.6 | 0.517 | 6.50 | 0.459 | 5.77 | 0.73 |
| 3 | 6.0 | 0.499 | 6.28 | n/a[a] | n/a[a] | n/a[a] |

[a]Batch 3 was not tested on the VSM due to limitations accessing the instrument.

### 1.3.3. MAP Data

The raw and calibrated data as well as the full results from the tests summarized in the main text of the paper are presented in the following sections.

**Sample Preparation**

For all MAP tests, the samples were prepared and tested using the same protocols. All iron oxide MNPs were suspended in water via thorough sonication for at least 5 min at temperatures below 30°C and mixed for at least 1 min via vortexing prior to testing in the MAP. Solutions were allowed to come to room temperature prior to MAP testing to avoid temperature-dependent effects on magnetic susceptibility.[12]

**Measurement Protocol**

Each solution was tested in a disposable poly(methyl methacrylate) (PMMA), standard 3.5 mL spectroscopy cuvette at a volume of 1 mL. The magnetic field was applied immediately when data collection began. Tests were run in the MAP for a duration of 3 minutes, which was determined to be a sufficient length to capture the full system response based on preliminary test steps with this material (iron oxide). Each sample was run in triplicate for reproducibility. Between each test, the sample was aspirated with a pipette at least 10 times to ensure that the MNPs were sufficiently resuspended.

**Conversion from Illuminance to Irradiance**

The light sensor used in the MAP reports the light intensity in units of illuminance: lux, or lumens per square meter. Illuminance is a photometric unit of light which weights the intensity of wavelengths of light based on the sensitivity of the human eye. The corresponding radiometric unit of light intensity is irradiance, or the radiant flux per unit area, in units of watts per square meter.

For convenience, we will convert the units of lux to the more common units of W/m$^2$. The illuminance $I$ is related to the spectral integration of the irradiance $E$ over the visible range to which the human eye is sensitive, from 380 nm to 780 nm. Specifically, the two metrics are related by the equation:

$$I = k_{max} \int_{380}^{780} V(\lambda) E(\lambda) d\lambda, \tag{S1.51}$$

where $\lambda$ is the wavelength in nm, $V(\lambda)$ is the phototopic luminous efficiency function, and $k_{max}$ is the maximum luminous efficiency, which has a value of 683 lumens/W [13].

For the LED selected in this work (SparkFun COM-00528), the datasheet lists a wavelength of 620-625 nm. To simplify the conversion, we assume that the LED is a monochromatic light source with a wavelength of 622.5 nm, the center wavelength of the range given by the datasheet. Thus, the integral is reduced to the irradiance at a single wavelength, $\lambda = 622.5\ nm$. Thus, eq S1.51 becomes

$$I = k_{max} V(622.5) E, \tag{S1.52}$$

or, when solved for the irradiance:

$$E = \frac{I}{k_{max} V(622.5)}. \tag{S1.53}$$

To find the value of the phototopic luminous efficiency at $\lambda = 622.5\ nm$, we linearly interpolate the given values of $V(\lambda)$ at $\lambda = 622\ nm$ and $\lambda = 623\ nm$ from the International Commission on Illumination dataset and find that $V(622.5) = 0.3508$. Substituting this value for $V(622.5)$ and $k_{max}$ into eq S1.53 yields the following conversion between irradiance and illuminance:

$$E = \frac{I}{239.6}. \tag{S1.54}$$

This conversion factor was used to transform all raw MAP data from illuminance to irradiance. All MAP light intensity data shown in this manuscript has been converted to the more common radiometric units of irradiance using this conversion factor. It should further be noted that the units of light intensity do not affect the process of determining the magnetic susceptibility of the particles from the data since the raw MAP data is transformed to optical attenuation as described in SI Section 1.2.2: Data Pre-Processing.

**Multi-Batch Data**

All three batches of iron oxide MNPs were tested in the MAP for comparison of the MAP-derived magnetic susceptibility to the SQUID and VSM values. Particles from each batch were suspended in water at a concentration of 100 µg/mL and tested in the MAP with the 4933 G surface field magnet. Controls of water and batch 1 MNPs tested without a magnet were also run. The intensity data, in radiometric units, is presented in Figure S1.6a-c. The data was transformed to concentration over time prior to analysis for magnetic susceptibility and is shown in Figure S1.6d-f. The effective magnetic susceptibility values derived from each MAP trial are given in Table S1.2.

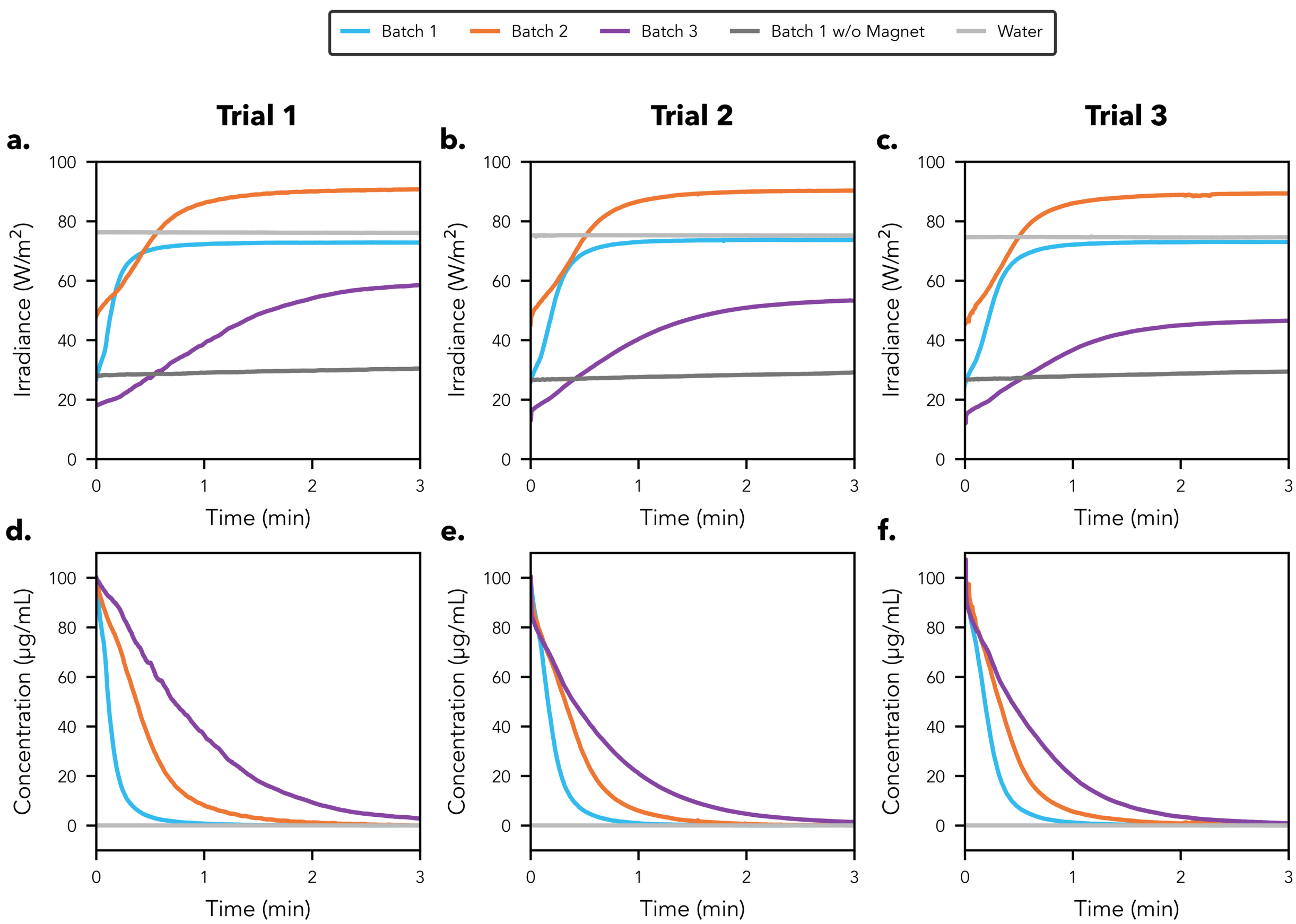

**Figure S1.6. a-c.** MAP light intensity data from trials 1, 2, and 3, respectively, using three different batches of iron oxide MNPs suspended in water at a concentration of 100 µg/mL. Two controls were added: water (0 µg/mL) and batch 1 not exposed to a magnetic field. Good reproducibility is observed between trials. **d-f.** The MAP data from trials 1-3, respectively, transformed into concentration over time. Note that the data from the batch 1 with no applied magnetic field were not included because the transformation protocol used in this work requires the signal to fully plateau.

**Table S1.2.** Effective Magnetic Susceptibilities[a] Determined by the MAP for Iron Oxide MNPs from Three Different Batches.

| **Batch** | **MAP $\chi_{eff}$** | | | **Avg. $\chi_{eff}$ ± Std. Dev.** |
|---|---|---|---|---|
| | **Trial 1** | **Trial 2** | **Trial 3** | |
| 1 | 5.15 | 5.64 | 5.15 | 5.31 ± 0.29 |
| 2 | 4.32 | 4.58 | 5.76 | 4.88 ± 0.77 |
| 3 | 3.40 | 3.56 | 3.48 | 3.48 ± 0.08 |

[a]All magnetic susceptibilities calculated in the SI system.

When comparing the MAP light intensity data between the three batches (Figure S1.6a-c), we can make several observations. First, the batches have different beginning and ending intensity values, which can be explained by the fact that these batches were all tested on different days (and thus the background intensity measured may vary) and the particle cross-sections vary between batches. More importantly, the intensity increases at different rates: batch 1 has the quickest increase in intensity and batch 3 the slowest increase. This is consistent across all three trials with each batch. This ordering corresponds to the effective magnetic susceptibilities measured by the MAP when compared across batches. Further, we see the same trend when comparing the SQUID and VSM results to the MAP results (Table S1.3).

**Table S1.3.** Magnetic Susceptibilities[a] Derived from MAP Measurements Relative to Those Derived from SQUID and VSM Measurements.

| **Batch** | **Avg. MAP $\chi_{eff}$** | **SQUID $\chi$** | **VSM $\chi$** |
|---|---|---|---|
| 1 | 5.31 | 6.86 | 6.22 |
| 2 | 4.88 | 6.50 | 5.77 |
| 3 | 3.48 | 6.28 | n/a[b] |

[a]All magnetic susceptibilities are calculated in the SI system. [b]Batch 3 was not tested on the VSM due to limitations accessing the instrument.

**Measurements of Varied Concentrations**

A dilution series was prepared from batch 1 of the iron oxide MNPs to characterize MAP performance across multiple concentrations. Serial dilutions were used to create solutions of iron oxide MNPs at concentrations of 1, 2, 4, 8, 16, 31, 63, 125, 250, and 500 µg/mL in water. A control of water was also tested. Figure S1.7a-c gives the irradiance data and Figure S1.7d-f gives the transformed concentration data. The effective magnetic susceptibility values for the concentrations in the linear working range are given for individual MAP trials in Table S1.4.

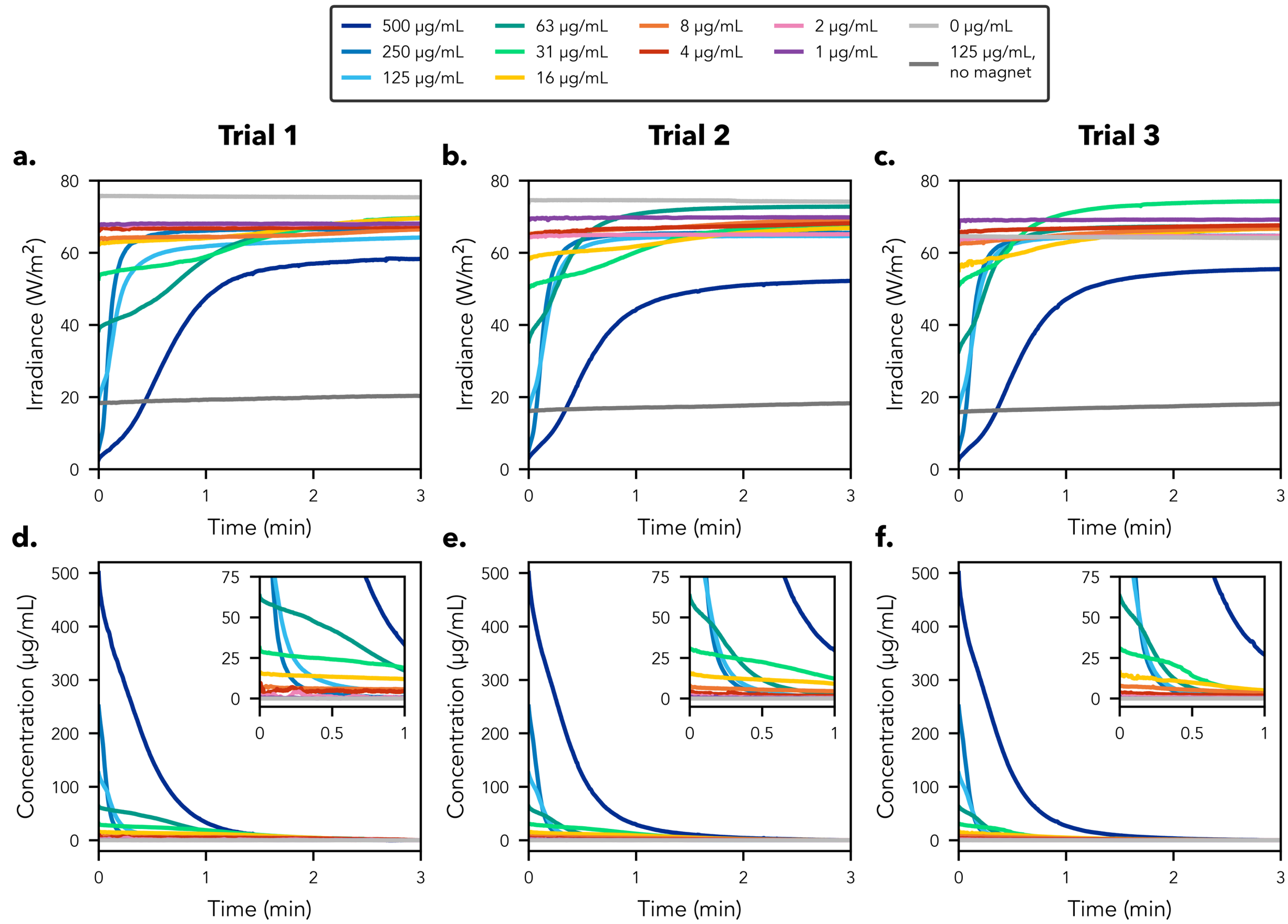

**Figure S1.7. a-c.** MAP light intensity data from trials 1, 2, and 3, respectively, using a range of MNP concentrations from 1 μg/mL up to 500 μg/mL. Two controls were added: water (0 μg/mL) and 125 μg/mL not exposed to a magnetic field. Note that the supernatant of the 500 μg/mL sample appeared translucent but still tinted brown, which could account for the lower final intensity values observed. **(insets)** The first minute of data from the lower concnetrations. **d-f.** The MAP data from trials 1-3, respectively, transformed into concentration over time. Note that the data from the 125 μg/mL with no applied magnetic field were not included because the transformation protocol used in this work requires the signal to fully plateau.

**Table S1.4.** Effective Magnetic Susceptibilities[a] Determined by the MAP for Iron Oxide MNPs of Varied Concentrations from the Linear Range.

| **Concentration (μg/mL)** | **MAP $\chi_{eff}$** | | | **Avg. $\chi_{eff}$ ± Std. Dev.** |
|---|---|---|---|---|
| | **Trial 1** | **Trial 2** | **Trial 3** | |
| 125 | 5.25 | 5.16 | 4.59 | 5.00 ± 0.36 |
| 63 | 4.14 | 4.62 | 5.25 | 4.67 ± 0.56 |
| 31 | 3.54 | 4.32 | 4.12 | 3.99 ± 0.41 |
| 16 | 3.73 | 3.84 | 3.43 | 3.67 ± 0.21 |

[a]All magnetic susceptibility values calculated in the SI system.

The MAP light intensity data shows the same trend shown in Figure 3 and discussed in the Measurements of Varied Concentrations section of the main text. The change in intensity over the duration of the MAP test corresponds to the initial starting concentration of the solution, even over the larger range of initial concentrations (Figure S1.7a-c). However, it should be noted that in the highest concentration tested, 500 μg/mL, the supernatant was translucent brown even after MAP testing, indicating that the solution is likely over saturated. This is reflected in the lower final intensity for all three trials of this concentration. We also note that some trials, such as the first trial with water (Figure S1.7a) show some noise in the signal, but the overall signal trend is not affected. This emphasizes the importance of using differential measurements, which minimizes the effects of this noise. Once the data is transformed to concentration over time, the trials appear much more consistent within the three trials of a single concentration (Figure S1.7d-f).

The effective magnetic susceptibilities given in Table S1.4 appear to show a slightly downward trend as concentration decreases. This is consistent with the scaling of effective magnetic susceptibility with changes in packing ratio and the theory that suspended nanoparticles will form chains in the presence of an external magnetic field, and as these chains grow, the alignment of dipoles within the chains would increase the magnetic susceptibility.[14,15] As the concentration decreases, the average distance between MNPs would increase, resulting in a lower likelihood of chains forming and a lower average chain length. This would account for the decreased effective magnetic susceptibility observed as concentration decreases. Because the MAP measures the effective magnetic susceptibility of suspended particles, it may be able to provide information specific to solutions (such as the effect of concentration) that the SQUID cannot.

**Measurements with Different Magnetic Fields**

In addition, we tested MNP solutions by changing the applied magnetic field using different permanent magnets. The MNP solution was prepared from batch 1 of iron oxide MNPs at a single concentration of 100 μg/mL. The magnets applied in this work were all neodymium block magnets sourced from K&J Magnetics and had surface fields ranging from 2108 to 4933 G. They are listed with their characteristics detailed in Table S1.5, including the linear approximation used in the model as described in eq S1.18. Note that all magnets will be referred to by their surface field for simplicity.

**Table S1.5.** Magnets used in this Work and Their Characteristics.

| K&J Magnetics ID | Dimensions (in.) | Grade[a] | Surface field (G)[a] | Residual Flux Density, $B_r$ (G)[a] | Field strength at center of sampling region (G)[b] | Linear approximation used in model, $az + b$ (T)[c] |
|---|---|---|---|---|---|---|
| BX0X03 | 1” x 1” x 3/16” | N42 | 2108 | 13200 | 658 | -6.677$z$+0.156 |
| BX0X04 | 1” x 1” x 1/4" | N42 | 2704 | 13200 | 815 | -8.209$z$ + 0.192 |
| BX0X06 | 1” x 1” x 3/8” | N42 | 3682 | 13200 | 1064 | -10.546$z$ + 0.249 |
| BX0X06-N52 | 1” x 1” x 3/8” | N52 | 4128 | 14800 | 1193 | -11.824$z$ + 0.279 |
| BX0X08-N52 | 1” x 1” x 1/2" | N52 | 4933 | 14800 | 1401 | -13.656$z$ + 0.324 |

[a]As reported by K&J Magnetics. [b]Computed via eq S1.17 based on the geometry of the MAP. [c]Described in eq S1.18. All linear fits had $r^2$ values > 0.9999. Note that the linear approximation used has units of Teslas; the model was completed in the SI system for consistency. Gauss converts to Teslas at a conversion rate of 10000 G to 1 T.

The MAP intensity data is presented in Figure S1.8a-c, as well as the transformed data (concentration over time) in Figure S1.8d-f. The effective magnetic susceptibility values for each MAP trial at each magnetic field strength except 0 G are given in Table S1.6.

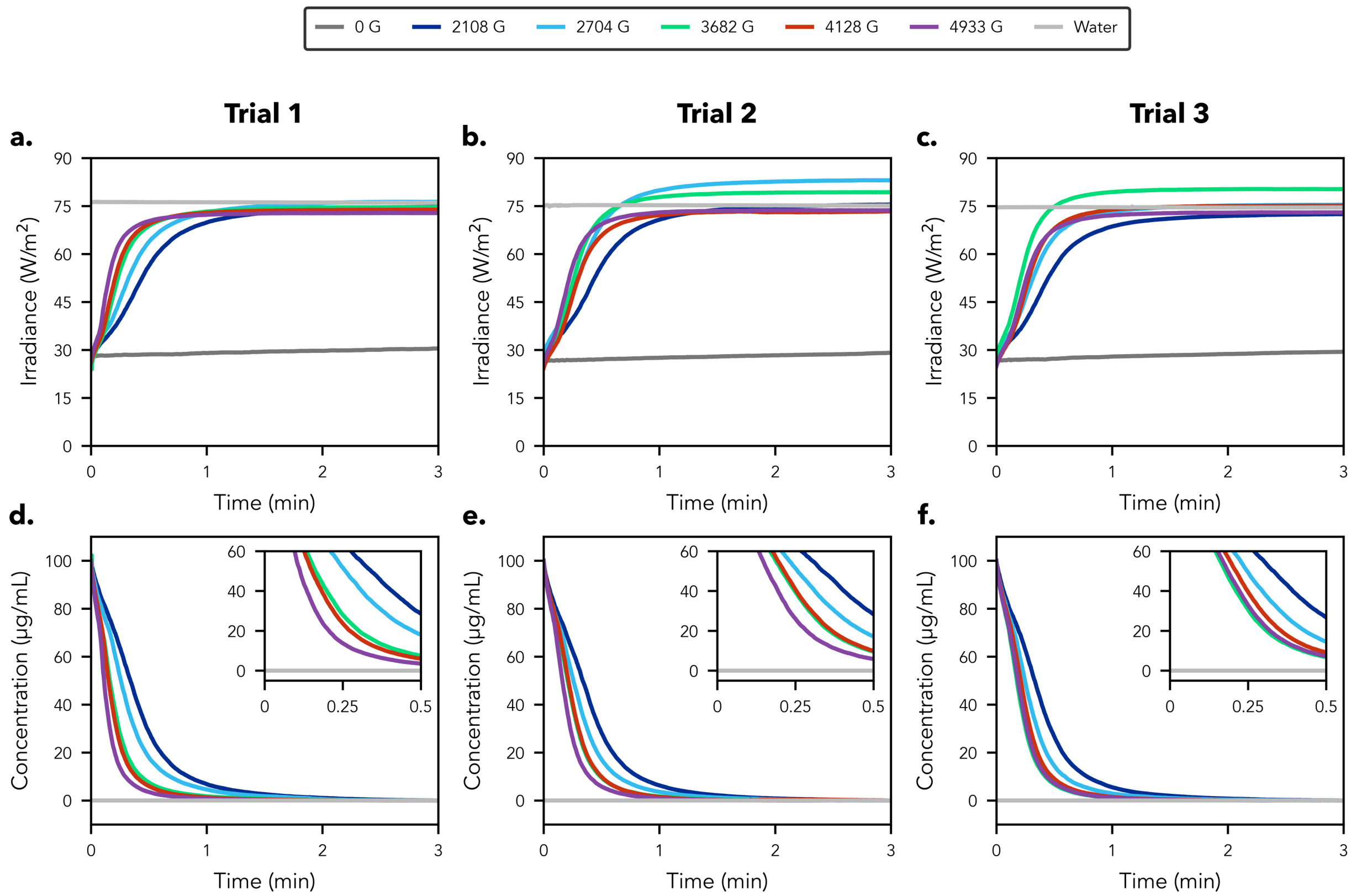

**Figure S1.8 a-c.** MAP light intensity data from trials 1, 2, and 3, respectively, of MNPs tested using a range of applied magnets with surface fields from ranging 2108 G to 4933 G. Two controls were added: no magnetic field (0 G) and water exposed to the 4933 G surface field. Each sample was tested at a concentration of 100 µg/mL suspended in water. **(insets)** The first minute of the data. **d-f.** The MAP data from trials 1-3, respectively, transformed into concentration over time. Note that the data from the 0 G trials were not included because the transformation protocol used in this work requires the signal to fully plateau.

**Table S1.6.** Effective Magnetic Susceptibilities[a] Determined by the MAP for Iron Oxide MNPs Tested with Different Applied Magnetic Fields.

| **Surface field of applied magnet (G)** | **MAP $\chi_{eff}$** | | | **Avg. $\chi_{eff}$ ± Std. Dev.** |
|---|---|---|---|---|
| | **Trial 1** | **Trial 2** | **Trial 3** | |
| 2108 | 5.30 | 6.47 | 4.67 | 5.48 ± 0.91 |
| 2704 | 5.20 | 6.14 | 5.78 | 5.71 ± 0.48 |
| 3682 | 4.26 | 5.02 | 5.87 | 5.05 ± 0.81 |
| 4128 | 3.86 | 5.54 | 5.12 | 4.84 ± 0.88 |
| 4933 | 5.15 | 5.64 | 5.15 | 5.31 ± 0.29 |

[a]All magnetic susceptibility values calculated in the SI system.

Examination of the MAP data reveals several trends. First, we see that the response time increases as the magnetic field strength decreases, a trend discussed in more detail in the next section. This follows intuitively from the definition of magnetic force given by eq S1.6 and eq S1.16 indicating that the force would scale with changes to magnetic field.

Second, we also see evidence of the two-phase response discussed in the Data pre-processing section. In many of the trials, there is a clear inflection point when the slope changes abruptly. The time point at which this occurs for each magnetic field strength is shown in Figure S1.9. We can see that as the magnetic field strength increases, this time decreases. This supports our hypothesis that the initial phase of the data is from the motion of particles from their resuspension and the inflection point represents the time at which the particle motion is uniformly downward towards the magnet. As the magnetic force increases, the random motion of the particles is overcome quicker, which results in the change of phase of motion occurs earlier in the trial.

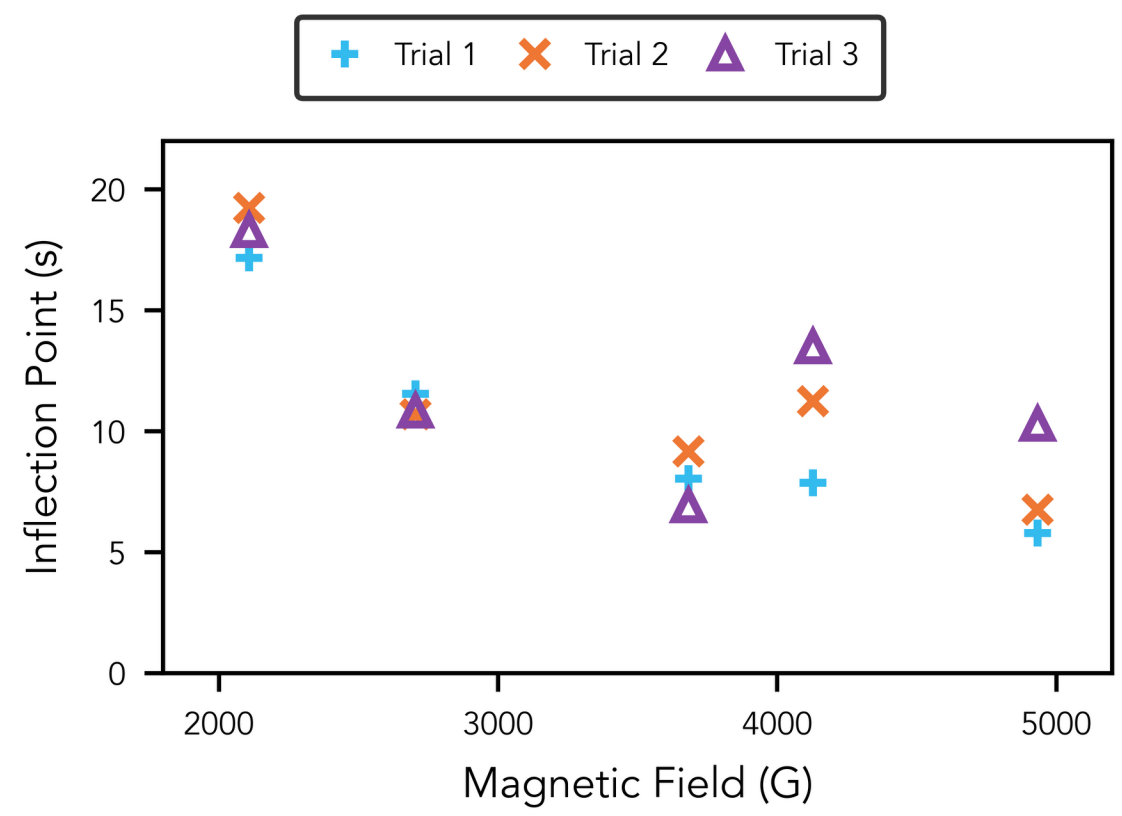

**Figure S1.9.** The magnetic field strength relative to the inflection point. The inflection point was determined to be the time point at which there was a minimum in the time derivative of the concentration data and the concentration began consistently increasing.

**Measurements with surface-coated particles**

To test the ability of the MAP to complete nondestructive measurements of magnetic susceptibility, we selected a standard commercial MNP and compared the ability of these MNPs that had either been tested or not tested in the MP to then bind to proteins. We used Dynabeads (ThermoFisher Dynabeads M-280 Streptavidin, #11205D), which consist of an iron oxide core with a polystyrene shell coated with a monolayer of streptavidin.

Biotin-4-fluorescein (fl-biotin, ThermoFisher Biotin-4-Fluorescein, #B10570) binds readily to streptavidin and provides a method to quantify the bioactivity of the Dynabeads through fluorescence. The fl-biotin was initially suspended in dimethyl sulfoxide (DMSO) at a concentration of 5 mg/mL. Aliquots of the fl-biotin were then serially diluted with additional DMSO to produce solutions with concentrations of 3.46 µM, 0.346 µM, or 0.0346 µM.

To show that the streptavidin proteins on the surface were not destroyed, we showed that Dynabeads that had undergone MAP testing could bind the same quantity of biotin as Dynabeads that have not undergone MAP testing (Figure 5a, main text). Aliquots of the original 10 mg/mL solution of Dynabeads were diluted down to 100 µg/mL with Dulbecco's phosphate buffer saline (DPBS, ThermoFisher #1404117) at a total volume of 1 mL. The solutions were washed once by separating the Dynabeads with a magnetic rack and removing the supernatant via pipette. DPBS was added to resuspend the solution at a volume of 1 mL at a concentration of 100 µg/mL and the solution was vortexed for 1 minute to resuspend the particles.

The aliquots were grouped into two groups: 1) Dynabead solutions to be tested in the MAP prior to incubation with fl-biotin (MAP-tested Dynabeads) and 2) Dynabead solutions to only be incubated with fl-biotin and not undergo prior MAP testing (untested Dynabeads). Within each of these two groups and in replicates of three, 1 mL solutions of Dynabeads were then incubated with one of three different quantities of biotin (1 pmol, 10 pmol, or 100 pmol) by adding 29 µL of one of the three fl-biotin concentrations (0.0346 µM, 0.346 µM, 3.46 µM, respectively). The solutions were vortexed for 30 seconds after the addition of the fl-biotin and then incubated for 1 hour with gentle rotation at room temperature.

After incubation, all solutions were resuspended by vortexing for 30 seconds and then 1 mL of each were tested in the MAP to separate the magnetic Dynabeads and bound fl-biotin from the supernatant containing the free fl-biotin. The top 10% of the solution (100 µL of the supernatant) was carefully removed from the cuvette, with care not to disturb the separated Dynabeads, and tested in a fluorometer (Molecular Devices SpectraMax M2e) with an excitation wavelength of 494 nm and emission measured at 523 nm. Fluorescence of each solution is given as black circles in Figure 5b in the main text, with the average fluorescence of the three trials represented by the height of the bar. The groups of MAP-tested Dynabeads and untested Dynabeads did not fluoresce with any significant difference from each other but both fluoresced significantly more than the free fl-biotin solution, indicating that MAP testing prior to incubation did not affect the Dynabeads' ability to bind proteins (Figure 5b, main text).

MAP data of the particles tested before biotin binding (from the MAP-tested Dynabeads) is given in Figure S1.10 and the effective magnetic susceptibility values are given in Table S1.7.

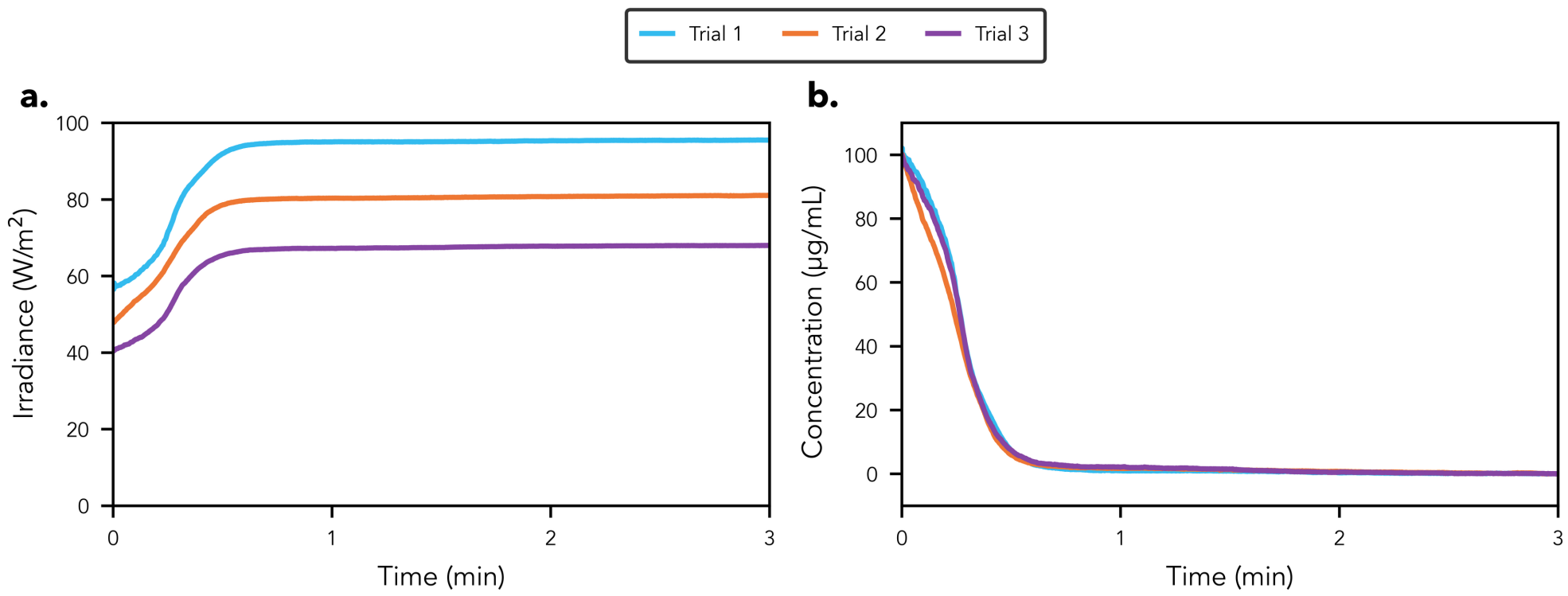

**Figure S1.10.** Dynabeads MAP data shown for 3 trials tested with the 4933 G surface field magnet. **a.** Raw MAP data. **b.** Calibrated data.

**Table S1.7.** Effective Magnetic Susceptibilities[a] of the Dynabeads Determined by the MAP.

| **Trial Number** | **MAP** $\chi_{eff}$ |
|---|---|
| 1 | 2.56 |
| 2 | 2.74 |
| 3 | 2.61 |
| **Avg.** $\chi_{eff}$ **± Std. Dev.** | 2.63 ± 0.09 |

[a]All magnetic susceptibilities are calculated in the SI system.

## 1.4. Sensor Characterization

Using the datasets from batch 1 of the iron oxide MNPs, we quantified a variety of common sensor characteristics for the MAP.[16] They are summarized in Table S1.8. The metrics and methods are described in more detail in this section.

**Table S1.8.** Summary of Sensor Characterization Metrics.

| Metric | Definition | MAP | Limiting Component(s) |
|---|---|---|---|
| Dynamic range | The total range of light intensity that can be detected | ~0-30 klux | Optical path (LED and/or light sensor) |
| Noise level | The level of noise in MAP signal as defined by 3 standard deviations from the mean | 16.2 lux (control: water)<br><br>21.6 lux (100 μg/mL MNP sample) | Optical path (LED and/or light sensor) |
| Working (linear) range | The range of concentrations for which the measured intensity change is a linear function of concentration | 16-125 μg/mL | Optical path (LED and/or light sensor), absorption of MNPs |
| Saturation Point | The concentration beyond which the measured intensity does not change | ~250 μg/mL | Optical path (LED and/or light sensor), absorption of MNPs |
| Limit of Detection | The lowest concentration that produces a change in measured intensity signal | 2 μg/mL | Optical path (LED and/or light sensor), absorption of MNPs |
| Temporal resolution | The average time between measurements | 141 ms | Light sensor, microcontroller |
| Dead time | The time after which the magnetic field is applied before a change in measured intensity is detected | 141 ms | Light sensor, microcontroller |
| Response time | The time after which the magnetic field is applied to the point at which 90% of the change in measured intensity has occurred | 22.2-60.8 s (varying magnetic field strengths)[a] | Magnetic field strength, magnetic susceptibility of MNPs |

[a]See Table S1.9 for individual response times for different magnetic field strengths.

### 1.4.1. Dynamic Range

The dynamic range is the total range of light intensity that can be detected by the MAP. This can be limited by a number of factors, including the dynamic range of the light sensor component itself and the intensity of light emitted by the LED, which is affected by the choice of resistors (and thereby the voltage difference across the LED) and manufacturing variation between specific LEDs. Based on our data, the dynamic range of our system was approximately 0-30 klux, but this may vary across the lifetime of the LED.

This is reported in lux instead of being converted to irradiance to reflect the raw output of the sensor.

### 1.4.2. Noise Level

The noise level quantifies the threshold for noise in MAP signal as 3 standard deviations above the mean measured lux value calculated using a steady state signal, also known as the $3\sigma$ noise. It is expected to vary with the LED intensity level and the individual light sensor. This value was quantified for three conditions: a control of just water and the 100 µg/mL MNP sample both with and without a magnet. Each signal was first fit to find the linear system drift. The data was adjusted and then fit to a Gaussian to calculate the $3\sigma$ noise.

For the control of water, a 3-minute collection period was used for a total of 1272 data points. The linear system drift was $-1.15 \times 10^{-3}$ W/m$^2$·s, and the $3\sigma$ noise was 0.0506 W/m$^2$. These values set the limit of detection of the system; measurements with a magnitude greater than 0.0506 W/m$^2$ is in fact signal.

For the MNP sample with a magnet, the steady-state signal from the last 1.5 minutes of a 3-minute data collection with a 100 µg/mL suspension of batch 1 $Fe_3O_4$ MNPs exposed to the 4933 G magnet. This dataset was a total of 636 data points. The linear system drift was $9.64 \times 10^{-4}$ W/m$^2$·s and the $3\sigma$ noise was 0.018 W/m$^2$. This was comparable to the control of water, which was expected since this should represent the system without MNPs (i.e. also just water).

For the MNP sample without a magnet, the signal reflects the motion of the MNPs downward due to gravity. This dataset was also taken with a 100 μg/mL suspension of batch 1 $Fe_3O_4$ MNPs, and the entire 3-minute dataset was used, for a total of 1277 individual data points. The linear system drift was $1.27 \times 10^{-2}$ W/m$^2$·s, which represents the change in signal due to gravitational, buoyant, and drag forces on the MNPs. The 3σ noise was 0.243 W/m$^2$, higher than the noise observed with just water, which is expected due to the presence of MNPs in this measurement.

### 1.4.3. Working (Linear) Range

The working range, or linear range, is the range of sample concentrations for which the output signal, or change in transmission, is directly proportional (linear) to the input concentration. This, along with the saturation point and limit of detection, can be dependent on a number of factors, including the system components (the specific LED and sensor), the wavelength of the light source, the MNP (particularly its cross-sectional size and absorption profile), and the solvent, so this metric could vary significantly. For our system, we found a linear range of about 16 μg/mL to 125 μg/mL. This also corresponded to the concentration range within which our model performed best (Table S1.8). This confers advantages in that the MAP only requires small amounts of sample to take magnetic susceptibility measurements.

This characteristic is shown in the sensor response curve in Figure 3b in the main text.

### 1.4.4. Saturation Point

The saturation point, also known as the full-scale output, is the sample concentration beyond which the MAP does not measure an increase in the overall change in intensity. We measured this value to be around 250 µg/mL for our system with batch 1 of our MNPs. It should also be noted that at higher concentrations, the MNP solution itself saturates and the MNPs do not fully suspend; this is why concentrations above 500 µg/mL were not tested. The supernatant appears translucent brown instead of clear after MAP testing in samples tested above 250 µg/mL as discussed in Figure S1.7a-c.

This characteristic is shown in the sensor response curve in Figure 3b in the main text.

### 1.4.5. Limit of Detection

The limit of detection (LOD) is the sample concentration below which the MAP does not measure any change in light transmission (i.e. generates no signal). We will define signal as a change in intensity level over the duration of the test that is greater than 3 standard deviations above the change observed for tests with no MNPs (i.e. just the solvent, water). Data from six trials with water are presented previously in the Validation data section of the main text; the first trial from the concentration sweep will be excluded due to the artifact at the initial part of that trial. The mean change in intensity observed in the remaining five trials is 70.3 ± 45.0 lux. Thus, the LOD must have a change in lux greater than 205.3 lux. While we did not optimize around this parameter, opting instead to optimize for cost and accessibility, for our system and nanoparticles, the LOD was approximately 2 µg/mL.

This characteristic is shown in the sensor response curve in Figure 3b in the main text.

### 1.4.6. Temporal Resolution

The temporal resolution, also referred to as the acquisition rate, is defined as the time between measured data points. This value was quantified for six files: 3 tests with water and 3 tests with the 100 µg/mL associated with batch 1 testing. The mean temporal resolution was 141 ms, while the median was 137 ms. The temporal resolution is determined by the integration time of the sensor, which is configured to the minimum value of 100 ms, and the processing time of the microcontroller. The minimum time between measurements actually observed was 131 ms. In contrast, within this subset of 12 data files, the maximum value ranged as high as 273 ms, which we attribute to sporadic increased processing time of the microcontroller. This may correspond to timepoints at which the microcontroller updated the display, which appeared to visually slow the microprocessor at times, but this variation in time between measurements did not appear to ultimately alter performance of the MAP. If necessary, an easy improvement would be to decrease the refresh rate of the display or to use a microcontroller with better processing performance.

### 1.4.7. Dead Time

The dead time is defined as the time after a stimulus (i.e. the magnetic field) is applied to when a change in signal is observed. In the case of the MAP, this is defined as the time between when the external magnetic field is applied and when the change in light intensity exceeds the noise level. Note that in our datasets, the magnet is applied manually at approximately t = 0 s, but because of human error, the exact timepoint at which the magnet is applied may vary. Even in the 100 µg/mL samples with the lowest applied magnetic field, a surface field of 2108 G, the light transmission increases at a rate faster than the acquisition rate of our system, which is 141 ms on average. Therefore, at this sample concentration and applied magnetic field strength the dead time is equal to (and limited by) the acquisition rate of 141 ms. We would expect the dead time to potentially increase with decreasing sample concentration or decreasing magnetic field strength.

### 1.4.8. Response Time

The response time is defined as the time period after which the stimulus (i.e. the magnetic field) has been applied to the point at which 90% of the change in measured light intensity has occurred. This value was quantified for each of the non-zero multi-magnet series trials testing a 100 µg/mL solution. The results are presented in Figure 4b in the main text, and in more detail in Table S1.9. Generally, the response time decreased with increasing applied magnetic field strength, which is as expected since the MNPs would experience a greater magnetic force. Importantly, because these response times are so short, they facilitate a short test duration. The duration of 3 minutes used in this work gives a wide margin from the response times to ensure capture of the full MNP response while still being significantly shorter than other magnetometer data collection protocols, but trial lengths of 2 minute or even 1 minute may still be sufficient to capture the full response in some cases.

**Table S1.9.** Response Times for Different Applied Magnets.

| **Surface Field of Applied Magnet (G)** | **Response Time (s)** | | | **Avg. Response Time ± Std. Dev. (s)** |
|---|---|---|---|---|
| | **Trial 1** | **Trial 2** | **Trial 3** | |
| 2108 | 60.8 | 59.1 | 56.2 | 58.7 ± 2.4 |
| 2704 | 51.0 | 47.3 | 42.6 | 47.0 ± 4.2 |
| 3682 | 32.6 | 35.9 | 30.8 | 33.1 ± 2.6 |
| 4128 | 29.1 | 35.8 | 35.0 | 33.3 ± 3.7 |
| 4933 | 22.2 | 28.8 | 31.8 | 27.6 ± 4.9 |

### 1.4.9. Thermal Effects

It was essential to avoid heating the nanoparticles or surrounding fluid during MAP tests to avoid effects on the optical signal as well as the magnetic properties of the particles themselves. Heating is not expected due to the low power optical source (the LED) used in this work. To confirm no heating occurred, a 1 mL sample of 100 µg/mL of iron oxide nanoparticles suspended in water was exposed to the LED for 90 minutes. Notably, this is 10x longer than the measurements in the main text. The temperature of the liquid suspension was measured using a FLIR One Pro camera at both the beginning and end of the 90-minute exposure period (Figure S1.11). Three points were measured for each timepoint. The measured sample temperature was 24.4 ± 0.1 °C at the beginning of the exposure and 24.7 ± 0.1 °C after 90 minutes of exposure. A control point measured 22.4 °C and then 22.2 °C, indicating some drift in either the ambient temperature or the camera measurement. It should be noted that this is within the reported accuracy of the FLIR One Pro camera of ± 3°C. Still, if we assume a baseline shift in the temperature between the two images, then we could estimate a change of approximately 0.5 °C over 90 minutes. To change the magnetic susceptibility of the nanoparticles, a much higher temperature change is needed. Therefore, over the course of the reported measurements, we do not expect any heating to occur.

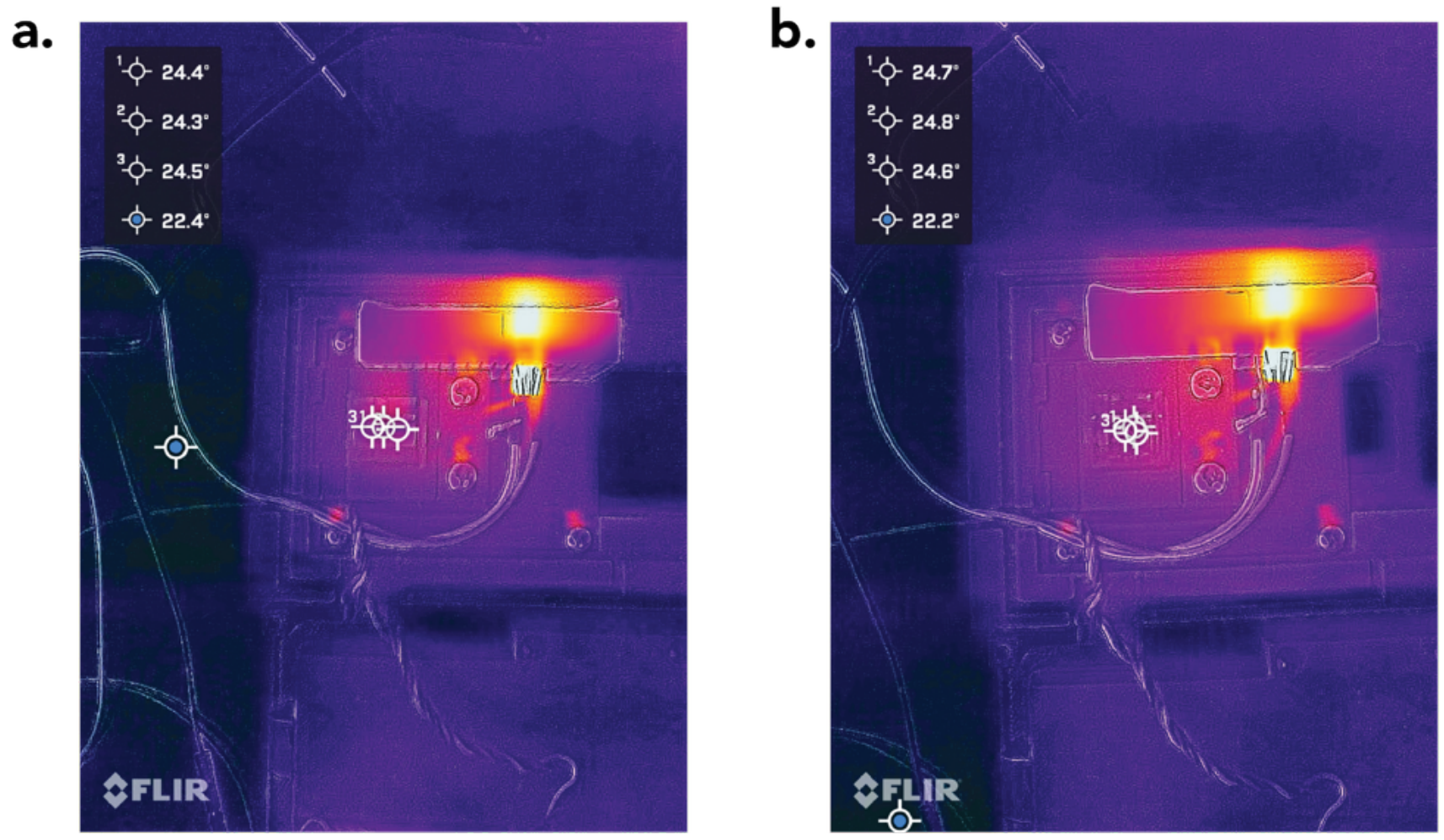

**Figure S1.11.** FLIR camera images and measurements of 100 µg/mL iron oxide MNPs suspended in water after **a.** 0 minutes and **b.** 90 minutes of direct exposure to the LED. The cover was closed during the 90-minute exposure, as it would be during MAP testing. The 3 open circle measurements are positioned on the MNP suspension while the circle with the blue center indicates the minimum measurement in the image, which was used as a control.

In addition, the optical signal was measured over a similar timespan to monitor for solvent heating. The results from these measurements are discussed in the Troubleshooting Section (SI section 1.4).

### 1.4.10. Reproducibility

To confirm reproducibility of MAP testing, we ran 10 consecutive 3-minute tests on the same 100 μg/mL sample of iron oxide MNPs suspended in water. This served both to evaluate the reproducibility of the tests and also to confirm that the MAP testing process does not cause any changes to the particles, such as thermal heating from the LED or permanent aggregation of the particles that could affect magnetic properties. The results are shown in Figure S1.12 and explicitly reported in Table S1.10.

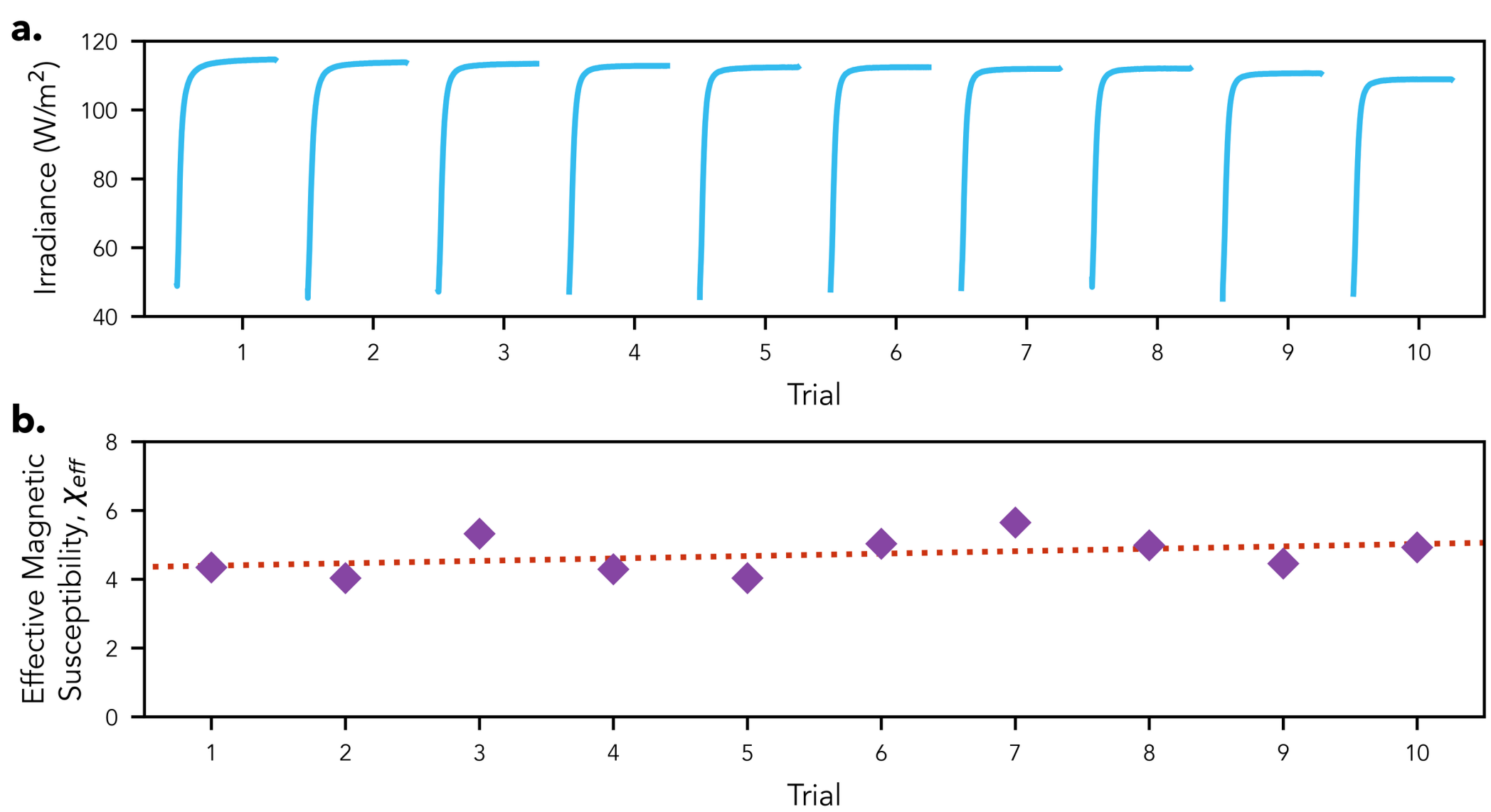

**Figure S1.12. a.** MAP light intensity data from 10 consecutive trials on the same sample of 100 μg/mL iron oxide MNPs suspended in water, tested for 3 minutes each with the 4933 G surface field magnet. **b.** The effective magnetic susceptibilities vs. the trial number, with a linear line of best fit (gray). The regression curve is $\chi_{eff} = 0.070n + 4.322$ where n is the trial number, with an $r^2 = 0.147$, indicating a very weak correlation.

**Table S1.10.** Effective Magnetic Susceptibilities[a] of Iron Oxide MNPs in 10 Consecutive MAP Tests.

| **Trial Number** | **MAP $\chi_{eff}$** |
|---|---|
| 1 | 4.34 |
| 2 | 4.03 |
| 3 | 5.32 |
| 4 | 4.29 |
| 5 | 4.03 |
| 6 | 5.03 |
| 7 | 5.65 |
| 8 | 4.98 |
| 9 | 4.46 |
| 10 | 4.92 |
| **Avg. $\chi_{eff}$ ± Std. Dev.** | 4.71 ± 0.53 |

[a]All magnetic susceptibilities are calculated in the SI system.

First, to evaluate the number of trials required for reproducible results, we evaluated all 120 subsets consisting of 3 data files from this group of 10 tests and compared the mean of those 3-trial groups to the mean of all 10 measurements using a Student's t-test. None of the 120 groups showed statistically significant differences in effective magnetic susceptibility from the full group of 10 measurements ($\alpha < 0.05$). Therefore, the use of three measurements is statistically sound, and based on the findings, the MAP provides a reproducible result for a given nanoparticle suspension.

Second, to confirm that the MAP testing did not cause any unintended changes to magnetic susceptibility such as through heating of the suspension, we fit a linear line of best fit to the effective magnetic susceptibility and found that $\chi_{eff} = 0.070n + 4.322$, where n is the trial number (Figure

S1.12b). However, this shows only a very weak correlation with $r^2 = 0.147$. Therefore, we can conclude that repetitive MAP testing causes no effects on the signal or magnetic susceptibility.

# Supplementary Information 2: Design Documentation

# Open-Source Benchtop Magnetophotometer (MAP) for Characterizing the Magnetic Susceptibility of Nanoparticles

*Alexis Scholtz[1,2], Jack Paulson[2,3], Victoria Nuñez[4], and Andrea M. Armani[1,2,3,*]*

[1]Alfred E. Mann Department of Biomedical Engineering, University of Southern California, Los Angeles, CA 90089, USA

[2]Ellison Medical Institute, Los Angeles, CA 90064, USA

[3]Mork Family Department of Chemical Engineering and Materials Science, University of Southern California, Los Angeles, CA 90089, USA

[4]Thomas Lord Department of Computer Science, University of Southern California, Los Angeles, CA 90089, USA

*aarmani@emila.org

## Table of Contents

## 2.1. Materials and Equipment

The assembly of MAP has three phases: (1) creating the data acquisition path (software), (2) building of the electronics and creating the user interface (hardware), and (3) assembly into the system packaging. Each phase is detailed in the subsequent sections. Full lists of equipment, tools, and materials needed to build this instrument are found in Tables S2.1-S2.3.

From the outset, the MAP was designed to be easy to assemble using readily accessible components. As a result, it does not require advanced equipment or tools beyond access to a consumer-grade 3D printer to create the packaging (Table S2.1). However, recently, 3D printing services have been established, so direct access to a 3D printer is not even needed.

All hardware materials (Table S2.2) are widely available from multiple sources; we opted for McMaster-Carr. In addition, all electronics (Table S2.3) were selected from companies who follow open-source principles. This philosophy allows greater access to documentation which is critical in the development of a new instrument for a teaching facility as well as broadening potential alternative parts should components become unavailable.

The system is designed to be compatible with any magnets smaller than 1” x 1” x 1/2”. We selected a series of N42 and N52 grade neodymium magnets (Table S2.4) from K&J Magnetics due to their wide variety, reliability, low cost, and excellent documentation and available resources, including modeling of the magnetic fields. However, the system housing could be easily modified to accommodate additional magnet geometries.

**Table S2.1.** Required Equipment and Tools.

| Equipment or Tool | Purpose |
|---|---|
| 3D printer | Produce the custom-designed hardware. Any 3D printer with a minimum layer height of 0.10 mm should suffice, although we used a Prusa i3 MK3S+. |
| Phillips head screwdriver #1 | Insert the #2-56 screws used for securing some of the electronics. |
| Phillips head screwdriver #3 | Insert the #4-40 screws used for securing some of the electronics and hardware. |
| MicroSD card reader | Connect the microSD card to a computer for data transfer. |
| 5 V USB Power Source | Power the LED and the control system. Note that two USB ports are required. |
| Xacto knife and/or razor blade | Remove support material and clean up 3D prints if needed (i.e. if there is stringing). |
| Needle nose pliers and/or flush cutters | Remove support material and clean up 3D prints if needed (i.e. if there is stringing). |
| Computer with Python 3 installed | Program the embedded system of the MAP and analyze data. |

**Table S2.2.** Required Hardware and Materials.

| Part name | Vendor | Vendor ID | Cost per unit[a] | Quantity needed | Total cost[a,b] |
|---|---|---|---|---|---|
| Hatchbox PLA filament[c] | Hatchbox | | 28.00/kg | 549.64 g | 15.39[d] |
| #2-56 x 3/16” Brass Pan Head Phillips Screw[e] | McMaster-Carr | 94070A076 | 0.063 | 4 | 0.25 |
| #2-56 x 3/8" Brass Pan Head Phillips Screw[e] | McMaster-Carr | 94070A079 | 0.088 | 4 | 0.35 |
| #4-40 x 1/4" Brass Pan Head Phillips Screw[e] | McMaster-Carr | 94070A106 | 0.084 | 16 | 1.34 |

[a]All prices are given by the listed vendor (the source) and are current as of June 2025. Costs shown in US Dollars (USD). [b]The total cost for all required hardware and materials is $17.33. [c]Any standard filament (i.e. PLA, PETG, ABS, etc.) may be used but we experienced smooth printing and sufficient mechanical properties with Hatchbox PLA. [d]This total cost is estimated based on the amount of filament required to print one full set of the 3D printed parts at the recommended infill percentages given in Table S2.5. [e]We recommend brass screws to avoid magnetic interactions with the neodymium magnets but other screws may also be used.

**Table S2.3.** Required Electronics.

| Part Name | Vendor | Vendor ID | Cost per Unit[a] | Quantity Needed | Total Cost[a,b] |
|---|---|---|---|---|---|
| ESP32-S2 TFT Feather – 4 MB Flash, 2 MB PSRAM, STEMMA QT | Adafruit | 5300 | 24.95 | 1 | 24.95 |
| TSL2591 High Dynamic Range Light Sensor – STEMMA QT | Adafruit | 1980 | 6.95 | 1 | 6.95 |
| LED – Super Bright Red | SparkFun | COM-00528 | 1.25 | 1 | 1.25 |
| Qwiic Openlog | SparkFun | DEV-15164 | 19.95 | 1 | 19.95 |
| Qwiic Button – Green LED | SparkFun | BOB-16842 | 4.95 | 1 | 4.95 |
| Qwiic Button – Red LED | SparkFun | BOB-15932 | 5.50 | 1 | 5.50 |
| Breadboard – Mini Modular (White) | SparkFun | PRT-12043 | 5.25 | 1 | 5.25 |
| 330 Ω Resistor[c] | SparkFun | PRT-14490 | 0.063 | 2 | 0.13 |
| Flexible Qwiic Cable – 100 mm | SparkFun | PRT-17259 | 1.95 | 3 | 5.85 |
| Flexible Qwiic Cable – 200 mm | SparkFun | PRT-17258 | 1.95 | 1 | 1.95 |
| Male-Male Jumper Wires | Adafruit | 1956 | 0.098 | 2 | 0.20 |
| SanDisk Ultra 16 GB Micro SD Card[d] | Sandisk | SDSQUAR-016G-GN6MA | 6.19 | 1 | 6.19 |
| USB Type A to Type C cable (3 ft)[e] | Adafruit | 4474 | 4.95 | 1 | 4.95 |
| USB Type A Plug Breakout Cable with Premium Female Jumpers – 30 cm long | Adafruit | 4448 | 1.95 | 1 | 1.95 |

[a]All prices are given by the listed vendor (the source) and are current as of June 2025. Costs shown in US Dollars (USD). [b]The total cost for all required electronics is $90.02. [c]A single resistor with a resistance of at least 130 Ω would serve as an alternative. [d]Any microSD card could be used instead; SparkFun recommends at least a class 6 card for use with the OpenLog. This capacity has proven sufficient to run over 10000 three-minute tests without running out of memory. [e]This cable will be used to program and power the Feather. The Feather has a Type C connector; the other side may be Type A or C depending on your power source and computer USB ports.

**Table S2.4.** Magnets Used in This Work.

| Part Name | Vendor | Vendor ID | Cost per Unit[a] | Quantity Needed | Total Cost[a,b] |
|---|---|---|---|---|---|
| 1” x 1” x 3/16” Nickel Plated (N42) Magnet | K&J Magnetics | BX0X03 | 6.91 | 1 | 6.91 |
| 1” x 1” x 1/4” Nickel Plated (N42) Magnet | K&J Magnetics | BX0X04 | 8.81 | 1 | 8.81 |
| 1” x 1” x 3/8” Nickel Plated (N42) Magnet | K&J Magnetics | BX0X06 | 12.61 | 1 | 12.61 |
| 1” x 1” x 3/8” Nickel Plated (N52) Magnet | K&J Magnetics | BX0X06-N52 | 16.55 | 1 | 16.55 |
| 1” x 1” x 1/2" Nickel Plated (N52) Magnet | K&J Magnetics | BX0X08-N52 | 21.60 | 1 | 21.60 |

[a]All prices are given by the listed vendor (the source) and are current as of June 2025. Costs shown in US Dollars (USD). [b]The total cost for all magnets used in this work was $66.48.

## 2.2. System Design

The MAP was designed with open-source principles in mind from the start. It has been designed to be easy to operate, easy to assemble, and inexpensive to improve accessibility. The design itself, including editable 3D design files, the code to run the embedded system, and the software to run the data analysis is all published on our Github.[1] This facilitates modification of the MAP should a user have a specific need for improvement or a new feature.

In this section, we provide an overview of the various hardware components of the MAP: the electronic components and the custom-designed packaging. For the electronic components, we have detailed additional features that were key in the selection of that item. In line with the principles of open-source hardware, we also provide possible alternatives should the specified parts no longer be available or important considerations should the user opt to replace a component with one of their own choosing. For the packaging, we provide print instructions for each of the component files.

We also give an overview of the software design for both the embedded system and the data analysis software. We detail the operation of the MAP and use of the standalone GUI to complete data collection as well as give a brief introduction to the features of the open-source data analysis program that implements our mathematical model.

### 2.2.1. Electronic Components

Based on our overall design constraints, the components all need to be off-the-shelf, inexpensive, and not require soldering. To satisfy this requirement, we chose to utilize $I^2C$ communication (Figure S2.1), which reduces the complexity of wiring multiple components together, removes the need to solder header pins onto boards, and makes the electronics system much easier to assemble for an inexperienced user. We chose to take advantage of the cross-compatible SparkFun Qwiic and Adafruit STEMMA QT ecosystems. Compatible boards have ports for the 4-wire cables to be connected serially, and the details of the $I^2C$ communication protocols are handled by manufacturer-provided code libraries. The libraries also provide convenient functions for the user to customize code. While modification is not necessary to use our device as initially designed, the open-source nature of both the physical hardware and the software libraries that we chose to employ allows the end user full control of their system should changes be required or other features be desired.

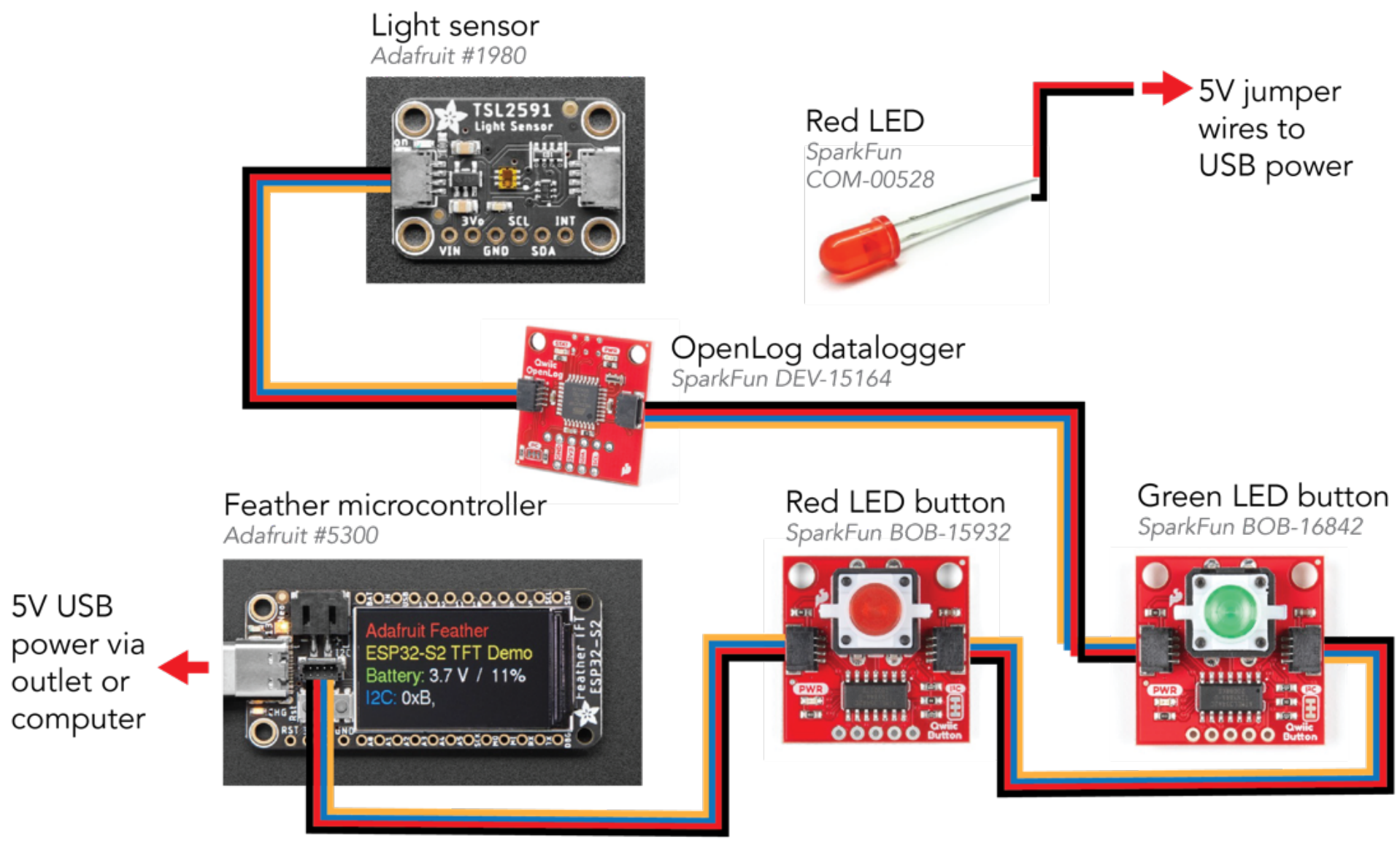

**Figure S2.1.** A schematic showing the wiring between various electronic components and the power sources for each part of the electronics system. Black, red, blue, and yellow lines represent the 4-wire QWIIC cables that connect various boards. The red and black wires to the LED circuit are the +5 V and ground wires. Note that the LED is powered independently from the rest of the electronics, which are powered through shared I$^2$C communication lines.

**Microcontroller and Display: Feather ESP32-S2 TFT**

The "brain" of the system, the Adafruit Feather, is a microcontroller family built around the open-source Arduino principles. Specifically, the Adafruit Feather ESP32-S2 TFT board (Adafruit #5300), known throughout the rest of the supplementary material as simply the "Feather," was selected because it boasts both a STEMMA QT I$^2$C port and a built-in TFT display.

The 240 pixel x 135 pixel display is the main output of the MAP's user interface, which is a crucial component of making the system self-contained. The USB-C port on the board provides power not only to the Feather itself but also to the rest of the components connected via the I$^2$C cables. The Feather is also compatible with both CircuitPython and the Arduino Programming Language. We chose to program the MAP using the Arduino C++ instead of Python to optimize library usage across all the chosen boards and maximize program execution speed to facilitate faster data collection.

It should be noted that this Feather could be replaced by using another Arduino-compatible board, such as a SparkFun Redboard, as long as it is also STEMMA QT- or Qwiic-compatible. If the microcontroller board does not include a screen, one must be added to allow for a truly embedded system. There is no required screen dimension. However, our code is optimized for the 240 pixel x 135 pixel display found on the Adafruit Feather ESP32-S2 TFT, so a 16:9 aspect ratio would allow for more seamless adaptation of the provided MAP code. The Adafruit ESP32-S3 Reverse TFT Feather (Adafruit #5691) would be an excellent alternative, although it was not available at the time of hardware development for this work, and slight modifications to the 3D-printed base and panel pieces would need to be made for this board.

**Light Sensor: TSL2591**

There are a wide variety of light sensors available even once $I^2C$ compatibility is considered as a filter. We chose the Adafruit TSL2591 (Adafruit #1980) for its high dynamic range and its sensitivity in the visible light region. It has two photodiodes for infrared and full-spectrum measurements that can be used to detect the optical signal. This chip measures intensity values from 188 µlux up to 88 klux, giving the MAP the capability to measure a wide range of solution concentrations and particle responses to magnetic field application. It also has the option of easily tuning the integration time and the gain of the TSL2591 sensor in the software. After some optimization, these were both set at their minimum values of 100 ms for the integration time and 1 for the gain to maximize the MAP's temporal resolution between measurements and sensitivity.

Alternate options for light sensors include a VEML7700 lux sensor (Adafruit #4162) and BH1750 ambient light sensor (Adafruit #4681), both of which are packaged on the same size breakout board and offer Qwiic/STEMMA QT compatibility, making them straightforward candidates for alternate light sensors that would fit right into the packaging of the MAP. However, we selected the TSL2591 as the superior option for our application.

**Light Source: Red LED**

Many options are available for light sources. The choice of SparkFun's Super Bright Red light emitting diode (LED) (COM-00528) represented an ideal mix of several factors. First and foremost, red LEDs are cheap as well as reliably and widely available. The wavelength of red light is within the high sensitivity range of the selected light sensor, the TSL2591. The availability of the "super bright" variety from SparkFun increases the maximum power of the LED source without resorting to a laser, which readily saturates the light sensor. By increasing the maximum intensity from the LED, more concentrated solutions may be used in the MAP.

Any infrared or visible LED would work as a replacement should the SparkFun COM-00528 no longer be available. As general design criteria, we suggest selecting a bright LED to maximize compatible concentration ranges. However, we recommend against using a white LED. These are typically made using three separate LEDs (red, green, and blue), which cycle on and off. This could cause small oscillations in the optical signal detected, introducing noise in the data.

A current-limiting resistor must always be placed in series with an LED to avoid damage to the device from over-drawing current (Figure S2.2). This resistor value $R$ (in Ohms) can be determined using the following equation derived from Ohm's law:

$$R \geq \frac{V_S - V_{LED}}{I_{LED}}, \tag{S2.1}$$

where $V_S$ is the voltage supply source in Volts (5 V from a USB supply as designed), $V_{LED}$ is the forward voltage drop of the LED in Volts (2.4 V for SparkFun COM-00528), and $I_{LED}$ is the current through the LED in Amperes (0.02 A or 20 mA for SparkFun COM-00528). We chose to use two 330 Ω resistors in parallel to produce an equivalent resistance of 165 Ω, but any resistor over 130 Ω would suffice for the COM-00528. This simple calculation described by eq S2.1 should be repeated if a different LED or voltage source is used to select an appropriate current-limiting resistor to avoid damaging the LED.

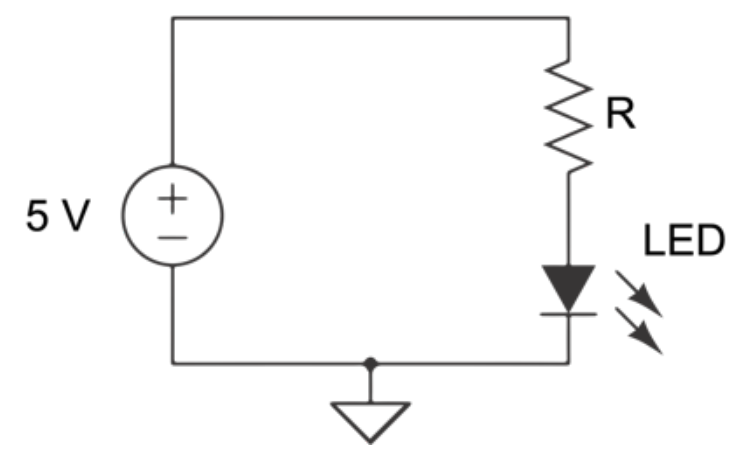

**Figure S2.2.** Schematic of the LED circuit, showing the 5 V voltage source, the resistor (R), and the LED.

**Hardware User Interface: Qwiic Buttons with Integrated LEDs**

Because this system is intended for benchtop use, the development of a fully embedded control and data acquisition system was prioritized. A key element of an embedded system is the user interface that allows the user to operate the system. The user interface of the MAP consists of the TFT display in the Feather as well as two of SparkFun's Qwiic Buttons, each of which have an integrated LED, one green (SparkFun BOB-16842) and one red (SparkFun BOB-15932). The two buttons form the simplest user interface possible with the capabilities required of the MAP, and the LEDs provide an additional measure of intuitive, instantaneous feedback to the user that the display cannot provide alone.

The Qwiic buttons provided the cleanest packaging for a button and LED. Of course, they could be replaced with individual components and wired using a breadboard.

**Data Acquisition: Qwiic OpenLog**

The integrated data acquisition system consists of a microSD card and the SparkFun Qwiic OpenLog (SparkFun DEV-15164), which handles writing to the microSD card. The data will be saved in tab-delimited format as a text (.txt) file that can be transferred to a computer for further analysis in a program of the user's choosing, such as Microsoft Excel or Origin, or via a script written in MATLAB or Python. The OpenLog is capable of writing with speeds up to 20 kilobytes per second, sufficient for use in the MAP.

As an essential piece of hardware in the embedded system, it may prove difficult to replace while retaining $I^2C$ compatibility. Adafruit offers a Feather "Wing" capable of writing to microSD cards, but it will require soldering header pins onto the Feather.

## 2.2.2. 3D-Printed Parts for System Packaging

Files for the system packaging were designed in Autodesk Fusion 360 and are available on our Github repository as both editable .step files and ready-to-print .stl files. Files for the 3D-printed parts (Table S2.5) can be found in the zip file named "3D_prints.zip" in the Github repository.[1] For the 3D printing itself, we chose to use a Prusa i3 MKS3+ 3D printer with Hatchbox PLA filament, but any common filament (i.e. PLA, PETG, ABS) and 3D printer with a minimum layer resolution of at least 0.10 mm and a build volume of at least 150 mm x 180 mm x 33 mm will suffice. All parts were designed without small features and with minimal need for print supports, except as noted. The recommended infill and maximum layer height for each file can be found in Table S2.5, and the recommended print orientation is shown in Figure S2.3. Note that although color of the parts should not matter, we recommend printing the parts in the vicinity of the testing stage in matte black filament to minimize light reflections and scattering that could interfere with measurements; these parts are noted in Table S2.5.

**Table S2.5.** 3D Printed Parts.

| File Name | Recommended Maximum Layer Height (mm) | Recommended Minimum Infill Percentage | Print Time Estimate[a,b] | Filament Estimate[a,c] (g) |
|---|---|---|---|---|
| System base | 0.20 | 50% | 27h 42m | 264.22 |
| Top panel[d] | 0.20 | 50% | 7h 56m | 93.85 |
| Testing stage[e] | 0.10 | 100% | 4h 15m | 26.07 |
| Sensor pinhole[e] | 0.10 | 100% | 0h 19m | 1.65 |
| LED cover[e] | 0.10 | 100% | 0h 22m | 1.19 |
| Stage cover[e] | 0.20 | 100% | 7h 21m | 88.18 |
| Magnet slider | 0.10 | 50% | 2h 32m | 15.50 |
| Magnet storage drawer | 0.20 | 50% | 2h 47m | 35.06 |
| Magnet storage lid (3 needed)[f] | 0.10 | 100% | 0h 33m | 3.04 |
| Magnet spacer – 3/16” | 0.10 | 100% | 0h 45m | 6.39 |
| Magnet spacer – 1/4" | 0.10 | 100% | 0h 36m | 5.12 |
| Magnet spacer – 3/8” | 0.10 | 100% | 0h 20m | 2.64 |
| Magnet storage latch | 0.20 | 50% | 0h 06m | 0.65 |

[a]Print times and filament estimates are based on a Prusa i3 MK3S+ 3D printer with the specified layer height, infill percentage, and orientations given in Figure S2.3. [b]Total print time is estimated to be 56 hours and 40 minutes. [c]Total mass of filament required is estimated to be 549.64 g. [d]Supports are recommended for minor elements of this print. Print time and filament estimates both reflect the use of organic supports. [e]For these pieces which border the optical path and enclose the testing area, we recommend they be printed in matte black filament at full infill to reduce the effects of reflection and scattering. [f]Note that 3 of these lids are needed to cover all slots in the magnet storage drawer. The individual listed print time and filament estimates are given for one individual magnet storage lid, but the total print time and mass of filament for the full MAP includes all 3 lids.

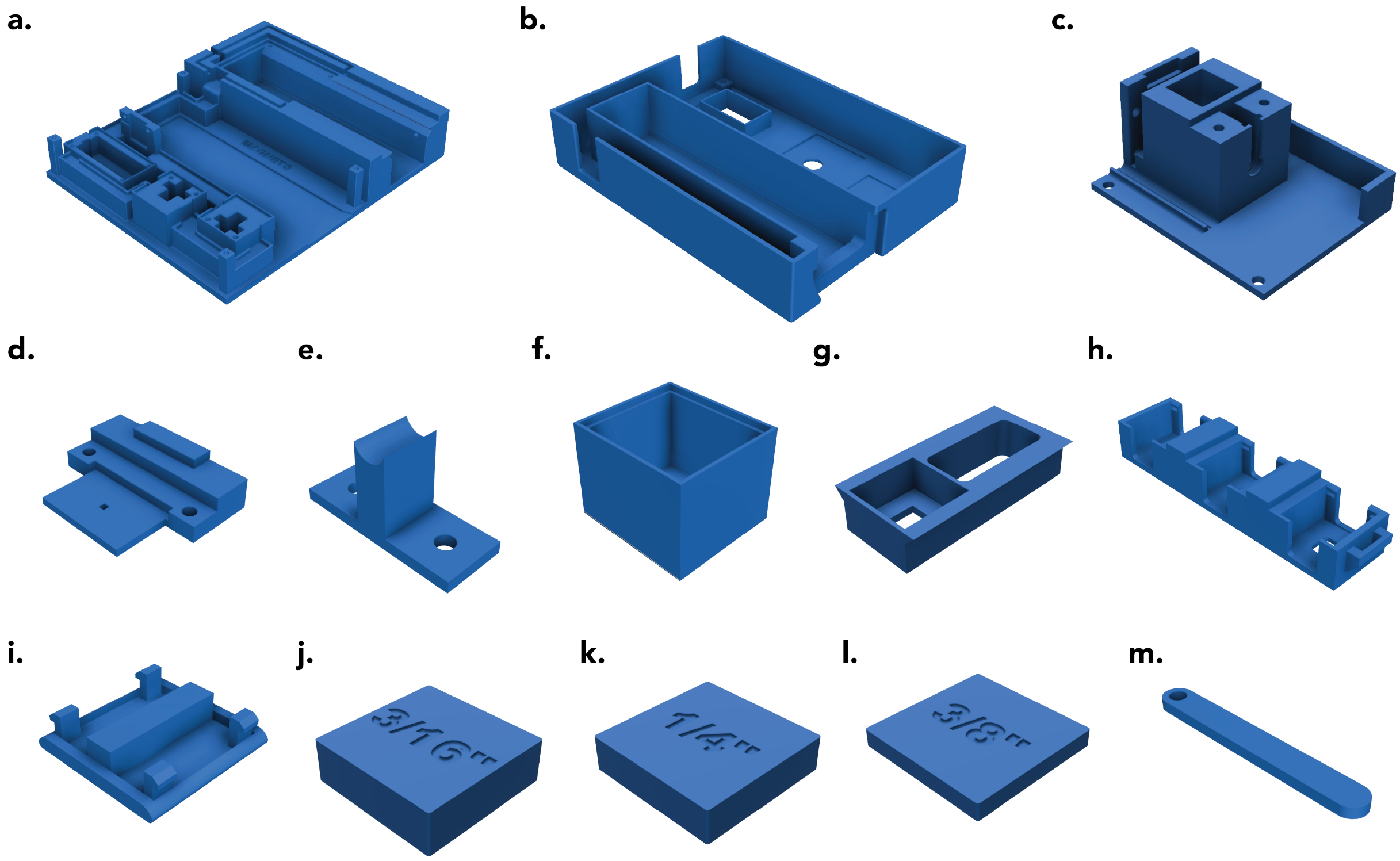

**Figure S2.3.** Print orientations for all 3D printed parts: **a.** system base, **b.** top panel, **c.** testing stage, **d.** sensor pinhole, **e.** LED cover, **f.** stage cover, **g.** magnet slider, **h.** magnet storage drawer, **i.** magnet storage lid, **j.** magnet spacer – 3/16", **k.** magnet spacer – 1/4", **l.** magnet spacer – 3/8", and **m.** magnet storage latch.

### 2.2.3. Software Design

We wrote custom, open-source software to complete the embedded data acquisition system on the MAP as well as scripts to visualize and analyze the data, which are all released on the Github repository.[1] The code running the MAP is written in the Arduino Programming Language and utilizes several libraries released by Adafruit and SparkFun. The program communicates with the hardware, provides a graphic user interface that allows the user to control the system, and collects and saves the data. The MAP data analysis program (MAP-DAP), written in Python, utilizes open-source packages often used in scientific data analysis.

**Control System and Graphic User Interface (GUI)**

The embedded system is fully coded in Arduino C++. The software of the control system is structured in a state machine to facilitate the user experience, with each state corresponding to a different screen that is displayed on the GUI (Figure S2.4) and to different backend functions. These states will be described in more detail, including descriptions of their display and functions, in an effort to describe the control flow utilized by the MAP.

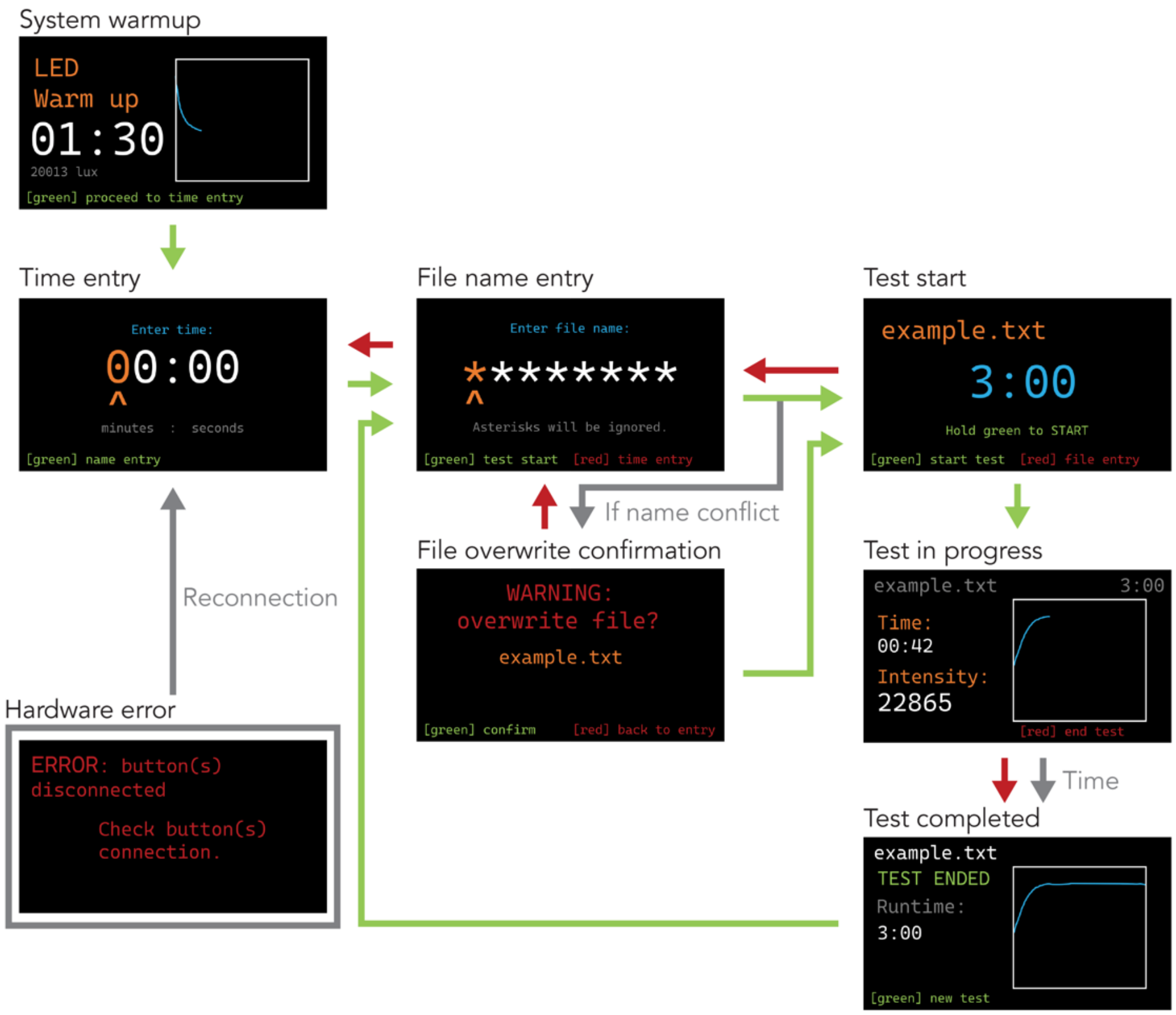

**Figure S2.4.** Renderings of the GUI displays with a simplified control flow shown. A green arrow indicates a long hold of the green button. A red arrow indicates a long hold of the red button. Gray arrows indicate a condition not related to user button presses. The user should hold the green button to proceed with MAP testing after system warmup. The user can navigate through time entry and file name entry by holding the green and/or red buttons. If there is a conflict with the entered file name, the MAP will prompt the user to confirm the file overwrite. Otherwise, the MAP will ask the user to confirm the entered test parameters, and holding the green button will begin the test. During the test, the MAP will display a preview of the data and the current intensity readings, updated approximately every second. At the conclusion of the test, the MAP will show the full data preview and if the user holds the green button, the MAP will reset to file name entry.

User interaction is primarily through the red and green buttons, with any system outputs shown on the display on the Feather. Both buttons have three different selection options based on the length of the button press, which are indicated by the brightness of the LED within the button in real-time. It should be noted that a press is not registered until the button is released. The modes of interaction are summarized in Table S2.6 but is described in more detail below.

**Table S2.6.** User Interaction Options.

| Description | Duration of Button Hold (s) | LED Behavior | Effect | |
|---|---|---|---|---|
| | | | Green Button | Red Button |
| Quick click | < 1 s | LED turns on | Increments current selection | Decrements current selection |
| Short hold | 1-2 s | LED flashes once to second brightness level | Advances the selection within the current screen | Returns the selection to the previous one within the current screen |
| Long hold | > 2 s | LED flashes twice to the third brightness level | Advances to the next screen | Returns to the previous screen |

The first option is a quick click, defined as the user holding down the button for less than one second, which will cause the LED in the button to turn on. This quick click is used to increment (green button) or decrement (red button) the currently selected character. The second is a short hold, where the user holds the button down for between one and two seconds and the button flashes once; this interaction option is used to advance the current selection to the next character or return the current selection to the previous character. The third and final option is the long hold, when a button is held down for at least two seconds and will flash twice, getting sequentially brighter. Generally, this long hold will be used to advance to the next screen (if the green button is held) or return to the previous screen (if the red button is held) within the GUI. We have included a visual quick guide that includes the instrumentation controls and the operating process flow as a one-page printable document on our Github.[1]

When the MAP is powered on, the screen will display a timer counting up indefinitely from zero. It is intended to be a timer for the user to allow the LED to stabilize; therefore, it is the user's responsibility to allow sufficient time for the LED to warm up and stabilize prior to taking their readings to avoid introducing artifacts to their measurements. The Troubleshooting section of this document discusses this in more detail. Once the LED signal has stabilized, after about 30 minutes in our experience, the user should execute the long hold on the green button to progress to setting test parameters.

The first test parameter to be set is the test duration. The time is entered in a "mm:ss" format, with a maximum time of 99 minutes and 59 seconds. Using quick clicks of the green and red buttons to increment and decrease digits respectively, and short holds to navigate between the digits, advancing and reversing one selection with the green and red buttons respectively, the user can enter the duration of the test. Once done, the user should perform a long hold of the green button to progress to the file name entry.

Entering the file name is performed in the same manner as setting the time. The file name may be up to 8 characters, which will be appended with ".txt". These characters are initially represented as asterisks but any unchanged asterisks will be removed from the final file name (i.e. an entry of "test****" will result in a file name of "test.txt"). Character options are the lowercase alphabet, digits 0-9, an underscore, and an exclamation point, in that order. Once the user is satisfied with their entered file name, the user should perform a long hold of the green button to progress to the data collection start. Once entered, the MAP will check to confirm that no file exists on the microSD card with that file name; if there is file of the same name, the MAP will prompt the user to confirm that they want to overwrite the previous file, and if not, will return them to the file name entry screen to modify their entry.

Once the file name has been entered, the screen will display the entered file name and test duration for confirmation. At this point, the sample should be resuspended, inserted into the MAP, and covered in preparation for the test. A long hold of the green button will begin the testing process. The display will prompt the user to push the magnet into place after a 3-second countdown, at which point data collection will begin automatically.

During data collection, the display will show the time elapsed from the beginning of data collection, the current intensity, and a preview of the data. The time and intensity measurements will update every second; the preview of the data will update at intervals corresponding to about every 1/100th of the total test duration to generate the preview of the whole data file (i.e. a 3-minute test would result in points being added to the plot every 1-2 seconds). However, note that the MAP is still logging measurements more frequently, on the scale of 100-150 ms, depending on the performance of the microcontroller. If the user wants to end the test early, a long hold of the red button will end data collection.

When data collection is finished either by completing the full test duration or due to user intervention, the MAP will display the full preview plot and the file name until the user performs a long hold of the green button to return to file name entry. At this point, a new test may be run. Note that the MAP will skip the step of time entry for ease of running consecutive tests, but if the test duration needs to be modified, a long hold of the red button during file name entry will return the MAP to the time entry stage.

If at any point during the use of the MAP, there is a hardware connection error, the MAP will notify the user and prompt them to check the $I^2C$ connections between the hardware components. Once connections have been properly re-established, the MAP will return to time entry.

Note that if the microSD card is removed from the MAP at any point, the MAP will need to be restarted (i.e. unplugged and powered back on) after the microSD card is re-inserted in order for data to be saved properly.

**Data Analysis Program**

We implemented the mathematical physics model into a Python script and further developed analysis software to facilitate easier access to magnetic susceptibility measurements by removing the need to rederive the physical mechanism behind our measurement. The MAP Data Analysis Program (MAP-DAP) has a graphic user interface (GUI) coded using the open-source Tkinter library for Python and runs on Python 3. It allows the user to batch select multiple files for visualization and analysis and easily plot the raw MAP data and analyze the files to find the magnetic susceptibility. MAP-DAP is published on our Github.[1] There is also a simpler command line version for quick individual file analysis.

## 2.3. Assembly Instructions

## 2.3.1. Software Setup

Some setup is required before the code can be successfully uploaded to the MAP's system. The required downloads are summarized in Table S2.7. To download and install all software requirements, on a laptop, follow the steps below. This approach is cross-platform and has been tested on Mac OS 14.2 and Windows 11.

**Table S2.7.** Required Downloads.

| **Name** | **Version**[a] | **Link** |
|---|---|---|
| Arduino IDE | 2.3.6 | https://www.arduino.cc/en/software |
| ESP32 Package for Arduino Board Manager | 3.2.0 | https://raw.githubusercontent.com/espressif/arduino-esp32/gh-pages/package_esp32_dev_index.json |
| SparkFun Qwiic Button Arduino Library | 2.0.6 | https://github.com/sparkfun/SparkFun_Qwiic_Button_Arduino_Library |
| SparkFun Qwiic OpenLog Arduino Library | 3.0.2 | https://github.com/sparkfun/SparkFun_Qwiic_OpenLog_Arduino_Library |
| Adafruit TSL2591 Library | 1.4.5 | https://github.com/adafruit/Adafruit_TSL2591_Library |
| MAP_main | 2.0 | https://github.com/armanilab/MAP |

[a]Version used in the MAP. Current as of June 2025.

1. Download and install the Arduino IDE if it is not already installed on the computer: https://www.arduino.cc/en/software
2. For the steps 3-5, follow the instructions given by Adafruit to setup the Feather and Arduino IDE. Major steps are reproduced in brief below, but full detailed instructions can be found at this link: https://learn.adafruit.com/adafruit-esp32-s2-tft-feather/arduino-ide-setup
3. Set up the Arduino IDE for use with the Feather.
    a. In the Arduino IDE Preferences, add the following URL to the text field for Additional Boards Manager URLs: https://raw.githubusercontent.com/espressif/arduino-esp32/gh-pages/package_esp32_dev_index.json and click OK to save settings.
    b. In the Arduino IDE, navigate to Tools > Board > Board Manager, search “esp32,” and click the Install button to install the esp32 package by Espressif Systems.

c. Plug the Feather into the computer. The computer and Arduino IDE should now recognize the Feather. To confirm, select the Feather as the board by navigating to Board > Adafruit Feather ESP32-S2 TFT.

4. Setup and confirm the Feather is ready to be programmed. Major steps are reproduced in brief below, but full detailed instructions can be found at this link: https://learn.adafruit.com/adafruit-esp32-s2-tft-feather/using-with-arduino-ide
   a. The first time the Feather is used, it needs to be placed into ROM bootloader mode. Hold down the DFU/Boot0 button, click the Reset button, and then release the DFU/Boot0 button.
   b. In the Arduino IDE, navigate to Tools > Port and select the one that says "ESP32S2 Dev Module."
   c. To confirm that the Feather is properly setup, open up the Blink sketch from Examples > 01. Basics > Blink and upload it to the Feather. If a warning appears, this is fine and press the Reset button on the Feather. The onboard LED (labeled LED 13) should repeatedly turn on for one second and then turn off for one second.
5. Install the Arduino libraries for the TFT display on the Feather. Major steps are reproduced in brief below, but full detailed instructions can be found at this link: https://learn.adafruit.com/adafruit-esp32-s2-tft-feather/built-in-tft
   a. Open the Library Manager from Sketch > Include Library > Manage Libraries…
   b. Search for "ST7789" and install the Adafruit ST7735 and ST7789 Library.
   c. Select "Install all" to include all dependencies. If you have not used Adafruit products before, these may include Adafruit GFX Library, Adafruit BusIO, and Adafruit seesaw Library; if you have some or all of these libraries installed already, you may not get this message.

6. Install the Arduino libraries for the STEMMA/QT hardware. These are open-source libraries provided by SparkFun and Adafruit that can be downloaded from their respective Github locations. Once downloaded, to install these in the Arduino IDE, navigate to Sketch > Include Library > Add .ZIP Library and select the .zip file for each library. Additional methods to install libraries may be found here if this method does not work: https://support.arduino.cc/hc/en-us/articles/5145457742236-Add-libraries-to-Arduino-IDE
    a. SparkFun Qwiic Button library: https://github.com/sparkfun/SparkFun_Qwiic_Button_Arduino_Library
    b. SparkFun Qwiic Openlog library: https://github.com/sparkfun/SparkFun_Qwiic_OpenLog_Arduino_Library
    c. Adafruit Sensor library: https://github.com/adafruit/Adafruit_Sensor
    d. Adafruit TSL2591 library: https://github.com/adafruit/Adafruit_TSL2591_Library
7. Download the MAP repository from: https://github.com/armanilab/MAP.[1] Unzip the file and save the folder at your chosen location. The unzipped directory should be titled "magnetophotometer." All code can be found in the "code" directory: "MAP_main" contains the Arduino code to run the full MAP system; "hardware_tests" contains code to test various hardware components; and "data_vis" contains the data analysis program.
8. Change the $I^2C$ address of the green button. By default, both buttons have an $I^2C$ address of 0x6F, but to connect both devices simultaneously to the Feather, they must have unique addresses. See Table S2.8 for the default and recommended $I^2C$ addresses of the various Qwiic/STEMMA QT devices.
    a. For this step, connect only the green Qwiic button to the Feather. All other $I^2C$ devices must be disconnected. Any Qwiic cable may be used.

b. In the Arduino IDE, navigate to the magnetophotometer > hardware_tests > change_button_i2c and open change_button_i2c.ino.

c. Upload the program to the Feather.

d. When prompted, enter “60” to change the $I^2C$ address of the green button to 0x60. Confirm that the device address was successfully changed.

e. Note that the $I^2C$ address may be changed to any other address as long as it is not taken by another device. If another address besides 0x60 is chosen, be sure to change the variable GREEN_I2C_ADDRESS in the MAP_main.ino file to reflect that.

**Table S2.8.** $I^2C$ Addresses.

| Device | $I^2C$ Address |
| --- | --- |
| SparkFun Qwiic Red Button | 0x6F |
| SparkFun Qwiic Green Button | 0x60[a] |
| SparkFun Qwiic OpenLog | 0x2A |
| Adafruit STEMMA QT TSL2591 | 0x28 and 0x29 |

[a]Changed from default.

9. (Optional but highly recommended) Confirm that the libraries and hardware are connected properly by running the following test programs in the hardware_tests folder:

   a. Open display_testing.ino in the Arduino IDE and upload it to the Feather. The display should turn on and show 3 lines of text: “display test”, “99:99,” and “Hold GREEN to start.” Note that pressing the green button will not actually have any effect.

   b. Connect both the red and green buttons to the Feather using any Qwiic cables. Open two_button_test.ino in the Arduino IDE and upload it to the Feather. If the buttons are connected and the $I^2C$ address has been changed properly, the Serial Monitor in the IDE should say the buttons were connected successfully. When a button is pressed, it should report the button color and whether the button was held for less than one second (clicked), held for one to two seconds (short hold), or held for more than two seconds (long hold).

c. Connect the light sensor to the Feather using any Qwiic cable. Open sensor_test.ino in the Arduino IDE and upload it to the Feather. If the TSL2591 is connected properly, the Serial Monitor will confirm successful connection to the sensor and will repeatedly print the lux readings.

d. Connect the OpenLog to the Feather using any Qwiic cable. Open openlog_test.ino in the Arduino IDE and upload it to the Feather. If the OpenLog is connected properly, the Serial monitor will confirm successful connection to the OpenLog and write a file called "test.txt." to the microSD card. When removed from the OpenLog and connected to the computer, the microSD card should contain a file called "test.txt" containing one line reading "hello world!"

The software setup is now complete; we recommend fully assembling the hardware before continuing with the final program upload.

## 2.3.2. Hardware Assembly

An abbreviated set of instructions are included as a one-page printable document on our Github,[1] but the full assembly instructions are detailed here. We recommend the following order of assembly:

1. Print all required packaging parts. See Table S2.5 for the names of individual 3D printed parts with suggested print settings and Figure S2.3 for suggested print orientations.
2. Connect the electronics with the Qwiic cables and orient them properly (Figure S2.1).
    a. Connect a 100 mm Qwiic cable to the Feather and the left port of the red button (when red button is oriented with the text facing the front).
    b. Connect a 100 mm Qwiic cable to the right port of the red button and the right port of the green button (when oriented with the text facing the front). Turn the green button so that the text now faces away.
    c. Connect the last 100 mm Qwiic cable to the left port of the green button and the right port of the OpenLog.
    d. Connect the 200 mm Qwiic cable to the left port of the OpenLog. This will be attached to the light sensor, but for now, leave it disconnected.
3. Attach the connected electronics to the system base (Figure S2.5a). The base should be oriented so that the text embossed into it is readable and facing the front (i.e. towards the user).
    a. Place the Feather in the bottom left mount in the system base marked with "FEATHER," oriented so that the USB-C port is on the left side. Attach it with four #2-56 3/16" screws (Figure S2.5b).
    b. Place the red button in the square mount marked "R," oriented so that the two screw holes align with those in the base. Attach it with two #4-40 1/4" screws (Figure S2.5c).
    c. Place the green button in the square mount marked "G," oriented so that the two screw holes align with those in the base. The text should appear facing away from the user, with the green button itself closer to the front of the base. Attach it with two #4-40 1/4" screws (Figure S2.5d).

d. Place the OpenLog in the vertical mount located in front of the text "OPEN LOG". The board should slide into the opening in the mount, oriented so that the microSD card slot is pointing up. Attach it with two #4-40 1/4" screws (Figure S2.5e).

e. (Optional but recommended) Use tape to secure the Qwiic cables in their channels. The 200 mm cable between the OpenLog and the sensor may need to be bundled up to fit nicely.

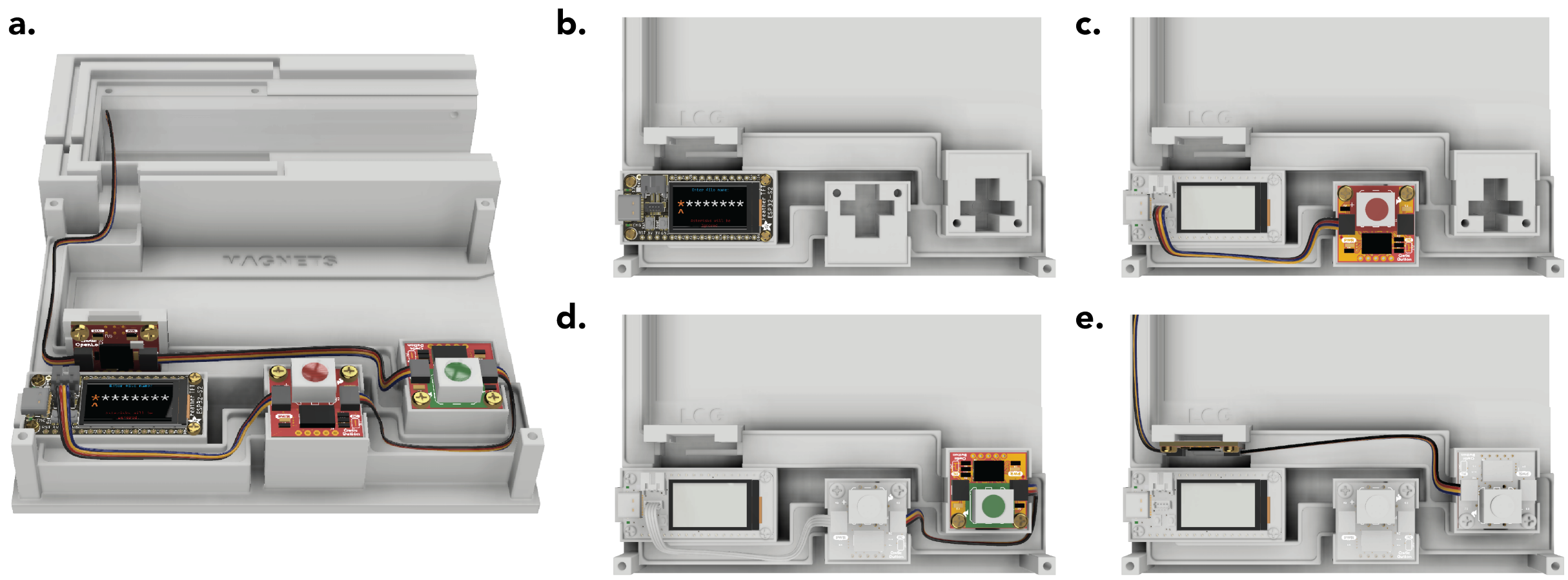

**Figure S2.5. a.** The MAP fully assembled after step 3, with all electronics secured to the base. All new parts are shown in color. **b.** Secure the Feather with four #2-56 3/8" screws. **c.** Secure the red button with two #4-40 1/4" screws. **d.** Secure the green button with two #4-40 1/4" screws. **e.** Secure the OpenLog with two #4-40 1/4" screws.

4. Build the LED circuit on the breadboard. See the recommended placement of the components as shown in Figure S2.6.

    a. Place the LED such that the two legs are on opposite sides of the "river" (center division) of the breadboard. This will make it easier to eventually fit the LED into the LED slot of the testing stage in a later step. We recommend placing the longer,

positive lead of the LED in breadboard hole 11f and the shorter, negative lead of the LED in breadboard hole 11d.

b. Place two 330 Ω resistors in parallel between the positive lead of the LED and an empty row on the breadboard. We recommend placing the legs of the two resistors in breadboard holes 11j and 15j, and 11i and 15i, respectively. The resistor legs may be trimmed so that the resistors sit lower on the breadboard (as shown in Figure S2.6), but it is not necessary.

c. Connect either two jumper wires or two 22 AWG wires with stripped ends to the red and black wires of the USB cable to jumper wire. Insert the jumper wire connected to the red positive wire in the same row as the currently unconnected end of the resistor (row 15), such as in breadboard hole 15f. Insert the wire connected to the black negative wire in the same row as the negative end of the LED (row 11), such as in breadboard hole 11a.

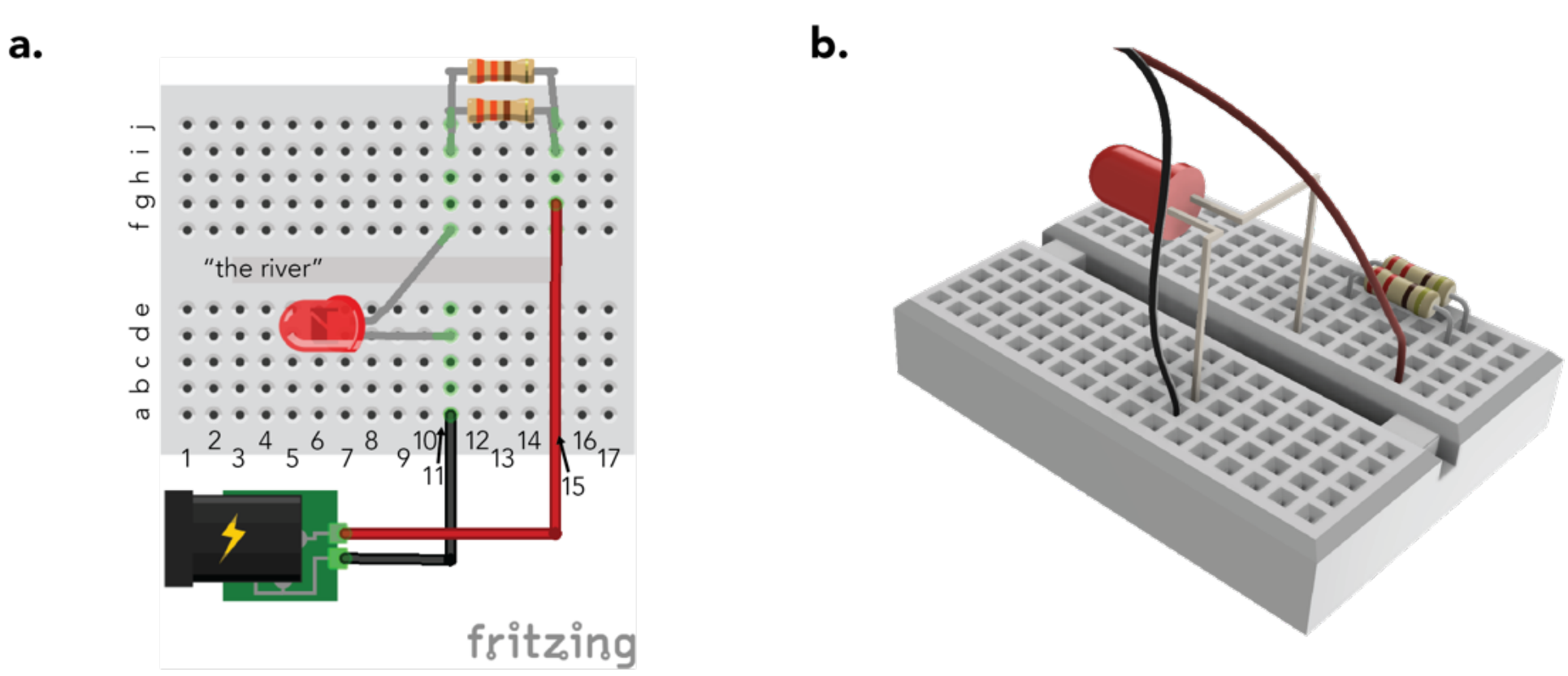

**Figure S2.6. a.** A schematic of the circuit made in Fritzing and labeled with relevant rows. The green highlights show breadboard rows that are active and connected. **b.** A rendering of the actual circuit. The red and black wires are the jumper wires, which will eventually be connected to the USB power cord.

5. Assemble the testing stage (Figure S2.7a).
    a. Slide the TSL2591 light sensor into the slot at the end of the 3D-printed testing stage next to the cuvette holder, oriented so that the text is upright (Figure S2.7b). Secure the sensor board with two #2-56 3/8” screws slotted through the lower two screw holes of the light sensor and into the bottom two holes of the testing stage (Figure S2.7c).
    b. Insert the 3D-printed pinhole between the light sensor and the edge of the cuvette holder portion of the testing stage. Note the tab that sits towards the edge of the stage; the top of the pinhole piece should be flush with the top surface of the testing stage. Secure the pinhole with two #2-56 3/8” screws slotted through the top two screw holes of the light sensor, the holes in pinhole, and the top two holes of the testing stage (Figure S2.7d).
    c. Place the breadboard with the completed LED circuit into the slot along the edge of the testing stage, with the LED sliding down into the bottom of LED slot (Figure S2.7e). The legs of the LED will need to be bent so that the LED sits at the bottom of the slot.
    d. Place the LED cover into the slot on top of the LED. Use two #4-40 1/4” screws to secure the cover in place (Figure S2.7f). Note that this piece is not symmetric; it is important that this cover is screwed so that the top is flush against the top surface of the testing stage, which will ensure that the LED is aligned with the TSL2591.
    e. Place the testing stage in the inset at the closed end of the slider rails on the base and secure it with three #4-40 1/4” screws (Figure S2.7g).
    f. Connect the free end of the 200 mm Qwiic cable from the OpenLog to the port of the light sensor closest to the OpenLog.

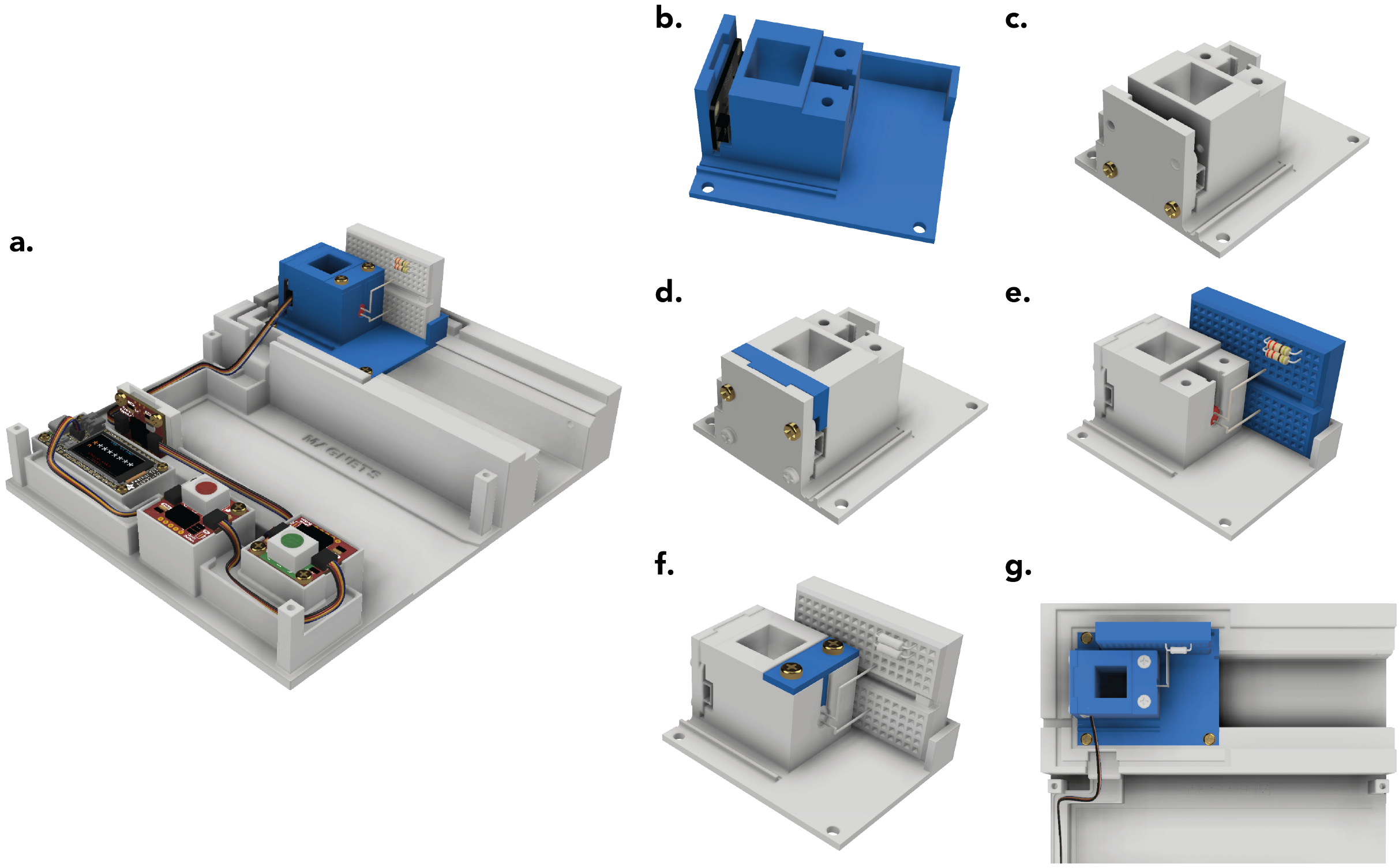

**Figure S2.7. a.** The MAP as assembled after step 5, with the addition of the testing stage. All new parts are shown in color. **b.** Insert the light sensor into the back of the testing stage, oriented right side up (i.e. text is upright). **c.** Use two #2-56 3/8” screws in the bottom two screw holes to secure the light sensor. **d.** Add the 3D-printed pinhole and secure it with two #2-56 3/8” screws through the pinhole and light sensor and into the testing stage. **e.** Add the breadboard with the completed circuit from step 3. The LED legs should be bent so that the LED can slide down to the bottom of the slot. **f.** Add the LED cover on top and secure it with two #4-40 1/4” screws. **g.** Secure the testing stage to the MAP base with three #4-40 1/4” screws and attach the Qwiic cable from the OpenLog to the light sensor.

6. Assemble the magnet storage drawer (Figure S2.8a).
    a. For each of the magnet slots, place a magnet and its corresponding spacer into the slot (Figure S2.8b). Snap the 3D-printed magnet cover into place over the slot (Figure S2.8c-d).

b. It is recommended to assemble each slot one at a time to avoid the magnets interacting with each other and to place the spacers on top of the magnets within each slot to minimize potential interactions.

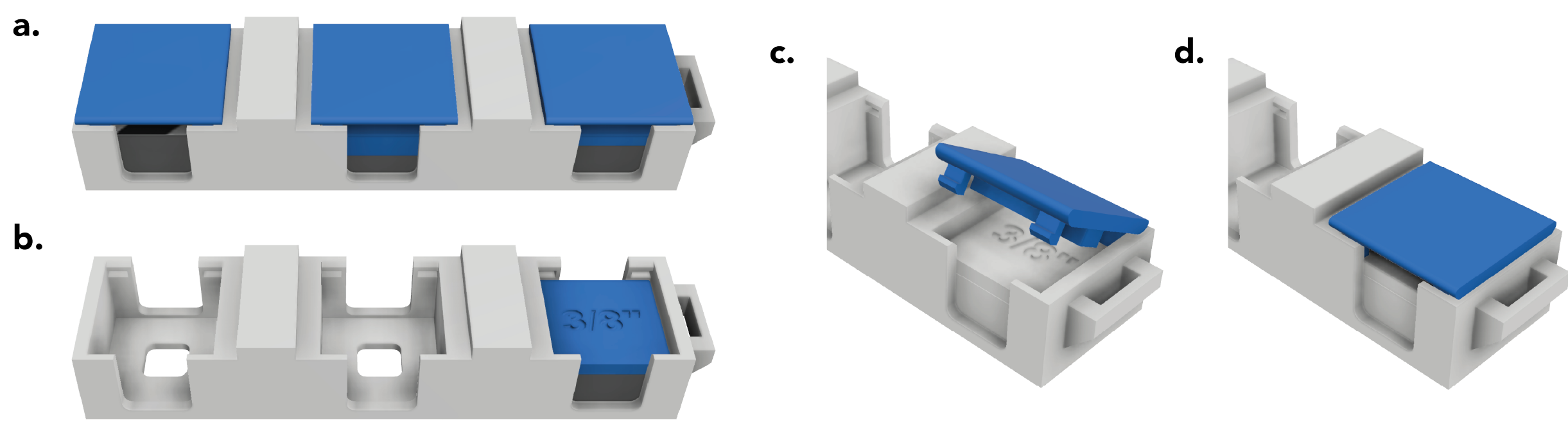

**Figure S2.8. a.** The magnetic storage drawer fully assembled with lids, spacers, and magnets. **b.** It is recommended to place the spacers on top of the magnets to minimize interactions between magnets as much as possible. **c.** To insert a lid, place the cantilevers on one side of the lid in their holes in the drawer at an angle and then **d.** push the lid down to snap it in place.

7. Assemble the full instrument (Figure S2.9a).
    a. Use one #4-40 screw to secure the 3D-printed magnet storage latch to the right side of the 3D-printed panel (Figure S2.9b).
    b. Use four #4-40 screws to secure the panel to the base (Figure S2.9c). The panel should be oriented so that the openings in the panel align with the microSD card slot of the OpenLog, the screen of the Feather, and the two buttons.
    c. Slide the magnet storage drawer into the MAP (Figure S2.9d).
    d. Place the removable 3D-printed cover on top of the testing stage so that it rests in the groove around the stage (Figure S2.9e). Ensure that the jumper wires (LED power wires) emerge from the opening on the left side of the MAP and that the Qwiic cable

between the light sensor and the OpenLog sits in the opening entering the panel (Figure S2.9e inset).

e. Place the 3D-printed magnet slider in the groove in the rails (Figure S2.9f). Optionally, a #4-40 screw may be threaded into the hole on the backside of the rails to prevent the slider from being removed from the instrument (for storage purposes only).

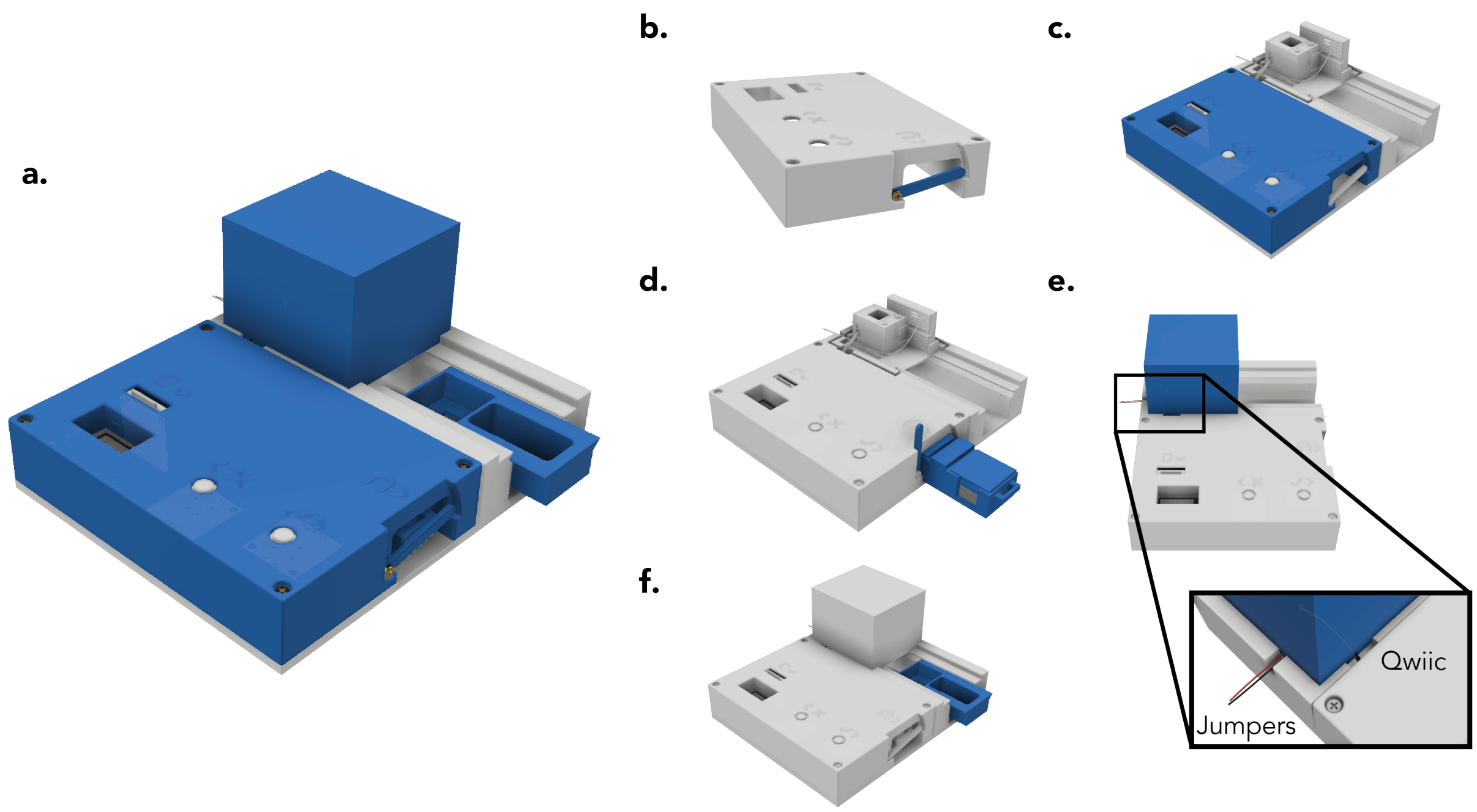

**Figure S2.9. a.** The MAP fully assembled after step 7. All new parts are shown in color. **b.** Attach the magnetic storage latch to the panel using one #4-40 1/4” screw, and **c.** then connect the panel to the MAP base with four #4-40 1/4” screws, aligning the openings with the electronics. **d.** Slide the magnet storage drawer into its storage location by rotating the latch. **e.** Place the cover over the testing stage. **(inset)** Ensure that the jumper wires and Qwiic cable both sit in their respective channels. **f.** Slide the magnet slider into the rails to finish MAP hardware assembly.

### 2.3.3. Final System Programming and Checks

1. Set up the MAP for use.
    a. Insert the microSD card into its slot.
    b. Connect the two LED jumper wires to the two power wires of the USB-to-jumper-wires cable. The jumper wire connected to the positive side of the LED should connect to the red jumper of the USB cable and the jumper wire connected to the negative side of the LED should connect to the black jumper of the USB cable. The blue and green jumpers of the USB cable will remain disconnected.
    c. Plug the end of the USB-to-jumper-wire cable into a USB block and plug the USB block into a wall outlet to power the LED.
    d. Plug the USB-C end of the USB-A-to-C cable into the Feather. Plug the other end into a laptop.
2. Upload the MAP code to the Feather and confirm the MAP is assembled properly.
    a. Open MAP_main.ino in the Arduino IDE.
    b. Select the Feather board and port and press Upload. The uploading process may take a minute.
    c. If successful, the screen on the Feather will turn on and begin counting the time since powered on, with the label “LED warming up” (Figure S2.4).
3. Remove the cover to visually confirm that the LED is emitting red light while the MAP is powered on.
    a. If it is not, it is likely that the LED is backwards in the circuit. Because LEDs are polarized components, the long leg must be on the positive side (on the side of the +5 V connection) and the short leg on the negative side (on the side of the ground connection). If the LED is not lighting up, try switching the orientation of the LED

within the circuit. If the LED still is not lighting up, check that the breadboard circuit was wired correctly.

4. Check that the TSL2591 is oriented correctly. The measured lux values reported on the screen should be around or above 15000 lux.
    a. If lux values are lower than that (i.e. below 5000), the TSL2591 should be removed and turned 180 degrees so that the text on the board is facing up when the TSL2591 is replaced in the stage (Figure S2.7b).
5. Try running a sample test, either with a cuvette filled with water or with no sample at all. Note that if you insert a cuvette of water, the measured lux value may increase.
    a. Confirm that the microSD card was inserted prior to the MAP being turned on. If not, insert the microSD card; then unplug and power on the MAP again.
    b. Set the test name and desired time. A short test (for example, 15 seconds) will be fine.
    c. Start the test and push the magnet in.
    d. Allow the test to finish. Turn the MAP off and remove the microSD card.
    e. Copy the file from the microSD card to a computer and view the text file. Confirm that the data was recorded properly. Be sure to insert the microSD card back into the MAP before it is powered on.
6. Congratulations! Your MAP system is now up and running properly. Please note the following points regarding MAP operation:
    a. Once the code has been uploaded, the MAP may be fully powered from a wall outlet.
    b. The microSD card must be inserted prior to the Feather being powered on for data to be recorded properly.

c. The LED needs time to warm up so that the incident light intensity stabilizes prior to testing. The listed SparkFun LED in the bill of materials requires an approximately 30-minute warmup period. See the Troubleshooting section for more information.

d. Note that if the microSD card must be removed between tests, the Feather must be restarted once the microSD card is reinserted. Note that this can be accomplished without turning off the LED by removing and reconnecting the USB-C cable from the Feather (to avoid the need for the LED to warm up a second time).

e. Do not use too much force when pushing the magnet slider into place; it might cause vibrations that will cause artifacts in the data.

f. When setting the test length, it must be long enough for the optical signal to plateau. Since this varies based on factors such as nanoparticle size, nanoparticle magnetic susceptibility, and solvent viscosity, we recommend running a few initial tests to determine the required minimum test duration. Fifteen minutes proved to be more than sufficient as a pilot trial for our iron oxide nanoparticles suspended in water.

g. We recommend removing the magnet storage compartment from the MAP prior to testing to avoid unintentionally introducing additional external magnetic fields to the measurement stage.

h. When adding or removing magnets in the storage drawer, we recommend being careful to only remove one cover at a time and keep the magnets themselves far away from each other until actively storing them. They will snap together rapidly, which can be a safety hazard or potentially break the magnets themselves.

i. We recommend storing the magnets with the plastic spacers on top of the magnets to avoid or minimize potential interactions when adding or removing magnets from the other slots.

j. The MAP can be sensitive to vibrations; we recommend not running equipment that can produce significant vibrations (i.e. a vortexer or centrifuge) on the same bench during data collection.

## 2.4 Troubleshooting

During our validation testing, we noticed and solved several possible issues. A short description of possible solutions to these problems are provided here.

### 2.4.1. LED Warmup and Stabilization

The optical source (LED) must stabilize before beginning to use the system; otherwise, artifacts will be present in the measurement as the initial incident intensity may rise independent of the signal from the MNPs. For the specified LED, SparkFun COM-00528, the stabilization time was about 30 minutes (Figure S2.10), but we recommend running an initial screen of an hour after the MAP is first assembled to determine how long the LED needs to warm up as this may vary for each LED. The warmup screen shown when the MAP is first powered can also be used to manually monitor the LED brightness as it warms up.

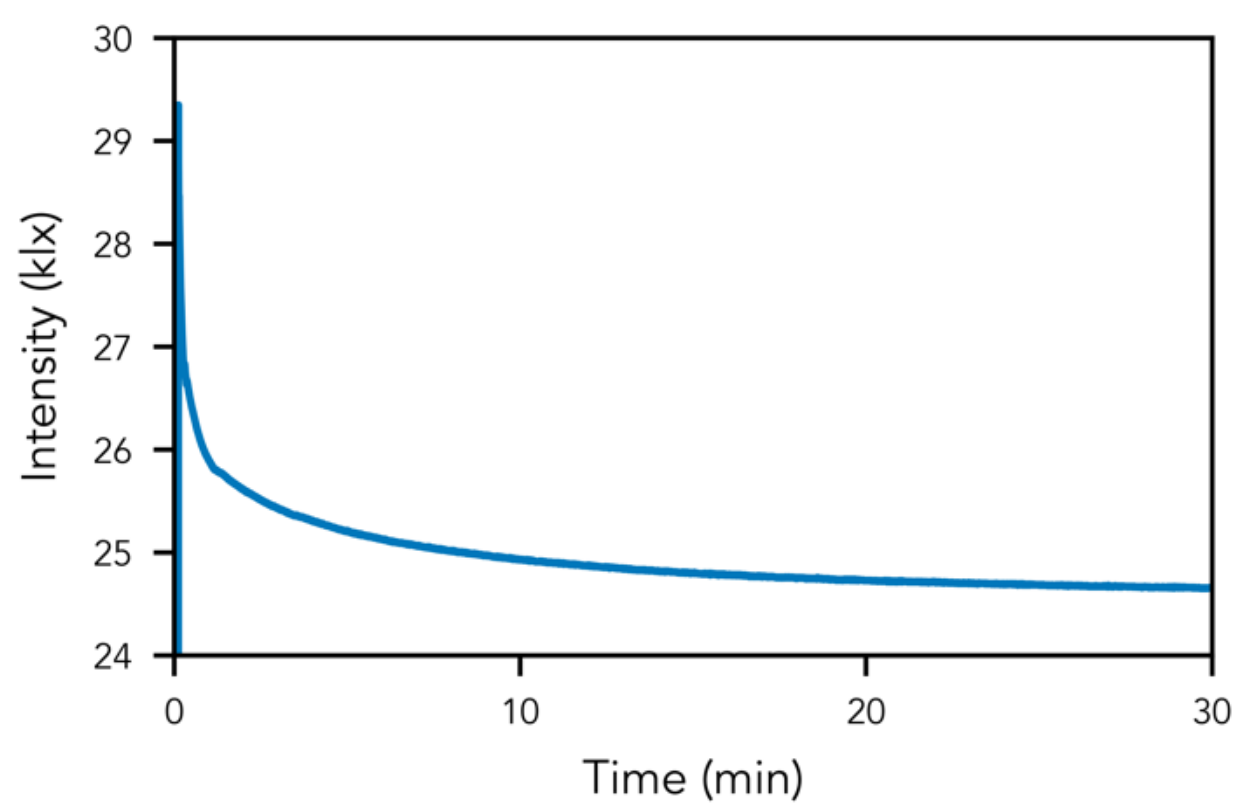

**Figure S2.10.** Measured LED intensity over time. The signal plateaus after about 30 minutes, in agreeance with the LED datasheet. Note that the initial jump is the LED being turned on after 5 seconds. The raw output of the MAP is shown here and is thus shown in photometric units.

We found that the basic two resistor current control circuit given in Figure S2.2 provided sufficient light intensity stability for our measurements. However, there are other options to obtain constant current control. One notable option is finding an integrated constant current driver, such as the CL2 component by Microchip, which comes in a through-hole package that easily integrates into the current design (Figure S2.11). Alternatively, the user may be able to construct a constant current circuit using bipolar junction transistors or metal oxide semiconductor field-effect transistors. These more complex current control options can provide further light intensity stability.

It should also be noted that the below measurement further confirms that the system does not undergo any thermal or heating effects.

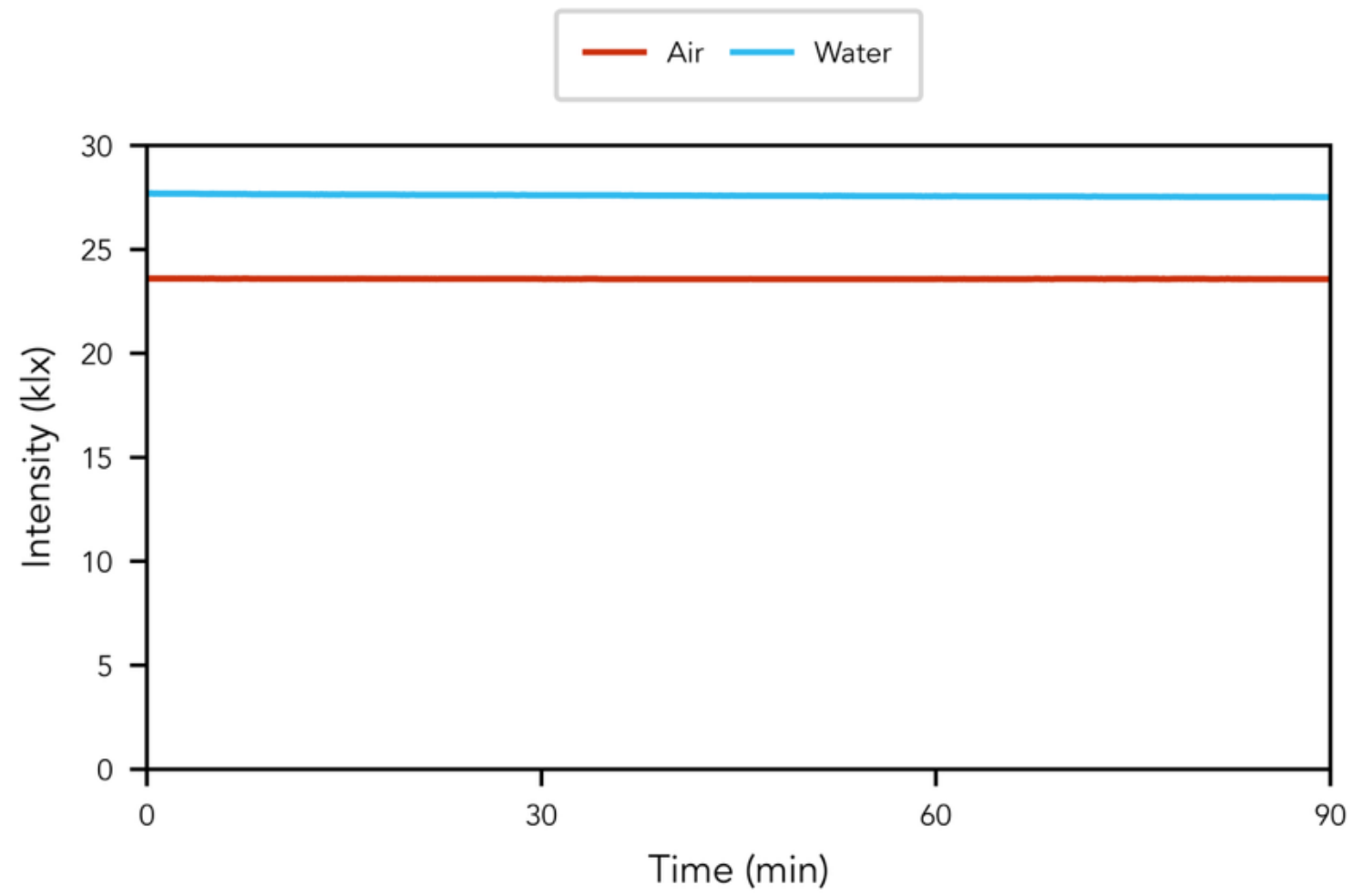

**Figure S2.11.** Measured LED intensity over an extended period of time (90 minutes) for both empty optical paths (no cuvette, red) and a cuvette of water (blue) with the CL2 current controller.

## 2.4.2. Low Optical Signal

If the light intensity signal is very low (less than 10k lux) without a solution of nanoparticles, there are a few possible causes.

If the MAP has just been assembled, it is possible that the light sensor is not oriented properly. It should be oriented so that the text on the board is right side up, as described in Step 5a of Hardware Assembly (Figure S2.7b).

If the LED light seems dim, it is possible that the current through the LED is not sufficiently high. Measure either the voltage through the LED with a multimeter and calculate the LED current or directly measure the current through the LED and compare it to the manufacturer's listed forward current for the LED. If it is lower, choose appropriate resistor values based on eq S2.1 and replace the resistors in the LED circuit (Figure S2.2). If the LED brightness is lower despite the ideal current flowing through the LED, then the LED may need to be replaced.

### 2.4.3. Nanoparticle Settling and Clumping

The measurement is dependent on the interaction of the MNPs with the optical field. Therefore, it is important that the nanoparticles are uniformly distributed throughout the optical field. As has been discussed in prior works, there are several mechanisms that can lead to non-uniform distributions, resulting in measurement artifacts.[2,3]

Among the potential mechanisms, the most commonly observed effects during the present work were MNP settling and clumping in the solutions. While particle clumping was able to be avoided by using optimized sample concentrations, MNP settling required active monitoring and mitigation controls. Specifically, whenever a solution is tested in the MAP, it was thoroughly shaken, vortexed, or sonicated to completely re-disperse the MNPs in solution prior to taking a measurement. This is particularly true when running a repeated test, as MNP clumping may occur when the MNPs are aggregated at the bottom of the solution.